\documentclass{aa}  
\usepackage{graphicx}
\usepackage{subfig}

\def\ox{[\ion{O}{II}]3727}

\usepackage{txfonts}
\begin{document}

   \title{The OTELO survey}

   \subtitle{Faint end of the luminosity function of \ox\, emitters at $\langle$z$\rangle$ $=$ 1.43}

   \author{Bernab\'e Cedr\'es
          \inst{1,2}
          \and
          \'Angel Bongiovanni \inst{3,4}
          \and
          Miguel Cervi\~no \inst{5}
          \and
          Jakub Nadolny\inst{1,2}
          \and
          Jordi Cepa \inst{1,2,4}
          \and
          Jos\'e A. de Diego\inst{6,1}
          \and
          Ana Mar{\'i}a P\'erez Garc{\'i}a\inst{5,4}
          \and
          Jes\'us Gallego\inst{7}
          \and 
          Maritza A. Lara-L\'opez\inst{8}
          \and 
          Miguel S\'anchez-Portal\inst{3,4}
          \and
          J. Ignacio Gonz\'alez-Serrano\inst{9,4}
          \and
          Emilio J. Alfaro\inst{10}
          \and
          Roc\'io Navarro Mart\'inez\inst{4}
          \and
          Ricardo P\'erez Mart\'inez\inst{11,4}
          \and
          J. Jes\'us Gonz\'alez\inst{12}
          \and 
          Carmen P. Padilla Torres\inst{1,2,13}
          \and 
          H\'ector O. Casta\~neda\inst{14,\dagger}
          \and
          Mauro Gonz\'alez\inst{1,2}
          }

   \institute{Instituto de Astrof{\'i}sica de Canarias (IAC), 38200 La Laguna, Tenerife, Spain\\
              \email{bcedres@iac.es}
         \and
             Departamento de Astrof{\'i}sica, Universidad de La Laguna (ULL), 38205 La Laguna, Tenerife, Spain
        \and
            Instituto de Radioastonom{\'i}a Milim\'etrica (IRAM), Av. Divina Pastora 7, N\'ucleo Central 18012, Granada, Spain
        \and
            Asociaci\'on Astrof{\'i}sica para la Promoci\'on de la Investigaci\'on, Instrumentaci\'on y su Desarrollo, ASPID, 38205 La Laguna, Tenerife, Spain
        \and
            Centro de Astrobiolog{\'i}a (CSIC/INTA), 28692 ESAC Campus, Villanueva de la Cañada, Madrid, Spain
        \and
            Instituto de Astronom\'ia, Universidad Nacional Aut\'onoma de M\'exico, Apdo. Postal 70-264, 04510 Ciudad de M\'exico, Mexico
        \and
            Departamento de F\'isica de la Tierra y Astrof\'isica, Instituto de F\'isica de Part\'iculas y del Cosmos, IPARCOS. Universidad Complutense de Madrid, E-28040, Madrid, Spain
        \and 
            DARK, Niels Bohr Institute, University of Copenhagen, Lyngbyvej 2, Copenhagen DK-2100, Denmark
        \and 
            Instituto de F\'isica de Cantabria (CSIC-Universidad de Cantabria), E-39005 Santander, Spain
        \and 
            Instituto de Astrof\'isica de Andaluc\'ia, CSIC, E-18080, Granada, Spain
        \and 
            ISDEFE for European Space Astronomy Centre (ESAC)/ESA, P.O. Box 78, E-28690, Villanueva de la Ca\~nada, Madrid, Spain
        \and 
            Instituto de Astronom\'ia, Universidad Nacional Aut\'onoma de M\'exico, 04510 Ciudad de M\'exico, Mexico
        \and 
            Fundaci\'on Galileo Galilei, Telescopio Nazionale Galileo,
            Rambla Jos\'e Ana Fern\'andez P\'erez, 7, 38712 Bre\~na Baja, Tenerife, Spain
        \and
            Departamento de F\'isica, Escuela Superior de F\'isica y Matem\'aticas, Instituto Polit\'ecnico Nacional, 07738 Ciudad de M\'exico, Mexico 
            }
   \date{}

 
  \abstract
   {}
   {In this paper, we aim to study the main properties and luminosity function (LF) of the \ox\, emitters detected in the OTELO survey in order to characterise the star formation processes in low-mass galaxies at $z\sim1.43$ and to constrain the faint-end of the LF.}
   {Here, we describe the selection method and analysis of the emitters obtained from narrow-band scanning techniques. In addition, we present several relevant properties of the emitters and discuss the selection biases and uncertainties in the determination of the LF and the star formation rate density (SFRD).}
   {We confirmed a total of 60 sources from a preliminary list of 332 candidates  as \ox\, emitters. Approximately 93\% of the emitters have masses in the range of $10^{8}<M_{*}/{\rm M_{\odot}}<10^{9}$. All of our emitters are classified as late-type galaxies, with a lower value of $(u-v)$\, when compared with the rest of the emitters of the OTELO survey.  We find that the cosmic variance strongly affects the normalisation ($\phi^*$) of the LF and explains the discrepancy of our results when compared with 
   those obtained from surveys of much larger volumes. However, we are able to determine the faint-end slope of the LF, namely, $\alpha=-1.42\pm0.06$, by sampling the LF down to $\sim1$\,dex lower than in previous works. We present our calculation of the SFRD of our sample and compare it to the value obtained in previous studies from the literature.}
   {}


   \maketitle
%

\section{Introduction}
 The luminosity function (LF) is a powerful tool used in characterising the distribution of star-forming galaxies at cosmological scales. By observing different emission lines, it is possible to trace the star-formation activity at different redshifts and, thus, at different epochs of the evolution of galaxies. Taking into account that the strongest emission line in a star-forming galaxy is usually H$\alpha$, this line can be considered the first probe for LFs (see e.g. \citealt{lilly}, where their redshift sample ranges at $0<z<1.3$; \citealt{geach} at $z=2.23$; \citealt{sobralref} at $0.40<z<2.23$; \citealt{sobralref2} at $z=0.8$; \citealt{khos} at $z=0.47;$  \citealt{ly} at $0.08<z<0.40;$ or more recently, \citealt{hayashi2} at $0.09<z<0.48;$ and \citealt{harish} at $z\sim0.62$). 

Other lines may be used when looking for other windows of observation in redshift, where H$\alpha$ is not available. For example, the next most luminous line of the Balmer series, H$\beta$ was employed for determining the LF
at $z\sim0.9$ by \cite{rocio}. Other works have used the combination of H$\beta$ and [\ion{O}{iii}], for example \cite{debarros} reaching as far as $z\sim 8$; \cite{khos2} up to $0.84<z<3.24$; and \cite{sobralref2} up to $z=1.4$. In addition, there are a series of works were the LF is obtained by employing the doublet [\ion{O}{iii}] $\lambda\lambda$4959,5007 alone, for example, in \cite{ly} at $0.41<z<0.84$; \cite{lum} at $z\sim0.83$; \cite{khos} at $0.91<z<1.10$;  or \cite{hayashi2} at $0.05<z<0.94$.

In order to sample the LF at very high redshift regimes, the Lyman $\alpha$\, emitters are the objects of choice. For example, \cite{ouchily}, \cite{sobrally}, and \cite{herenzly} obtained LFs at $3.1<z<5.7$,\, $2<z<6,$\, and $3<z<6$, respectively.

Going back to intermediate redshift regimes, the [\ion{O}{II}]$\lambda\lambda$ 3726,29 doublet, it is also a good candidate for the study of the LF. It is located in the bluemost part of the spectrum in rest frame, when compared with H$\alpha$, [\ion{O}{iii}]$\lambda\lambda$ 4959, 5007, and H$\beta$, allowing us to reach larger values of the redshift from ground observations and avoiding the less transparent parts of the atmosphere in the infrared regime. The LF has been derived for this doublet by a number of authors who have employed spectroscopy or narrow-band techniques (\citealt{gallego} for the local universe; \citealt{khos2} for $1.47<z<4.69$; \citealt{khos} for $1.57<z<1.9$; \citealt{drake} for $0.35<z<1.64$; \citealt{ly} for $0.89<z<1.47$; \citealt{comparat} for $0.1<z<1.65$; or \citealt{hayashi2} for $0.41<z<1.60$). It has also been simulated from $z=0.1$ to $z=3.0$  by \cite{park}.

Nevertheless, at $z>1,$ the faint end of the \ox\ LF has not been properly observed. Indeed, almost all studies have no been capable of going beyond a $\log(L\ox)\simeq41$ [erg\,s$^{-1}$] in the best of cases (see e.g. the discussion in \citealt{khos2}). We believe that it is paramount to reach as low as possible in terms of luminosity in order to study the evolution of low-mass galaxies. For example, according to \cite{sobral}, the environment may be responsible for the slope of the faint end of the LF, being steeper for low-density regions and shallower for high-density fields. Moreover, the correct determination of the slope of the faint end of the LF influences the integration of the LF itself and, therefore, may vary the determination of the cosmic star formation history (SFH).

In this work, we take advantage of the sources detected in the OTELO survey (see \citealt{otelo}) to generate a catalogue of \ox\ emitters at $z\sim1.43$. We study the physical properties of these sources and we extend the study of the faint end of the LF for the \ox\, line to constrain the faint end slope and. In addition, we obtain the star formation rate density (SFRD) at this redshift.

The paper is organised as follows. In Section 2, we describe the OTELO survey and the selection of the sources. In Section 3, we present several of the derived properties from the data, including the extinction, the line fluxes, the morphology, and the stellar masses. In Section 4, we discuss the derived equivalent width of the emitters, in addition to studying a handful of high equivalent-width galaxies and comparing our results with those from the literature. Section 5 is devoted to our derived LF and the SFRD, along with a comparison with similar data from the literature. In Section 6, we summarise our main conclusions.
For all the calculations carried out in this work, we assumed a standard $\Lambda$CDM cosmology with $\Omega_{\Lambda}=0.7$,\, $\Omega_{m}=0.3,$ and $h_0=0.7$.

\section{The OTELO survey data and selection of the sources}
\label{sec:sample}
\subsection{The OTELO survey}
In this work, we take advantage of the OTELO survey catalogue and its data products (for an in-depth description, see \citealt{otelo}). As a summary, the OTELO survey is a 2D spectroscopic blind survey covering a region of 56 arcmin$^2$ of the Extended Groth Field. It has a resolution of R$\sim$700, covering a 230\,\AA\, of a selected atmospheric window (around 8950-9300 \AA) that is relatively free of sky emission lines. The catalogue contains a total of 11237 sources and it is 50\% complete at an AB magnitude of 26.38. From these sources, 5322 preliminary emission line candidates were selected according to the following criteria: (i) at least one point of the pseudo-spectrum lies above a value defined by $f_c+2\times\sigma_c$, where $fc$ is the flux of the pseudo-continuum, defined as the median flux of the pseudo-spectra, and $\sigma_c$ is the root of the averaged square deviation from $f_c$ of the entire pseudo-spectrum; and (ii) there is an adjacent point with a flux density above $f_c+\sigma_c$ (\citealt{lum}). 

The catalogue also presents complementary reprocessed data from several other surveys. This ancillary data covers from X-ray to far-infrared (FIR), both photometric and spectroscopic.
Photometric redshifts (known as z$_{\rm best}$) were determined through the LePhare code (\citealt{ilbert}), using libraries for normal and starburst galaxies, QSOs, Seyferts and stars, and fitted to the OTELO photometric data. The z$_{\rm best}$ is designated as the \texttt{z\_BEST\_deepY} from \cite{otelo}\footnote{It means, the best galaxy/starburst solution including the OTELO survey data as an additional photometric point.}

\subsection{Selection of the sources}

Taking into account that the [\ion{O}{II}]$\lambda\lambda$ 3727,29 doublet must lie comfortably within the OTELO observing window of 8950-9300 \AA, we selected, from the preliminary emission line candidates list, all the sources in the range of $1.40<z_{\rm best}<1.50$.
Using this criteria, we selected a first subset of 332 emitters.

Following the work carried out by \cite{lum} for the [\ion{O}{iii}]{\ensuremath{\lambda}}4959,5007 emitters, and in order to produce a robust list of candidates, this first selection was visually screened by at least three researchers, looking for misidentifications, artifacts, and any other anomalies, such as truncated lines, overlapping sources, or multiple unlikely emission lines structures which appear in low signal-to-noise pseudo-spectra. From these cases, only those sources with truncated line in pseudo-spectra are considered true--positive emitters, despite their having been excluded from the final sample. However, this effect is included as a part of the sample completeness estimation described in Section \ref{sec:compl}. The resulting list was composed of 60 candidates. Table \ref{sel} summarises the classification determined for each one of the 332 candidates.

\begin{table*}
\centering
\renewcommand{\arraystretch}{1.5}
\begin{tabular}{c c c}
\hline
 & Number of objects & Percentage\\
\hline
Total candidates & 332 & 100\%\\
Selected \ox\, sources & 60 & 18\%\\
\hline\hline
\multicolumn{1}{c}{Rejection reason}\\
\hline
Multiple structures in low S/N pseudo-spectra & 136 & 41\%\\
Overlapping sources & 62 & 18.7\%\\
Truncated lines & 55 & 16.6\%\\
Misidentifications and other causes & 19 & 5.7\%\\
\hline             
\end{tabular}
\caption{Classification summary of emission line candidates.}
\label{sel}
\end{table*}

Employing the online tool developed for the OTELO database\footnote{\tt http://research.iac.es/proyecto/otelo/pages/otelo.php}, at least three researchers selected a new $z_{\rm\, guess}$ by visual inspection that is closer to the real value of the redshift than the $z_{\rm best}$. This $z_{\rm\, guess}$ was employed as an input for the deconvolution program. To obtain the line fluxes from the pseudo-spectra, we employed the so-called inverse deconvolution, introduced in \cite{belito} and developed in detail in \cite{kuba1}. Briefly, a series of simulations (on the order of $10^6$) are performed and then a $\chi ^2$ minimisation method is used to obtain the probability density function of the possible solutions for each source, taking the mode of the distribution as the best solution and the confidence intervals around the mode of all the associated parameters: fluxes, equivalent widths, redshifts, and continuum fluxes. In this paper, we quoted the uncertainties as the interval covering the 68\% of such confidence interval. After all the candidates were deconvolved, a new visual inspection was carried out, this time to check the goodness of the fitted models. A catalogue  with flux, equivalent width, redshift, and continuum flux was then generated.
Following the deconvolution, $z_{\rm\, OTELO}$  is derived from the OTELO pseudo-spectra for each emitter. This new redshift is a refinement of $z_{\rm\, guess}$.

Considering that we may have little contamination by AGNs  due to our selection criteria in redshift (we only took galaxies fitted by a starburst or star-forming galaxies template), we decided not to correct for AGN contamination. Moreover, \cite{drake} infer a contamination fraction about six percent in a sample of emitters at $z=1.6$ and they suggest that AGN activity is usually associated with star-forming processes, so removing AGNs would lead to an over-correction.

We ran {\tt LePhare} code a second time, but with redshift fixed to $z_{\rm\, OTELO}$ and restricted to star-forming galaxies templates with the aims of obtain a more accurate value of reddening. The applied galaxy templates are
\cite{kinney} and \cite{bc03} -- with a continuous star formation law.

\begin{figure}[]
   \centering
   \includegraphics[width=\hsize]{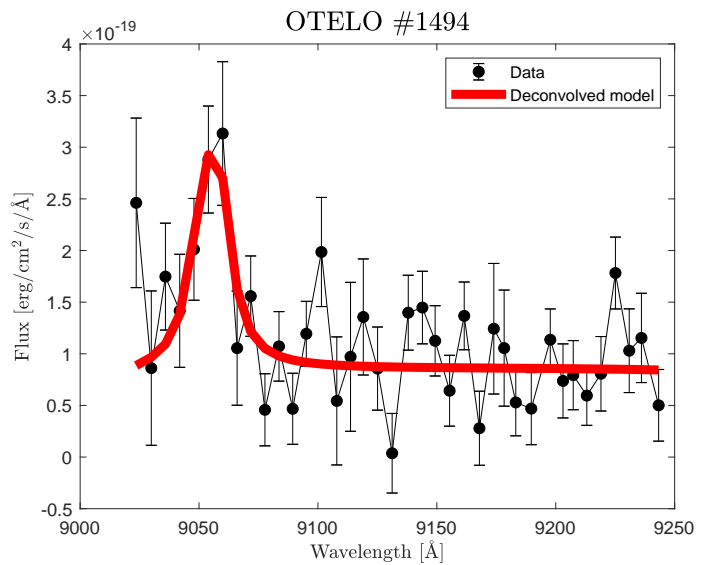}
      \caption{Pseudo-spectra and deconvolved spectra for the {\tt id:1494} emitter. We note the noisy aspect of the pseudo-spectra. However, in these circumstances, the model is capable of obtaining a good measure of the \ox\, lines. 
              }
         \label{esp1494}
   \end{figure}

\begin{figure}[]
   \centering
   \includegraphics[width=\hsize]{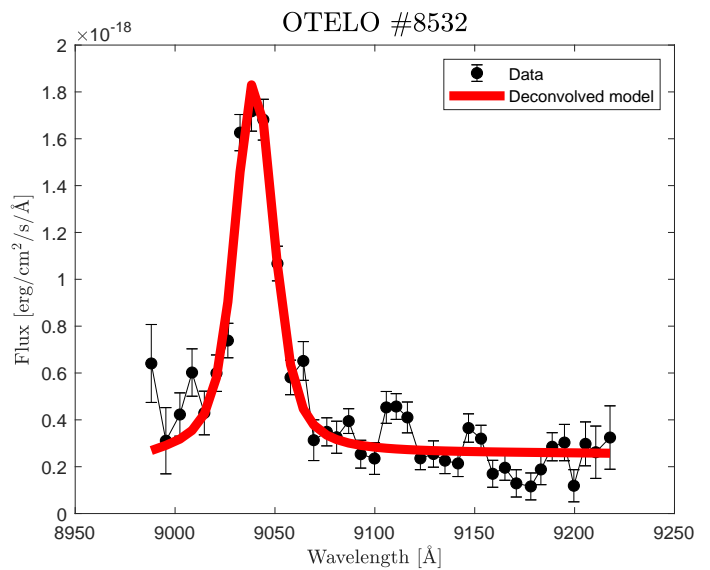}
      \caption{Pseudo-spectra and deconvolved spectra for the {\tt id:8532} emitter. The signal-to-noise ratio is better than in Fig. \ref{esp1494}
              }
         \label{esp5090}
   \end{figure}

One example of a deconvolved pseudo-spectrum is shown in Fig. \ref{esp1494}. Even if the source is noisy, the fitted model is good enough to extract the required information. On the other hand, in Fig. \ref{esp5090}, we  represent the pseudo-spectrum and deconvolving lines for an object with a much better signal-to-noise ratio (emitter {\tt id:8532}). In Fig.\ref{primera} in the appendix, we represent the pseudo-spectra of all selected emitters in the catalogue.


\section{Derived properties}
\label{sec:properties}

\subsection {Extinction correction}
The obtained fluxes were corrected via internal extinction following the law presented in \cite{calzetti} and employing the reddening $E(B-V)$ obtained as an output from the second fitting of the emitters using the {\tt LePhare} code (see \citealt{otelo}). We also took into account the inherent extinction of the templates in \cite{kinney} when deriving our value for the extinction.

\begin{figure}[]
   \centering
   \includegraphics[width=\hsize]{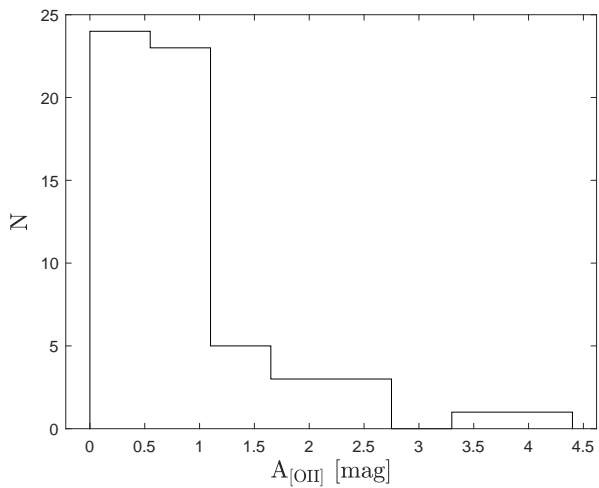}
      \caption{Histogram of the value \textbf{of $A_{\rm \ox}$} for all the emitters in the catalogue.}
         \label{histoext}
 \end{figure}

In Fig. \ref{histoext}, we have represent the histogram of the derived value of $A_{\rm \ox}$ for all the emitters. The mean result is $\langle A_{\rm \ox} \rangle=0.88$\,mag, which is equivalent to $\langle A_{\rm H\alpha} \rangle =0.5$\,mag, with a standard deviation of $\sigma=0.5$. This is lower than the normally employed $A_{\rm H\alpha}\sim1$\,mag. This value has been experimentally obtained, for example, by \cite{ibaha}, who reported an $A_{\rm H\alpha}=1.0\pm0.2$\,mag for objects at $z=1.47,$ employing a combination of far-IR and H$\alpha$ data. On the other hand, our result seems to agreed with the trend observed in other works and cited in \cite{hayashi}, where the authors found that \ox\,emitters selected by narrow-band (NB) techniques are likely to be dust-poor systems. For example, both \cite{hayashi} and \cite{khos} give $A_{\rm H\alpha}\sim0.35$\,mag for emitters at $z\sim1.47$, \cite{hayashi3} found $A_{\rm H\alpha}=0.63$\,mag for $z=1.5$, and \cite{khos2} obtains $A_{\rm H\alpha}=0.55\pm0.12$\,mag for emitters at $z=1.59$. Nevertheless, the large dispersion of the mean value derived makes difficult to obtain a clear conclusion about the absorption nature of our emitters. As  \cite{garnha} have pointed out, the individual extinctions vary significantly between galaxies, which is clearly visible in the histogram of Fig. \ref{histoext}.

\subsection{Line fluxes}

\begin{figure}[]
   \centering
   \includegraphics[width=\hsize]{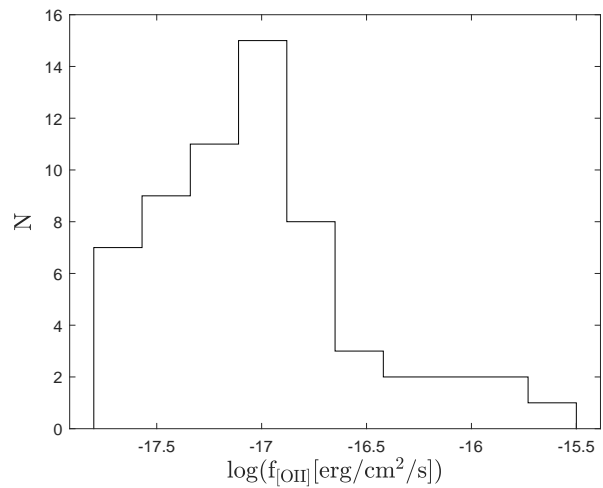}
      \caption{Histogram of the line flux (extinction corrected) observed for all the selected emitters.
              }
         \label{histoflux}
\end{figure}

In Fig. \ref{histoflux}, we represent the histogram of the measured flux for all the emitters, corrected for extinction. The lowest flux is $\log (f_{\ox\,\mathrm{min}})=-17.72$\,[erg/cm$^2$/s], the maximum is $\log (f_{\ox\,\mathrm{max}})=-15.56$\,[erg/cm$^2$/s], and the mean value is $\langle \log(f_{\ox}) \rangle =-17.02$\,[erg/cm$^2$/s]. We can observe that only the 15\% of the emitters have a flux higher than $\log ( f_{\ox})=-16.75$\,[erg/cm$^2$/s].

\subsection {Morphology}

Following \cite{kauffmann}, we define the concentration index $C$ as the ratio of the radius enclosing the 90 percent of the luminosity in a defined band over the radius enclosing the 50 percent of the luminosity in the same band. In our case, we employed the high-resolution images from the bands F814W(I) and the F606W(V) from HST/ACS. However, only 45 emitters from our sample had data in those filters. The median value for the I filter is $C_{I}=1.92$. However, as shown, for example, in \cite{strateva} and in \cite{morfo}, this parameter may only serve to establish a crude classification between early-type (ET) and late-type (LT) galaxies.

In \cite{morfo} the intensity spatial distribution of the emitters from the OTELO survey is fitted using a parametric form in a \cite{sersic} profile, defined as:
\begin{equation}
I (r) = I_{e} \exp \{-b_{n}[(r/r_{e})^{1/n} - 1]\},
\end{equation}
where $r_{e}$ is the effective radius (i.e. radius containing 50\% of total flux), $I_{e}$ is the intensity at $r_{e}$ in the filter selected (F814W or F606W from HST/ACS in this case), $n$ is the S\'ersic index, and $b_{n}$ is a function of $n$ as defined in \cite{ciotti}. For a detailed description of the fitting process, see \cite{morfo}.
The obtained S\'ersic index $n$ is an useful tool in order to classify the morphology of the emitting sources. For disc-dominated LT galaxies, the expected value of this parameter in the I filter is $n_\mathrm {I}<4$ (\citealt{morfo}).
The median values obtained for the \ox emitters is $n_{\rm I}=2.1\pm1.7$. This value appears to be slightly larger than the mean value obtained for all the emitters with enough data for the OTELO survey \cite{morfo} ($n_\mathrm{I}=1.3$). However most of the emitters show $n_\mathrm {I}$ values below the threshold for LT galaxies. On the other hand, our results are well within the range obtained by \cite{paumor}, who give  $n_{\rm I}=1.16^{+1.62}_{-0.72}$\, for the emitters at $z=1.47$ from High-Z Emission Line Survey (HiZELS). 

\begin{figure*}[]
   \centering
   \includegraphics[width=17cm]{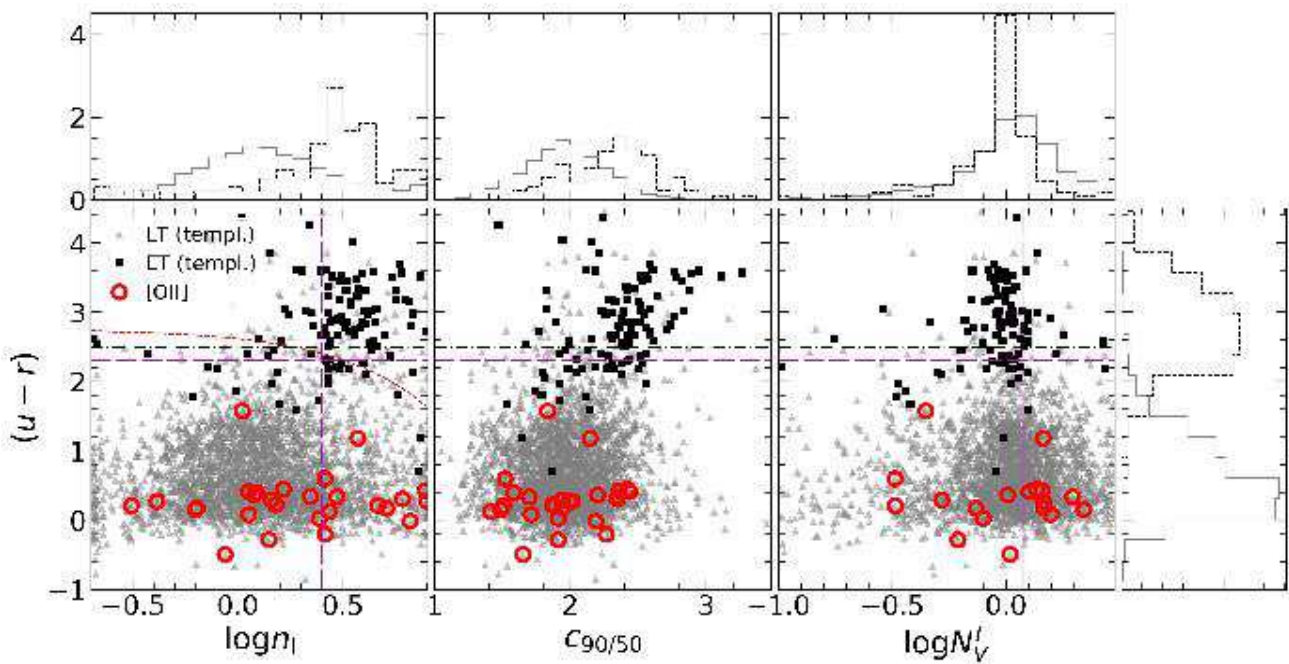}
      \caption{Observed $(u-r)$ colour versus morphological parameters for all the sources of OTELO. From left to right: Logarithm of the S\'ersic index in the filter F814W (I), concentration index for the I filter too, and the logarithm of the wavelength-dependent ratio of the S\'ersic indices for I and V (F606W).Triangles and solid lines in grey (histograms) represent LT, squares and dashed lines in black (histograms) represent ET galaxies. Red circles represent the \ox emitters from our sample with HST/ACS ancillary data.Top histograms corresponds to the respective value, as indicated in x-axis label, while right-hand histogram show $(u-r)$ distribution. All histograms represents density distributions. Horizontal dot-dashed line in black shows $(u-r) = 2.5$. Red dashed line shows the results from \cite{dnn}. Dashed lines in magenta represent limits from \cite{marina}: horizontal cut in $(u-r) = 2.3$, while vertical dashed-lines in magenta represent $\log(n_\mathrm{I}) = 0.4$ and $\log(N^{I}_{V}) = 0.08$. Figure adapted from \cite{morfo}.}
         \label{urmor}
\end{figure*}

We also derived the ratio of S\'ersic indices in both bands $N^{I}_{V}=n_{I}/n_{V}$  (introduced by \cite{marina}. This parameter is sensitive to the internal structure of the galaxy when paired with a colour term. In this case, the median value obtained was $N^{I}_{V}=1.3\pm0.7$, which is well inside the error margin for the whole OTELO database, where the value was $N^{I}_{V}=1.1\pm0.4$ for the LT galaxies (\citealt{morfo}). In Fig. \ref{urmor}, we reproduce the results from \cite{morfo}.

The red dashed line marks the linear discriminant analysis developed by \cite{dnn}, employing deep learning methods for morphology classification. In the figure we can see that all the selected emitters (indicated by red circles) are comfortably distributed in the late-type locus. Moreover, almost all detected emitters are somewhat bluer than the mean of the all OTELO ET galaxies. 

\subsection{Derived masses}

Following \cite{sanjuan} we obtained an estimation of the stellar mass for 48 of the emitters in the catalogue (for a detailed description of the process, see \citealt{lum}). The remaining 12 emitters with no stellar mass determination presented a large error in several of the ancillary bands that made it impossible to derive a meaningful value for the mass.

\begin{figure}[]
   \centering
   \includegraphics[width=\hsize]{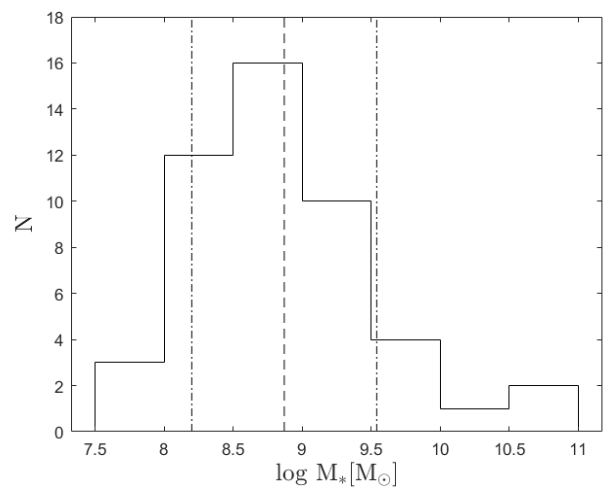}
      \caption{Histogram of the derived stellar mass (in solar masses) for the emitters. The dashed line marks the position of the median value at $\log\,({M_{*}}/{\rm M_\odot})=8.87$ and the point-dashed lines mark $\pm1\sigma$ over the median value.}
         \label{histomasa}
 \end{figure}

The median value derived for the stellar mass was $\log\,({M_{*}}/{\rm M_\odot})=8.87$, with a dispersion of $\log\,({M_{*}}/{\rm M_\odot})=0.67$. The largest mass derived was $\log\,({M_{*}}/{\rm M_\odot})=10.93$, meanwhile, the lowest mass was $\log\,({M_{*}}/{\rm M_\odot})=7.89$. If we define the low-mass population as the galaxies with $M_{*}<10^{10}\, {\rm M_{\odot}}$ (\citealt{lum}), only three emitters present are out of the low-mass regime. This means that about 93\% of the emitters belong to the low-mass criterion. In \cite{lum}, the percentage of low-mass galaxies reached 87\% in a catalogue composed by 171 emitters at a $\langle z\rangle$ $=$ 0.83; meanwhile for \cite{kuba1}, the whole sample of H$\alpha$ emitters was in the low-mass regime, with a 63\% being very low-mass galaxies ($M_{*}<10^{9}\, {\rm M_{\odot}}$). These two results seems compatible with the stellar masses derived in our sample, indicating that the OTELO survey favours the detection of low-mass galaxies. This is a direct consequence of an ultra-deep pencil-beam designed survey, as the OTELO survey is, in fact (\citealt{otelo}). The mass distribution for the emitters is represented in Fig. \ref{histomasa}.

\section{\ox\, Equivalent width}
 
\begin{figure}[]
   \centering
   \includegraphics[width=\hsize]{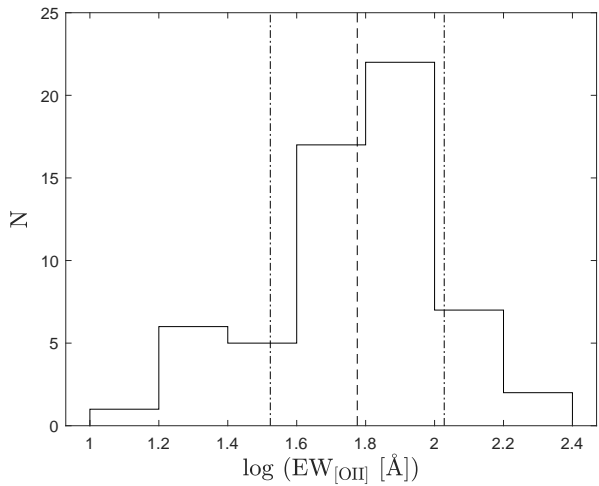}
      \caption{Histogram of the line equivalent width for all the selected \ox\ emitters in OTELO. The dashed line marks the position of the median value, and the point dashed lines mark $\pm1\sigma$ over the median value.}
         \label{histoew}
 \end{figure}

One of the parameters derived from the deconvolution is the equivalent width of the \ox\, doublet. 
In Fig. \ref{histoew}, we represent the histogram of the logarithm of $EW_{\rm \ox}$. The minimum value obtained in our emitters is 10$\pm$5\, \AA and the maximum is 185$\pm$16\, \AA. The median value is $EW_{\rm \ox}$=65\, \AA, with a $1\sigma$ dispersion of $EW_{\rm \ox}$=37\, \AA. Following \cite{marinanuestra1}, in $EW_{\rm H\alpha}$, the minimum value for detection with a probability $p\leq0.95$ was $\sim10.5$\, \AA. In order to estimate our detection limit, we have to translate our limit in $EW_{\rm H\alpha}$ into a limit in $EW_{\rm \ox}$. For this reason, we can employ the relationship given by \cite{robertin}, which is summarised in Eq. \ref{equi}: 
\begin{equation}
    {EW_{\ox}}=0.4\times {EW_{(\mathrm{H}\alpha +[\ion{N}{ii}])}}.
    \label{equi}
\end{equation}
However, this relationship was obtained for nearby galaxies and it also presents a large dispersion for galaxies with very strong emission lines. Nevertheless, employing \cite{lam} data, we were able to calculate this relationship for galaxies with $0.2<z<1.0$. In this case, the resultant slope was $0.5\pm0.1$, very similar to the 0.4 value from \cite{robertin} relationship. Based on this premise, the minimum value we would be able to detect with $p\leq0.95$ is $\sim4.2$\,\AA. In this case, it is clear that all our emitters are well over this threshold.

 \begin{table*}[]
     \centering
     \renewcommand{\arraystretch}{1.5}
     \begin{tabular}{c c c c c c}
         \hline
         OTELO ID. \# & $EW$(\AA) & $C_{\rm I}$ & $\log$ $n_{\rm I}$ & $\log N^{\rm I}_{\rm v}$ & $\log (M_*{\rm [M_\odot]})$\\
         \hline
         3077 & 153$\pm36$ & N/A & N/A & N/A & 8.17$\pm0.10$\\
         3345 & 140$\pm39$ & N/A & N/A & N/A & <7.3\\
         6445 & 180$\pm46$ & N/A & N/A & N/A & \textbf<8.6\\
         8532 & 185$\pm16$ & 2.02 & 0.3$\pm0.5$ & 0.4$\pm0.6$ & 9.41$\pm0.01$ \\
         \hline
     \end{tabular}
     \caption{Emitters with high \ox\ equivalent width}
     \label{altaanch}
 \end{table*}
 
  In \cite{huang}, several galaxies with high equivalent width in the [\ion{O}{iii}]$_{\lambda 5007}$ (>3000 \AA\, at restframe) lines are studied. They found that those galaxies can be classified as: low-mass, strong starbust, and compact. Following the models of \cite{miguelin}, we can assume a maximum equivalent width of the \ox\, line for 'normal'
galaxies at 150\,\AA. Indeed, \cite{kong} found that galaxies with $EW$\ox>100\,\AA\, are in the 90th percentile and \cite{thomas}, based on data from the SDSS-III/BOSS, showed that galaxies with $EW$\ox>100\,\AA\, are located outside the limits of their figures. Being conservative, we can then select for our galaxies a minimum value for high equivalent width galaxies at $\sim$150\,\AA.

Based on a first approximation, taking into account the models presented in \cite{miguelin}, those high equivalent-width galaxies in [\ion{O}{iii}]$_{\lambda 5007}$ may be an analogous to our detected emitters with high \ox\,equivalent width.
  
In Table \ref{altaanch}, we summarise the properties of the emitters with high \ox\, equivalent width. We note we include emitter {\tt id:3345}, with $EW_{\rm \ox}$=140\,\AA. This is smaller than the selected value of 150\,\AA, however, the error in the equivalent width associated with this emitter makes it compatible with a galaxy with high equivalent with in \ox.
  
\begin{figure*}
  \subfloat{\includegraphics[width=.20\linewidth]{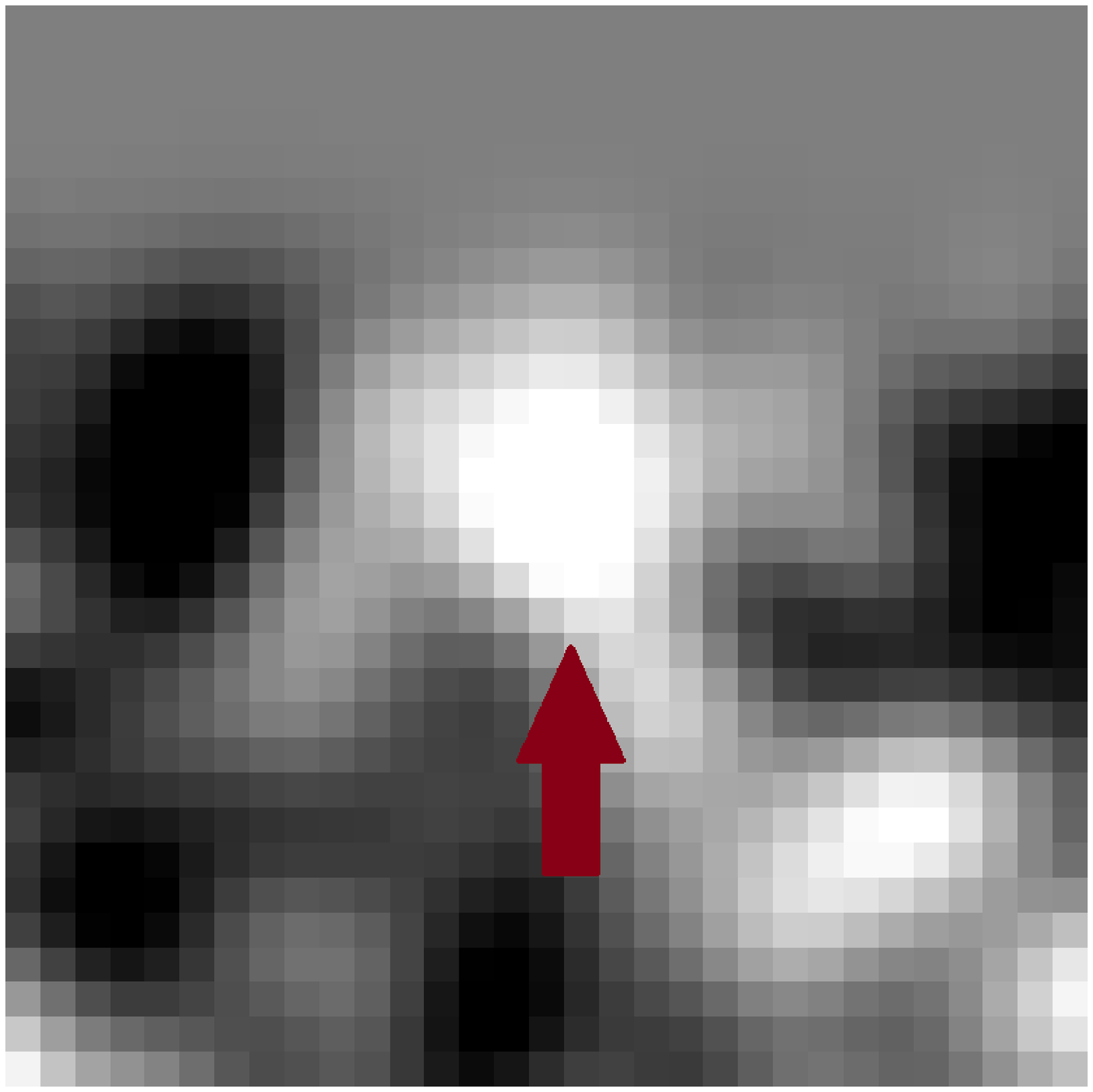}}
\subfloat{\includegraphics[width=.20\linewidth]{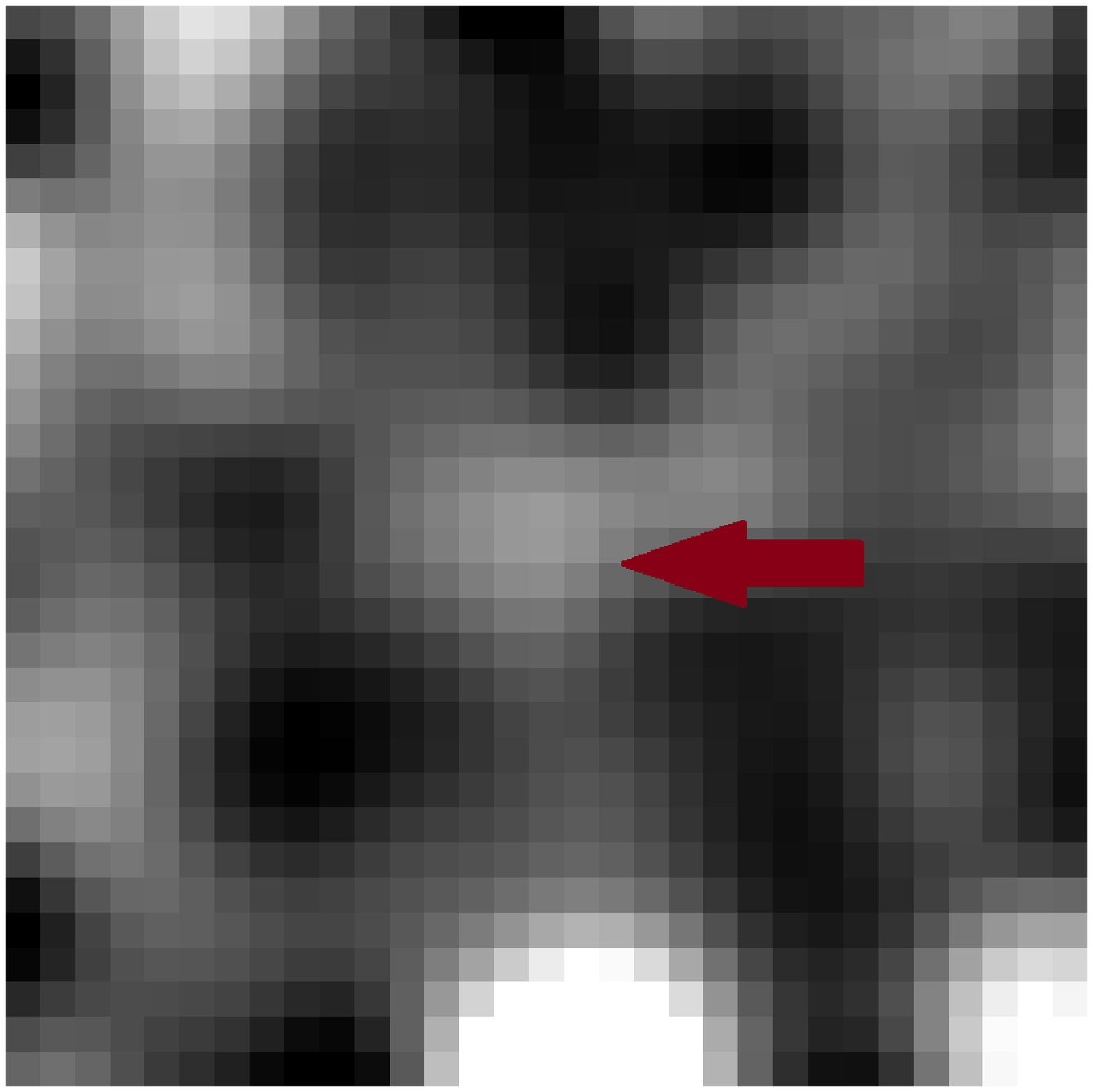}}
\subfloat{\includegraphics[width=.20\linewidth]{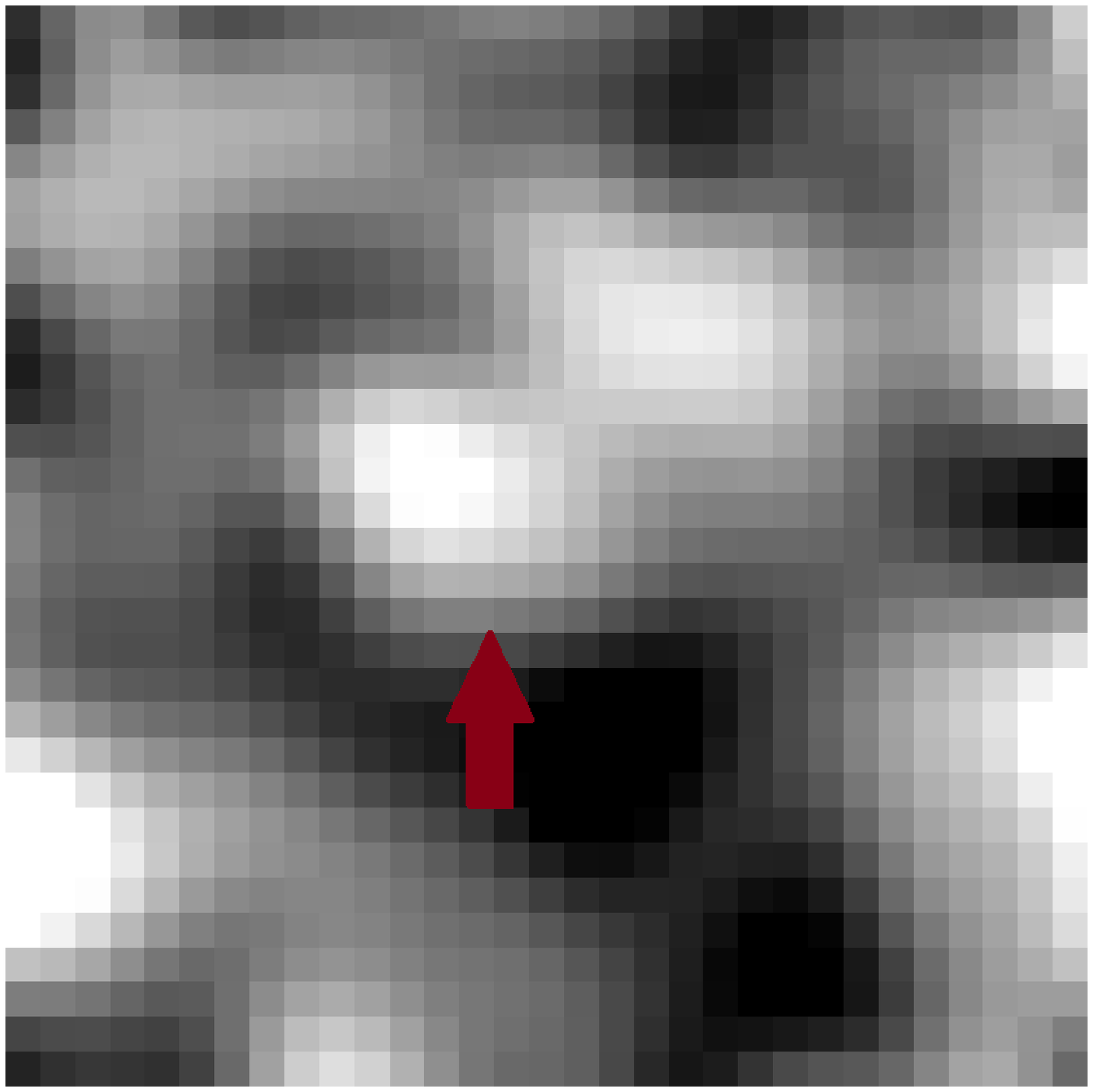}}
\subfloat{\includegraphics[width=.20\linewidth]{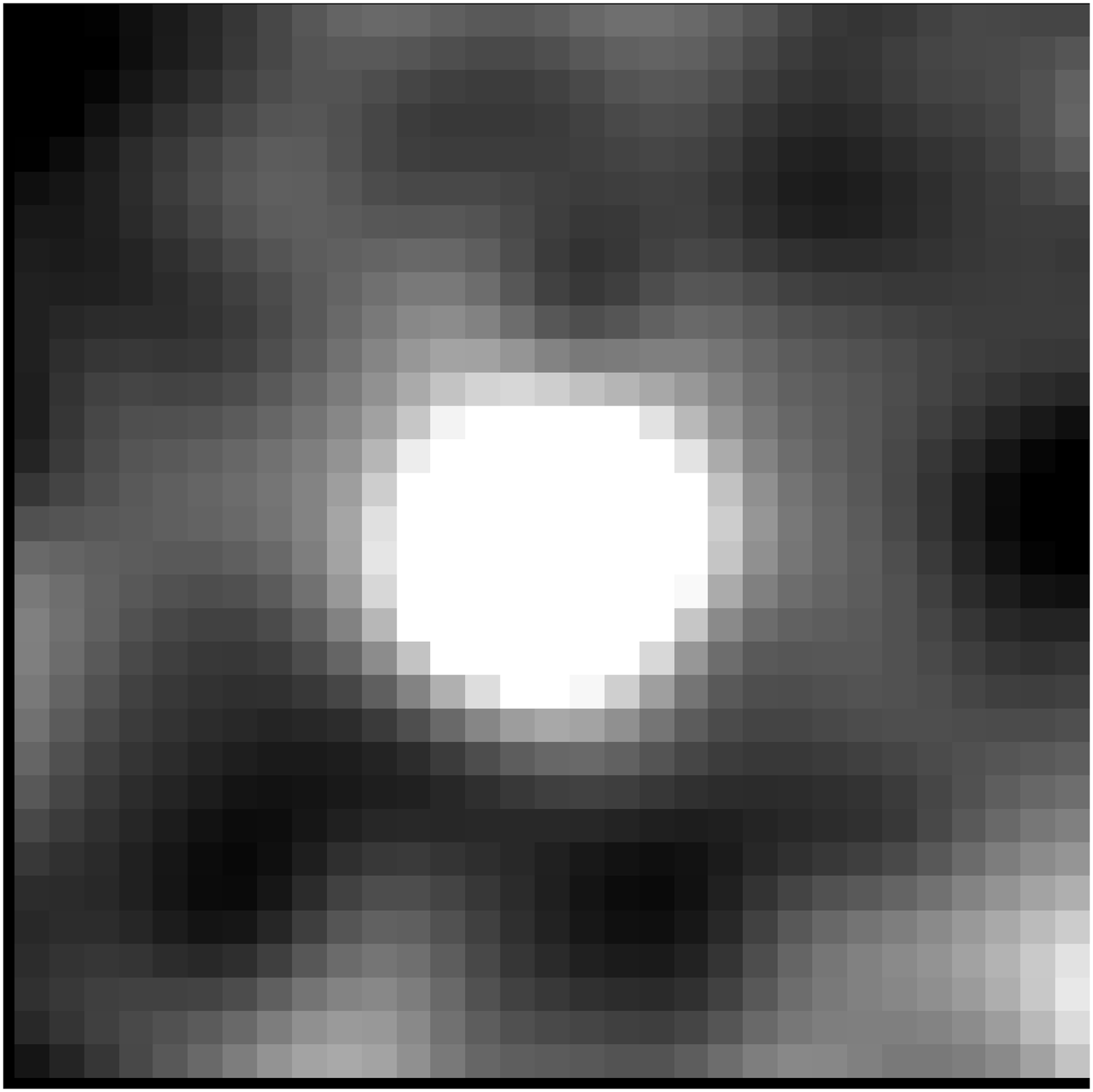}}
    \caption{Thumbnails in the OTELO composite filter (obtained adding all the slices of the scan) for the sources with high equivalent width. The red arrow indicates the location of the source. Each thumbnail has $8\times8$ arcsec size. From left to right:  {\tt id:3077},  {\tt id:3345},  {\tt id:6445,} and  {\tt id:8532}.}
    \label{tumb}
\end{figure*}

The emission at the line \ox\, is influenced by metallicity and the star formation history (see e.g. \citealt{anders}). It was also noted in \cite{miguelin} that $EW_{\ox}$ presents an important mass and metallicity dependence for ages lower than $\sim$4.5 Myr, and with no dependence at older ages. On the other hand, the $EW_{[\ion{O}{iii}]5007}$ has a negligible dependence with the mass. For this reason, a region with large $EW_{\ox}$ may not simultaneously have  a large $EW_{[\ion{O}{iii}]5007}$, and they may not be the same type of object.  
From Table \ref{altaanch}, it is clear that the emitters {\tt id:3077} and {\tt id:8532} have both masses that put them in the low-mass galaxy regime. Moreover, we have complete morphology data for emitter {\tt id:8532} and it seems to point out towards a disc-like emission galaxy. This may also indicate also that a high [\ion{O}{iii}] equivalent-width objects are not the same type of emitters as the ones with high \ox ~equivalent width. Unfortunately, we do not have enough information to establish the correct morphology for all our emitters nor their masses. Nevertheless, we are able to establish a upper limit for the masses of {\tt id:3345} and {\tt id:6445}, both of them well inside the low-mass regime. In Fig. \ref{tumb} we have represented the cutouts for the emitters with high \ox\,equivalent width in the composite OTELO band, obtained by adding all the slices of the scan (see \citealt{otelo} for an in-depth description). 
  
\begin{figure*}[!]
   \centering
   \includegraphics[width=17cm]{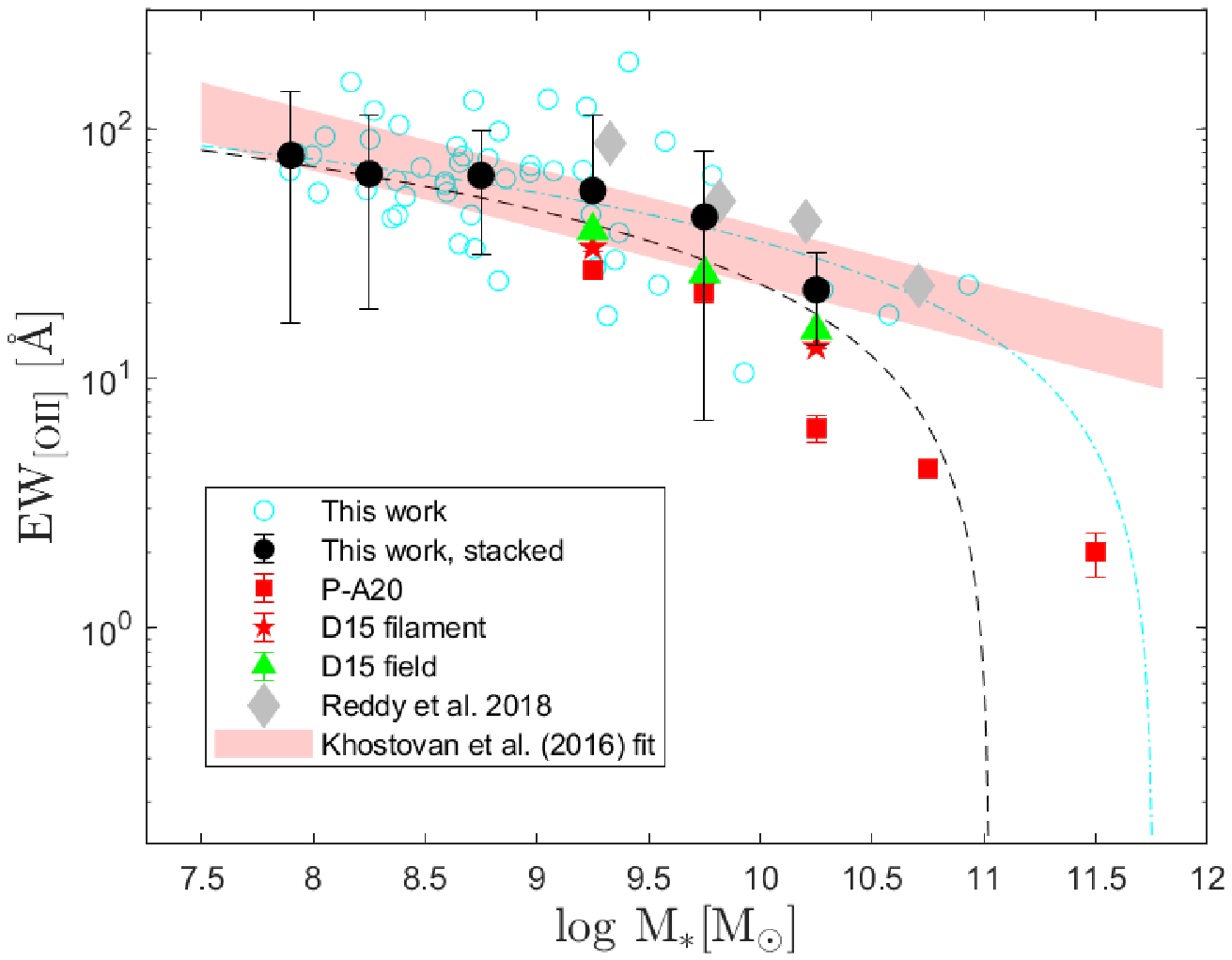}
      \caption{Relationship between the \ox\, equivalent width and the derived stellar mass for our emitters (cyan open circles). The black diamonds are the median of our emitters for bins 0.5 dex in $\log (M_*/{\rm M_{\odot}})$, and the data from \cite{paulino} (red closed squares), \cite{darvish} (filled purple stars for filament galaxies and green triangles for field galaxies), and \cite{reddy2} (grey diamonds). The black dashed line represents the best fit for all the binned data and the cyan point-dashed line represent the best fit for the only the binned data of the emitters detected in this work. The pink band represents the power law fit with uncertainties from \cite{khosew} for \ox\, emitters at $z=1.47$.
              }
         \label{ewvsmasa}
\end{figure*}

In Fig. \ref{ewvsmasa}, we represent the \ox\ equivalent width in \AA\ versus the logarithm of the derived mass of our emitters. We stacked our data in bins of 0.5 dex and we added data from other authors: \cite[][P-A20]{paulino} with galaxies $z\sim$0.8, \cite[][D15]{darvish} with $z\sim$0.5, and \cite{reddy2} with $z\sim1.5$. We also fitted a simple function using least squares to the data in the form of the following equation:

 \begin{equation}
    \centering
     {EW\ox}=p_1 \times \log({M_*/{\rm M_{\odot}}})+p_2.
     \label{ewm}
 \end{equation}
 
 The results of the fitting are given in Table \ref{constantes} and we also represent the fit obtained from \cite{khosew} for the \ox, emitters at $z=1.47$, with its uncertainties, as a pink band. This fit is a power law with the form $EW_{\ox}=k\times M_*^{\,\beta}$, where $\beta=-0.23$ and $\log(k)=3.79$. As pointed out by \cite{paulino}, higher stellar mass galaxies have lower equivalent widths in \ox, independently of the environment. They suggest that this trend is mostly a consequence of the underlying main sequence of star-forming galaxies. Our data seems to follow this behaviour.
 On the other hand, there is a large difference between our data and those of \cite{paulino} at higher stellar masses.
 We must point out that we are sampling lower-mass galaxies than \cite{paulino}, \cite{darvish}, and \cite{reddy2}. On the other hand, \cite{reddy2} seem to have a large value for the equivalent width when compared with other authors (\citealt{paulino}, \citealt{darvish}), but their results are inside the error margins of ours -- and this is not surprising. As \cite{khosew} have shown, there is also a dependence of the $EW_{\ox}$ with the redshift, with larger values for $EW_{\ox}$ when increasing $z$, up to $z\sim4$, followed by a decreasing of the equivalent width at higher redshifts. This may be the reason for the lower value of $EW_{\ox}$ at high stellar masses obtained in \cite{paulino}, as well as in \cite{darvish}, when compared with those presented in \cite{reddy2}, \cite{khosew}, and this work.
 
\cite{cava} studied two groups of \ox\, emitters: one at $z\sim$0.84 and other at $z\sim$1.23. They reached a result a bit below $\log (M_*/{\mathrm M_\odot})<9$, while for $z\sim$1.23 and at low mass galaxies, there is little correlation between the equivalent width and the derived stellar mass. This is similar to our results for the non-stacked and stacked sources at low stellar mass. On the other hand, at z$\sim$0.84, \cite{cava} found a clear correlation between the masses and the equivalent width as that presented in \cite{paulino}, down to $\log (M_*/{\mathrm M_\odot})\simeq8.5)$. This discrepancy may be due to the low number of sources detected at larger redshifts, so we may assume that the correlation should appear with a set of enough emitters. Nevertheless, our results closely follow the fit from \cite{khosew} down to $\log (M_*/{\mathrm M_\odot})<8.5$, even when \cite{khosew} data has few galaxies below that threshold. Moreover, if we calculate the exponent $\beta$ for our binned data, we obtain $\beta=-0.20\pm0.04$, which agrees, within uncertainties, with the one obtained in \cite{khosew}, $\beta=-0.23\pm0.01$.

\begin{table}[]
    \centering
    \renewcommand{\arraystretch}{1.5}
    \begin{tabular}{c c c}
         \hline
         Data & $p_1$ & $p_2$ \\
         \hline
        This work & -20.03 & 235.5\\
        This work + P-A20 + D15 & -23.2 & 255.8\\
        \hline
    \end{tabular}
    \caption{Parameters fitted for the $EW$--stellar mass relation in Eq.\ref{ewm}.}
    \label{constantes}
\end{table}

\section{The \ox\, luminosity function}
\label{sec:lf}

Using the data obtained for all the emitters in the catalogue, we derived the LF for the \ox\, line emitters at $z\sim$1.43 after taking into account the main sources for uncertainties and selection effects: the completeness of the sample and the cosmic variance (CV).

\subsection{Completeness}
\label{sec:compl}

\begin{figure}[!]
   \centering
   \includegraphics[width=.7\linewidth,angle=-90]{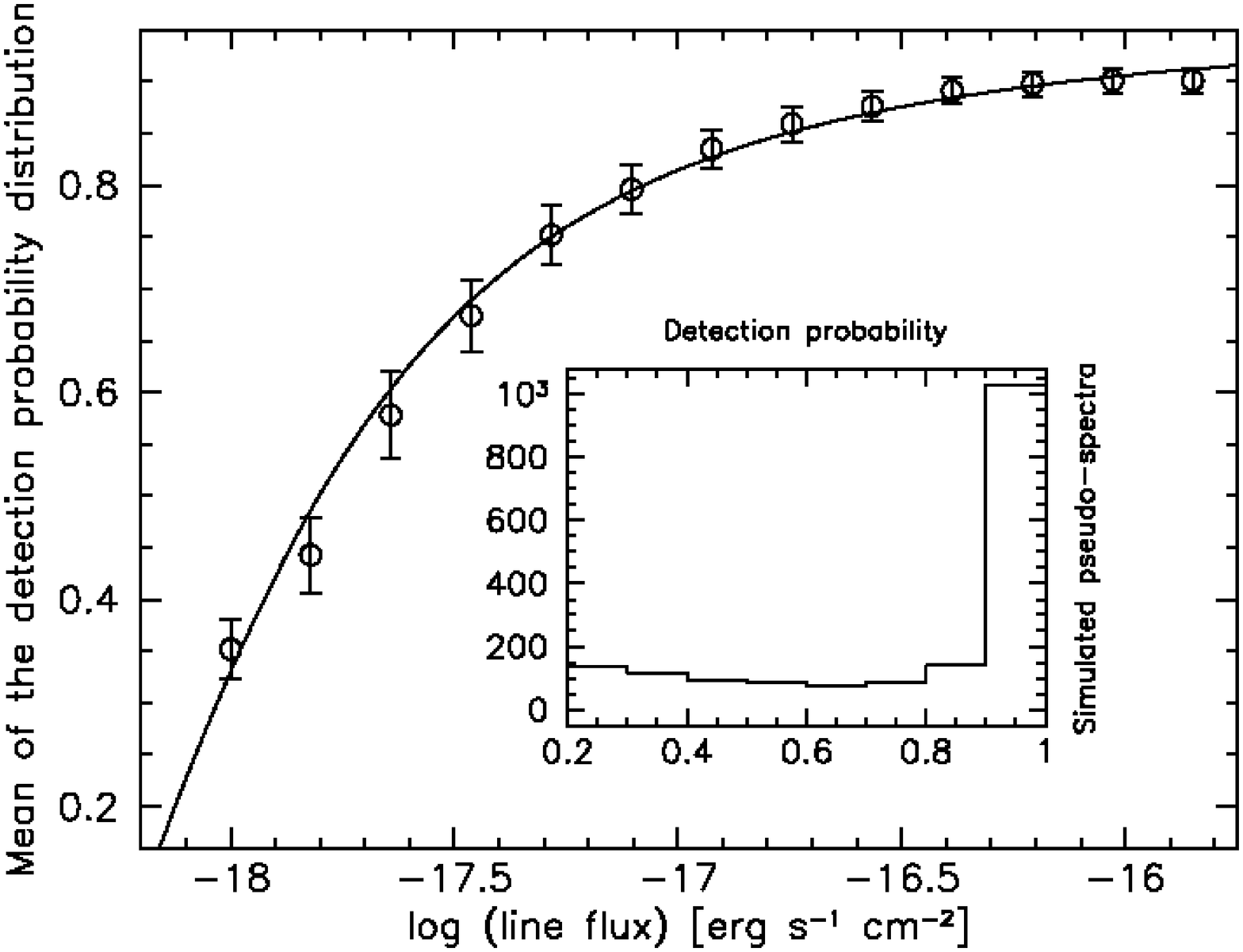}
      \caption{Mean of the detection probability of the \ox emitters at $\langle$z$\rangle$ $\sim$ 1.43. As described in \cite{lum}, the inset shows the detection probability distribution obtained from the simulations. The curve represents the least-square weighted fit of the sigmoid function. The bars indicated the mean standard error.}
         \label{comp}
 \end{figure}

To correct for completeness, we have followed the methods described in \cite{lum}. To summarise, several simulations were carried out to calculate the detection probability function from emission-line sources in the OTELO survey. These simulations helped to calibrate the detection probability as a function of the \ox\, line flux. This probability function was then fitted by a sigmoid algebraic function (Eq. \ref{sigm}). 
\newline
\begin{equation}
    d=\frac{aF}{\sqrt{c+F^2}},
    \label{sigm}
\end{equation}

\noindent where $d$ is the fitted mean detection probability, $F=\log(f_{\ox})+b$, and $a=0.961\pm0.010$, $b=18.297\pm0.032$, $c=0.660\pm0.099$ are the fitted parameters. In Fig. \ref{comp}, we show the completeness correction function and the mean detection probability for the emitters at $z\sim1.43$.
\newline

\subsection{Cosmic variance}
\label{sec:cv}

The comoving volume covered by OTELO at the redshift of this catalogue is about $\sim1.0\times10^4$\,Mpc$^3$, along with an explored sky area of about 0.015 deg$^2$. Because of this small volume (when compared with other surveys; see Table \ref{constantes}), the effects of CV are remarkable. Indeed, \citealt{stroe} found that volumes smaller that $3\times10^4$\,Mpc$^3$ are affected in such a way for CV that the results in the determination of the LF could lead to errors up to 100\% in the parameters employed to fit the such LF. This is confirmed in \cite{marinanuestra1}, where it was derived a global root of the CV of $\sigma_{v}\simeq0.73$, based on the prescription of \cite{Somerville2004}, for a comoving volume of $1.4\times10^3$\,Mpc$^3$ and a redshift of $z\sim0.40$. And in \cite{lum}, for a mean volume of $6.63\times10^3$\,Mpc$^3$ and a a sample of objects at $z\sim$0.83, the root CV obtained was $\sigma_{\rm CV}=0.396$, by using the code\footnote{{\tt http://www.usm.uni-muenchen.de/people/moster/home/download.html}} described in \cite{moster11}. It is thus made clear that obtaining an estimation of the CV is a necessary condition if we want to correctly characterise the \ox\, luminosity function.

According to the mean number density of \ox\ line emitters ($1.4\times10^{-3}$\,Mpc$^{-3}$), it is expected that the CV effects are more pronounced than the estimated for previous OTELO samples at lower redshifts, despite the increased cosmic volume explored at $z\sim$1.43 as compared to those cases. On this basis, we examined the contribution of the CV to the uncertainty of the LF from each approach used in the OTELO papers referred above. On one side, \cite{Somerville2004} provide a recipe to estimate the CV based on a known average redshift and number density in deep surveys, but unknown clustering strength. The mean root CV obtained for the science case presented here is $\sigma_{v}\simeq0.61$. On the other hand, the approach given by \cite{moster11} allows us to compute CV as a function of mean redshift, redshift bin size, and the stellar mass of the subject galaxy population. By following this prescription, we were able to estimate an uncertainty function attributed to the CV, $\sigma_{\rm CV}(M_\star)$ for six bins in the stellar mass range 8.5 $\leqslant$ $\log({M_\star/{\rm M}_{\odot}})$ $\leqslant$ 11.5 for the \ox\ sample, and the mean uncertainty obtained for our sample at $\langle z \rangle$=1.43 is $\langle \sigma_{\rm CV} \rangle$=0.304. In view of the noticeable difference between these estimations, we decided to adopt the \cite{Somerville2004} recipe for being the most conservative one. In the final part of Section \ref{sec:lfcalc}, we further discuss the reliability of this approach for the science case presented in this work.

\subsection{\ox\, luminosity function}
\label{sec:lfcalc}

The LF computes the number of galaxies ($\phi$) per unit of volume and per unit of luminosity. The \cite{lfs} function is the usual parameterisation method:
\begin{center}
    \begin{equation}
        \phi(L)\, {\rm d}L = \phi^\ast (L/L^\ast)^\alpha \exp(-L/L^\ast)\, {\rm d}(L/L^\ast),
        \label{sch}
    \end{equation}
\end{center}

\noindent where $L$ is the luminosity of the emission line (\ox\, in our case), $\phi^\ast$ is the density number of galaxies, $L^\ast$ is the characteristic luminosity, and $\alpha$ defines the faint-end slope of the LF.

The value of $\phi$ was calculated according to \cite{lum} as follows:
\begin{center}
    \begin{equation}
    \label{otelophi}
        \phi[\log L(\ox)]= \frac{4\pi}{\Omega}\Delta[\log L(\ox)]^{-1}\sum_\ensuremath{i} \frac{1}{V_\ensuremath{i} d_\ensuremath{i}},
    \end{equation}
\end{center}

\noindent where $d_\ensuremath{i}$ is the detection probability for the $i$th galaxy, obtained with Eq. \ref{sigm}, $V_\ensuremath{i}$ is the comoving volume for the $i$th source, $\Omega$ is the surveyed solid angle ($\sim 4.7\times10^{-6}$ str), and $\Delta[\log L(\ox)]=0.4$ is the adopted luminosity binning. 

The \ox\ luminosities based on extinction-corrected line fluxes (Section \ref{sec:properties}) were sampled in seven bins distributed in the range $40.13 < \log L(\ox) < 42.93$. For each bin, $\phi$ values were obtained using Eq. \ref{otelophi}. The total uncertainty of each value mainly come from the quadratic combination per luminosity bin of the Poisson error obtained from the number of galaxies, which mean contribution is about 40\% (increasing from a 23\% for lower luminosity bins up to 70\% for the upper-side bin) and the uncertainty associated with the CV. Following the recipe provided by \cite{Somerville2004}, the estimated root of the CV, $\sigma_{v}$, increases from 0.6 at the $\log L(\ox) = 40.33$ bin, to 0.91 at the $\log L(\ox) = 42.73$ one. Total uncertainty in each bin also includes the contribution of the mean standard error of the completeness correction (\ref{sec:compl}) and the probability of incorrect emission-line sources identification (which also includes the fraction of bona fide emitters lost) based on \cite{lum}, amounting together to $\sim$6\%. The resulting uncertainties from such combinations, along with the LF data are given in Table \ref{oii_lf_values}.

\begin{table}[ht]
\begin{center}
\renewcommand{\arraystretch}{1.5}
\begin{tabular}{c c}
\hline 
$\log L(\ox)$  & $\log \phi$   \\
$ $[erg s$^{-1}$]      & [Mpc$^{-3}$ dex$^{-1}$]  \\
\hline  
40.33 & -2.54 $\pm$ 0.31  \\ 
40.73 & -2.22 $\pm$ 0.25  \\
41.13 & -2.27 $\pm$ 0.27  \\
41.53 & -2.55 $\pm$ 0.30  \\
41.93 & -2.96 $\pm$ 0.37  \\
42.33 & -3.27 $\pm$ 0.46  \\
42.73 & -3.57 $\pm$ 0.59  \\
\hline
\end{tabular}
\caption[Luminosity function values]{Binned values of the observed \ox\ luminosity function obtained from
Equation \ref{otelophi}. The computation of the total uncertainties is described in the text.
} 
\label{oii_lf_values}
\end{center}
\end{table}
\normalsize

In Table \ref{constantess}, we summarise the Schechter parameters derived for the fit of the LF, as well as the parameters for other works at similar redshifts and with the \ox\, line (\citealt{drake}, \citealt{ly} and \citealt{hayashi}). 
The Schechter model function of Eq. \ref{sch} was fitted to the data given in this table using a weighted least-squares minimisation algorithm based on the Levenberg--Marquardt method\footnote{\tt https://lmfit.github.io/lmfit-py/index.html}. Weights are defined as the reciprocal of the squared total uncertainties of LF data.

We used a fixed value for $\log L^\ast$ obtained as the average of the well-known values from the works presented in Table \ref{constantess}, that is, $\log L^\ast = 42.44$, with a standard deviation of $\sigma=0.09$. This was done because our sampling over $\log(L^\ast) \gtrsim 42.5$ is poor, and consequently, $\log L^\ast$ is completely unconstrained in the fitting process. Assuming this standard deviation as a measure of the $\log L^\ast$ uncertainty and that it is normally distributed, we estimated the standard errors of the fitted parameters $\phi^{*}$ and $\alpha$ from 10$^5$ realisations of the fitting procedure described above. The uncertainties thus obtained amount 50.3\% and 4.2\% of the fitted $\phi^{*}$ and $\alpha$ values. Due to the limited cosmic volume explored by OTELO, large variances would be expected in the normalisation of the LF after fitting, as confirmed by the error estimate of $\phi^{*}$. The opposite occurs with the uncertainty linked to the faint-end slope of the LF since our sample extends far beyond the lower limits reached by other, similar surveys.

\begin{figure*}
   \centering
    \includegraphics[width=17cm]{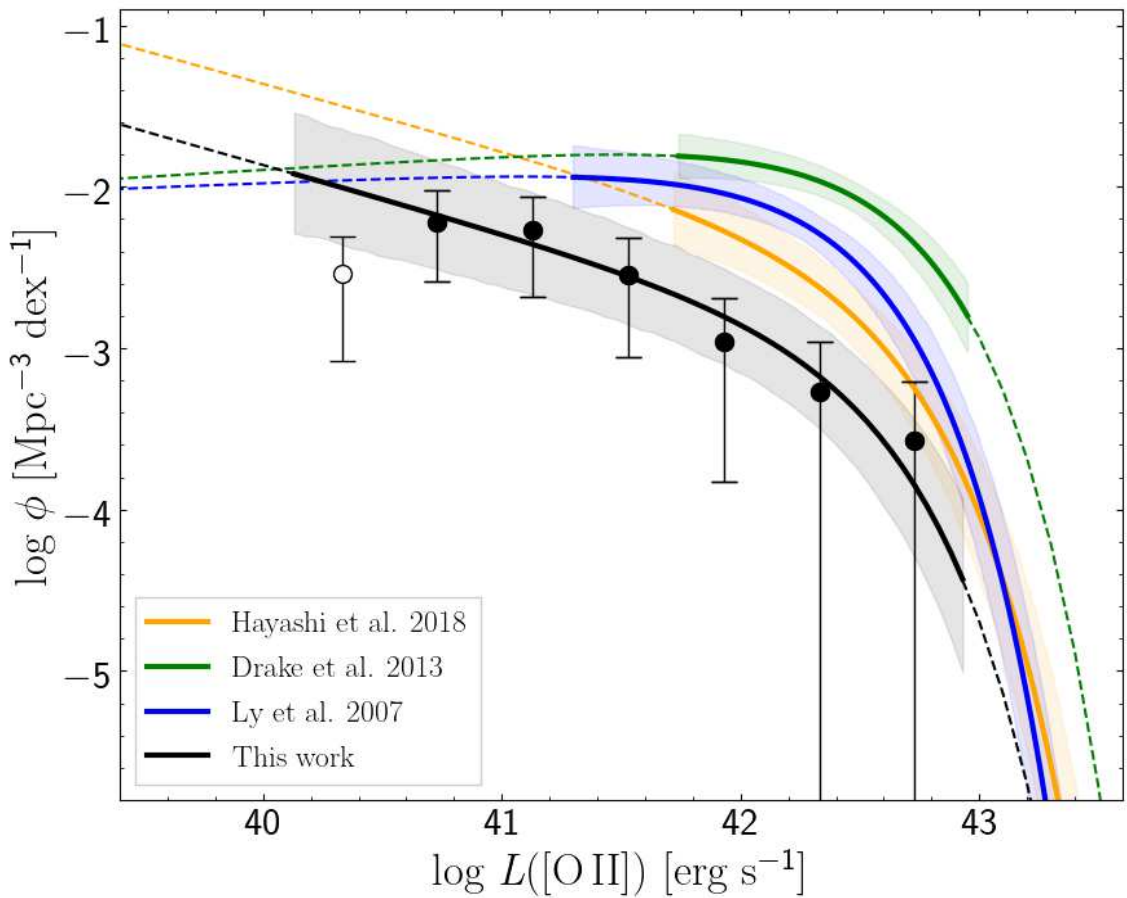}
      \caption{Luminosity function for the \ox emitters at z$\sim$1.43 (black circles), with extinction and completeness corrected. The error bars correspond to the total uncertainties given in Table \ref{oii_lf_values}, and obtained as described in the text. The open circle marks the lowest luminosity bin and was not employed in the fitting. The black line is the best error-weighted fit of  \cite{lfs} function. The green, blue, and orange lines are the LFs from \cite{drake}, \cite{ly}, and \cite{hayashi} respectively. The shaded areas represent the propagation of 1$\sigma$ uncertainties of the tabulated LF after 10$^4$ Monte Carlo realisations. For the OTELO data fitting, this propagation includes the standard deviation of the mean $L^*$ value. In each case, the solid line extend over the sampled luminosity range (see the right side columns in Table \ref{oii_lf_values}) and the dashed line is the extrapolation of the corresponding best fit. 
              }
         \label{lf}
   \end{figure*}

In Fig.\ref{lf}, we represent the LF data for our emitters and their best fittings. The bin with the lowest luminosity ($\log L\ox=40.33$\,[erg/s]) and marked with an open circle, has not been included in the calculations because the method employed to derive the completeness underestimates the number of emitters in the bin, so the errors there are also underestimated. It is clear that our data (represented as black dots) has a lower value for $\log\phi$ at lower luminosities when compared with other works. 
Consequently, our value of $\phi^*$ is lower when it is set against those presented in Table \ref{constantess}, and it is a clear effect of the uncertainties on the CV.
Nevertheless, as pointed out in \cite{drake}, the detection fraction of emitters is very sensitive to the limit of equivalent width in a NB survey, which is of the order of the filter width. So, in this case, as noted in the discussion of \cite{lum}, the lower limit of equivalent width with OTELO survey is about 6 \AA, which is lower than those obtained by the classic narrow-band studies. There, the lower limit for the equivalent width can be over 50 \AA, depending on the technique. This fact comparatively favours the detection of low-mass emitters, despite the CV effects.

\begin{table*}[t]
    \centering
    \resizebox{\textwidth}{!}{
         \renewcommand{\arraystretch}{1.5}
         \begin{tabular}{c c c c c c c c c c}
         \hline
         Source & Method & Number of & $\langle$z$\rangle$ & $\langle$V$_c\rangle$ & $\log \phi^{*}$ & $\log L^*$ & $\alpha$ & $\log L_{min}$ & $\log L_{max}$\\
         & & emitters & & $10^3 {\rm Mpc}^3$ & [${\rm Mpc}^{-3}{\rm dex}^{-1}$] & [erg s$^{-1}$] & &  [erg s$^{-1}$] &  [erg s$^{-1}$] \\
         \hline
        \cite{ly} & NB & 951 & 1.47 & 88.48 & -2.20$\pm$0.06 & 42.31$\pm$0.05 & -0.94$\pm$0.12 & 41.30 & 43.60\\
        \cite{drake} & NB & 2218 & 1.46 & 230.9 & -2.03$^{+0.04}_{-0.05}$ & 42.52$\pm$0.05 & -0.91$\pm$0.11 & 41.74 & 42.95\\
        \cite{hayashi} & NB & 11016 & 1.47 & $\sim 5\times10^3$ & -2.74$\pm$0.10 & 42.48$\pm$0.06 & -1.41$\pm$0.14 & 41.72 & 43.48\\
        This work & NB scan & 60 & 1.43 & 10.21 &  -3.25$\pm$0.21 & 42.44 (fixed) &  -1.42$\pm$0.06 & 40.40 & 42.50\\
        \hline
    \end{tabular}
    }
    \caption{Schechter parameters of the LF(\ox) from the recent literature and the best fit of the corresponding OTELO LF, according to data given in Table \ref{oii_lf_values}. NB stands for narrow-band. Further details about this fitting are given in the main text.}.
    \label{constantess}
\end{table*}

It is also worthwhile noting that we sampled the faint end of the LF by almost one dex lower when compared with other authors. We obtained a value for $\alpha$ equals to $-1.42\pm0.06$. This result is similar to that reported by \cite{hayashi}, but different from those reported in \cite{drake} and \cite{ly}. It must be taken into account that the properties of the sources that serve to create the LFs, such as the luminosity and the extinction, are going to generate different values for $\alpha$. For example, \cite{sobral2} found very different slopes for the LFs for H$\alpha$ and \ox\, for galaxies at the same redshift ($z\sim1.5$) in the HiZELS survey, where $\alpha_{\ensuremath{H}\alpha}=-1.6\pm0.4$ is steeper than  $\alpha_{\ensuremath{\ox}}=-0.9\pm0.2$. Only when calculating the star formation rate (SFR) in an analytical Schechter-like approximation, as \cite{smit} has suggested, the derived $\alpha_{\ensuremath{SFR}}$ should be similar among studies at the same redshift.

We may explain, in this case, the fair coincidence of our value of $\alpha$ and the one derived in \cite{hayashi}, and the difference with \cite{ly} and \cite{drake} following \cite{sobral}, where it is suggested that the slope at the faint en of the LF is also affected by the environment. In this case, a stepper slope indicates a low density field. If we made a crude calculation of the density of galaxies presented in the comoving volume, we find that for us it is 5.87\,Mpc$^{-3}$, and for \cite{hayashi}, it is even lower, 2.20\,Mpc$^{-3}$. Meanwhile, for \cite{ly} and \cite{drake} is 10.71\,Mpc$^{-3}$ and 9.60\,Mpc$^{-3}$, respectively.

\begin{figure}
   \centering
   \includegraphics[width=\hsize]{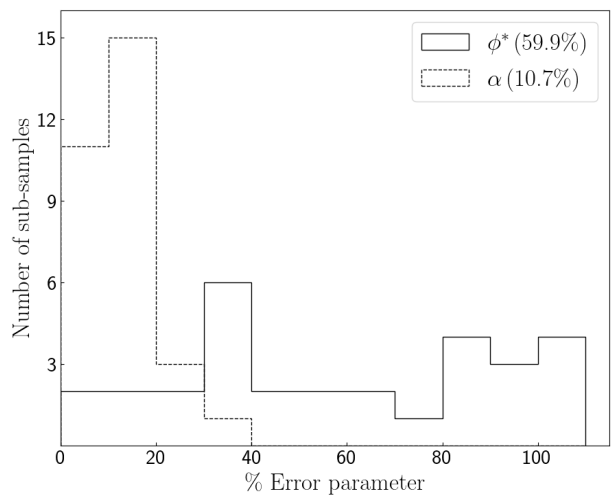}
      \caption{Fractional error distribution of $\phi^*$ and $\alpha$ after resampling the the OTELO field in 30 contiguous data cubes and estimate the corresponding luminosity function for testing the cosmic variance effects on these parameters. Values in the upper right corner correspond to the median values of these distributions.
            }
         \label{varexp}
   \end{figure}

Finally, in order to further test the possible contribution of the CV to explain the discrepancies related to the normalisation of the LF, we resampled the OTELO field in 30 $i$-contiguous cubes containing the 50\% of the surveyed cosmic volume and recomputed the LF for each sub-sample (as described above) to obtain $\log\phi^*_i$ and $\alpha_i$. After that, we calculated the ratio of the standard deviation of the $i$-parameters and those actually reported in Table \ref{constantess}. The distribution of these ratios are presented in Fig. \ref{varexp}. The median fractional error of $\phi^*$ and $\alpha$ obtained by this way, 59.9\% and 10.7\%, respectively, are consistent with the uncertainties associated with these parameters as given in Table \ref{constantess}, despite the volume considered is a half of the explored one. 
However, taking into account the upper limits of the factional error distribution of $\phi^*$ (about 100\%), it is clear that the CV could explain the discrepancies observed when our results are compared with those of other surveys that explored much larger cosmic volumes than OTELO, as comprehensively tested by \cite{sobralref2}. But on the other hand, on the basis of this test it is also evident that (i) the theoretical approach adopted in Section \ref{sec:cv} is enough educated for the purposes of this work; and (ii) the CV effects are less substantial regarding the faint-end slope. This reinforces our appreciation of the sensitivity of OTELO for fairly constraining this parameter.

\subsection{Star formation rate density}
\label{sec:sfrd}

The LF may be integrated to obtain the total luminosity density. According to \cite{lfs}, the integral is equal to:
\begin{equation}
    \mathcal{L}=\int_{0}^{\infty}\phi(L)L\,{\rm d}L=\phi^{\star}L^{\star}\,\Gamma(\alpha+2),
\end{equation}
\noindent where $\Gamma$ is the gamma function.

After the integration, we obtain a value for the total luminosity density of $\log\,(\mathcal{L})=39.38^{\,+0.16}_{\,-0.35}$ \, erg\,s$^{-1}$\,Mpc$^{-3}$. This luminosity density can be translated to the SFRD via a \cite{ken} equation:
\begin{equation}
    \rho\,(\mathrm{M_\odot\,yr^{-1}\,Mpc^{-3})}=(1.4\pm0.4)\times10^{-41}\mathcal{L}\mathrm{\ox}\,\mathrm{(erg\,s^{-1}\,Mpc^{-3})},
\end{equation}

\noindent which is derived employing a \cite{salpeter} initial mass function (IMF). In order to convert this SFRD from the Salpeter IMF to the \cite{kroupa} IMF, we only need to multiply by a constant factor of 0.67 (\citealt{madau}). It has to be noted that the direct conversion of \ox\, luminosity to SFR may presents several problems, including  metallicity and reddening dependencies (\citealt{khos2}). Nevertheless, it seems that the reddening is the main factor when deriving the SFR from \ox\, lines, as noted in \cite{zhu} and \cite{sfr}.

\begin{figure*}[!]
   \centering
   \includegraphics[width=17cm]{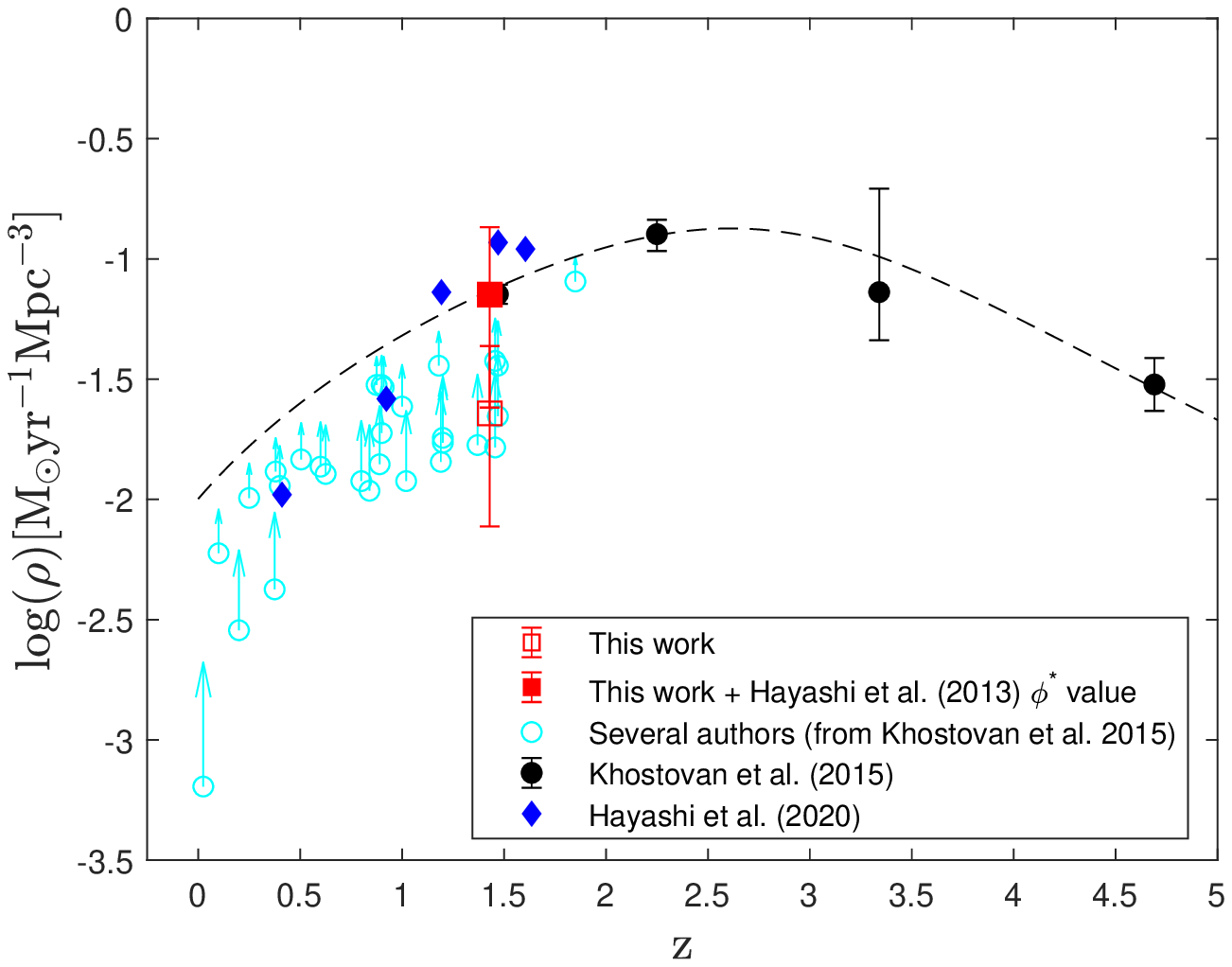}
      \caption{SFRD evolution for several authors. The cyan circles are SFRD derived from \ox, compiled by \cite{khos2} in the Table C1 of their paper, and includes data from \cite{ciardullo}, \cite{sobral2}, \cite{bayliss}, \cite{ly}, \cite{zhu}, \cite{takahashi}, \cite{glaze}, \cite{teplitz}, \cite{gallego}, \cite{hicks}, \cite{hogg}, and \cite{hammer}. We note that these points are not extinction-corrected, so they are a lower limit for the SFRD, as indicated by the cyan arrows.The filled black circles are data from \cite{khos2}. The filled diamonds are data from \cite{hayashi2}. The result of this work is represented by the open square. The red filled square represents the results of the SFRD from our LF but integrated with the $\phi^{*}$ from \cite{hayashi}. The black dashed line is the \ox\, fit to the \cite{khos2} data, following the parametrisation of \cite{madau}.}
         \label{sfh}
   \end{figure*}

We obtained a final result of $\log\,(\rho)=-1.65^{+0.28}_{-0.47}$\,(erg\,s$^{-1}$\,Mpc$^{-3}$). In Fig. \ref{sfh} we plotted our derived value for the SFRD (red filled square) as a function of the redshift. In order to illustrate the SFRD evolution, we complemented the figure with results from the literature. All the data has been converted to a Kroupa IMF. 
The black dashed line is the fit done by \cite{khos2} following the parametrisation of \cite{madau}:
\begin{equation}
    \log\,(\rho)=a\frac{(1+z)^b}{1+[(1+z)/c]^d},
\end{equation}
\noindent where $a=0.015$, $b=2.26$, $c=4.07,$ and $d=8.39$.

It is clear from Fig. \ref{sfh} that our result is below the curve, in a locus with the \ox\, data from the compilation. It must be noted that the SFRDs derived from \cite{khos2} compilation are not extinction-corrected, unlike our derived value for the SFRD.
The data from \cite{khos2} and \cite{hayashi2} are located about $\sim0.25$\,dex higher. 
This difference may be attributed to the low value of $\phi^*$ due to the large uncertainties in the CV, as discussed in Section \ref{sec:lfcalc}. Thereafter, this creates a lower value for the luminosity density when compared with their results, and, consequently, a lower value for the SFRD.

If we integrate our LF, but change the value derived for our $\phi^{*}$ by the one obtained in \cite{hayashi}, the new derived SFRD (indicated in Fig.\ref{sfh} by the filled red square) is now located over the line defined by \cite{madau} and has almost the same value as the SFRD derived by \cite{khos2} at the same redshift. 

\section{Summary and conclusions}
\label{sec:finale}

We created a catalogue of 60 \ox\, emitters at $\langle z\rangle=1.43$, selected from data of the OTELO survey. We derived, from the pseudo-spectra, the redshifts, fluxes, and equivalent widths of all emitters. In Table \ref{emisores}, we have catalogued all the emitters of the sample presented in this work. This table will be electronically available at CDS.

Taking advantage of the ancillary data of the OTELO survey, we were able to fit galaxy templates to our emitters through the {\tt LePhare} code. Thanks to this, we were able to derive extinctions for all the emitters and masses (or at least their upper bounds) for all the emitters. In total, 93\% of the emitters were low-mass galaxies.

We were also able to obtain morphology parameters (S\'ersic and concentration indices) for 44 of the emitters. The s\'ersic index derived for all of the emitters presented values compatible with disc galaxies. The majority of the \ox\, emitters are located in the bluer zone of the $(u-r)$ versus morphology diagrams when compared with the rest of the galaxies of the OTELO survey. Also, all of them were classified as LT using the discriminant developed in \cite{dnn}. We detected four emitters with unusually high $EW$ in \ox\, (\citealt{kong}). One of them, {\tt id:8532}, appears to be a low-mass disc galaxy.

The derived equivalent widths, having been binned and stacked, seem to follow the trends derived in \cite{paulino} and \cite{khosew} with the stellar mass. However, both \cite{paulino} ($z\sim0.8$) and \cite{darvish} ($z\sim0.5$) present lower values for the equivalent width at larger masses when compared with \cite{reddy}, \cite{khosew} and our data (at $z\sim1.5$, $z=1.47$ and $z=1.43$ respectively). This may be due to the evolution of the equivalent width with the redshift, as suggested in \cite{khosew}. Also, the lack of a clear correlation at lower masses for non-stacked sources detected in our data may be due the scarcity of emitters, as appears to  be the case in \cite{cava}.

After we applied corrections for the completeness and the cosmic variance, among others, we obtained the LF for the \ox\, line, reaching almost 1\,dex fainter than the works presented in literature. Nevertheless, our data present a lower value on $\log\phi^*$ when compared with other works. Because of the small cosmic volume sampled by OTELO when compared with modern NB surveys, the cosmic variance effects of the total uncertainties of LF data are quite large, and this propagates to the large variance obtained on this parameter. This effect can explain such differences with data provided in the literature.

On the other hand, the value $\alpha=-1.42\pm0.06$ obtained suggests a stepper slope on the faint end of the LF when compared with other authors (\citealt{drake} or \citealt{ly}). This parameter is much less sensitive to the cosmic variance effects than the normalisation ($\phi^*$) of the LF. However, we obtain practically the same result as the one derived in \cite{hayashi}. As suggested by \cite{sobral}, this low value of $\alpha$ may indicate that we are (as well as \citealt{hayashi}) sampling a low density field.

We have derived the SFRD by integrating the LF, employing the \cite{ken} relationship and \cite{kroupa} IMF. We have found that the data from \cite{khos2} and from \cite{hayashi2} are at least $\sim0.25$\,dex higher. This difference seems to come for our low value of $\phi^{*}$ when compared with other results from literature (\citealt{ly}, \citealt{drake} or \citealt{hayashi}) and may be caused by the uncertainties presented in the normalisation of the LF due to CV effects.

The existence of a population of low-mass star-forming galaxies has been confirmed at $z=1.43$. These galaxies present different characteristics when compared with the more massive populations at the same redshift that have been studied in several other works. This indicates that subsequent surveys with the same scope as demonstrated by OTELO are needed in order to complete a census of the galaxy populations at different epochs of the Universe.

\begin{acknowledgements}
This paper is dedicated to the memory of our affable colleague and friend H\'ector Casta\~neda, who passed away on Nov. 19th., 2020.

The Authors thank the anonymous referee for her/his feedback and constructive suggestions, which have contributed to significantly improve the manuscript.

BC wishes to thank Carlota Leal \'Alvarez for her support during the development of this paper.

JAdD thanks the Instituto de Astrof\'isica de Canarias (IAC) for its support
through the Programa de Excelencia Severo Ochoa and the Gobierno de Canarias for the Programa de Talento Tricontinental grant.

This  work  was  supported  by  the  Spanish  Ministry  of  Economy  and
Competitiveness  (MINECO) under  the  grants
AYA2013\,-\,46724\,-\,P,
AYA2014\,-\,58861\,-\,C3\,-\,1\,-\,P,
AYA2014\,-\,58861\,-\,C3\,-\,2\,-\,P,
AYA2014\,-\,58861\,-\,C3\,-\,3\,-\,P,
AYA2016\,-\,75808\,-\,R,
AYA2016\,-\,75931\,-\,C2\,-\,1\,-\,P,
AYA2016\,-\,75931\,-\,C2\,-\,2\,-\,P and
MDM-2017-0737 (Unidad de Excelencia Mar\'ia de Maeztu, CAB).

This work was supported by the project Evolution of Galaxies, of reference 
AYA2017\,-\,88007\,-\,C3\,-\,1\,-\,P, within the "Programa estatal de fomento de la 
investigaci\'on cient\' ifica y t\' ecnica de excelencia del Plan Estatal de 
Investigaci\'on Cient\'ifica y T\'ecnica y de Innovaci\'on (2013-2016)" of the 
"Agencia Estatal de Investigaci\'on del Ministerio de Ciencia, Innovaci\'on y 
Universidades", and co-financed by the FEDER "Fondo Europeo de Desarrollo Regional".

This article is based on observations made with the Gran Telescopio Canarias (GTC) at Roque de los Muchachos Observatory of the Instituto de Astrof\'isica de Canarias on the island of La Palma.

This study makes use of data from AEGIS, a multi-wavelength sky survey conducted with the Chandra, GALEX, Hubble, Keck, CFHT, MMT, Subaru, Palomar, Spitzer, VLA, and other telescopes, and is supported in part by the NSF, NASA, and the STFC.

Based  on  observations  obtained  with  MegaPrime/MegaCam,  a  joint  project  of the  CFHT  and CEA/IRFU, at the Canada--France--Hawaii Telescope (CFHT) which is operated by the National Research Council (NRC) of Canada, the Institut National des Science de l'Univers of the Centre National de la Recherche Scientifique (CNRS) of France, and the University of Hawaii.  This work is based in part on data products produced at Terapix available at the Canadian Astronomy Data Centre as part of the Canada-France-Hawaii Telescope Legacy Survey, a collaborative project of NRC and CNRS.

Based on observations obtained with WIRCam, a joint project of CFHT, Taiwan, Korea, Canada, France, at the Canada--France--Hawaii Telescope (CFHT), which is operated by the National Research Council (NRC) of Canada, the Institute National des Sciences de l'Univers of the Centre National de la Recherche Scientifique of France, and the University of Hawaii. This work is based in part on data products produced at TERAPIX, the WIRDS (WIRcam Deep Survey) consortium, and the Canadian Astronomy Data Centre. This research was supported by a grant from the Agence Nationale de la Recherche ANR-07-BLAN-0228.
\end{acknowledgements}

%
%

\bibliographystyle{aa} 
\bibliography{oxig} 

\appendix
\section{Catalogue and pseudo-spectra of the \ox\ emitters}
In Table \ref{emisores}, we summarise the main properties of our emitters. The distribution of columns is as follows: the first column is the OTELO identification number, the second and third columns are the coordinates of the sources in degrees, the fourth column is the redshift derived from the deconvolved model ($z_{\rm\, OTELO}$), the fifth column is the extinction in mag, the sixth column is the extinction corrected flux in units of $10^{-17}$\,erg/cm$^2$/s, the seventh column is the \ox\, equivalent width in \AA, and the last column is the derived stellar mass. This table will be available electronically on CDS.\\

\begin{figure*}
    \centering
\subfloat{\includegraphics[width=.20\linewidth]{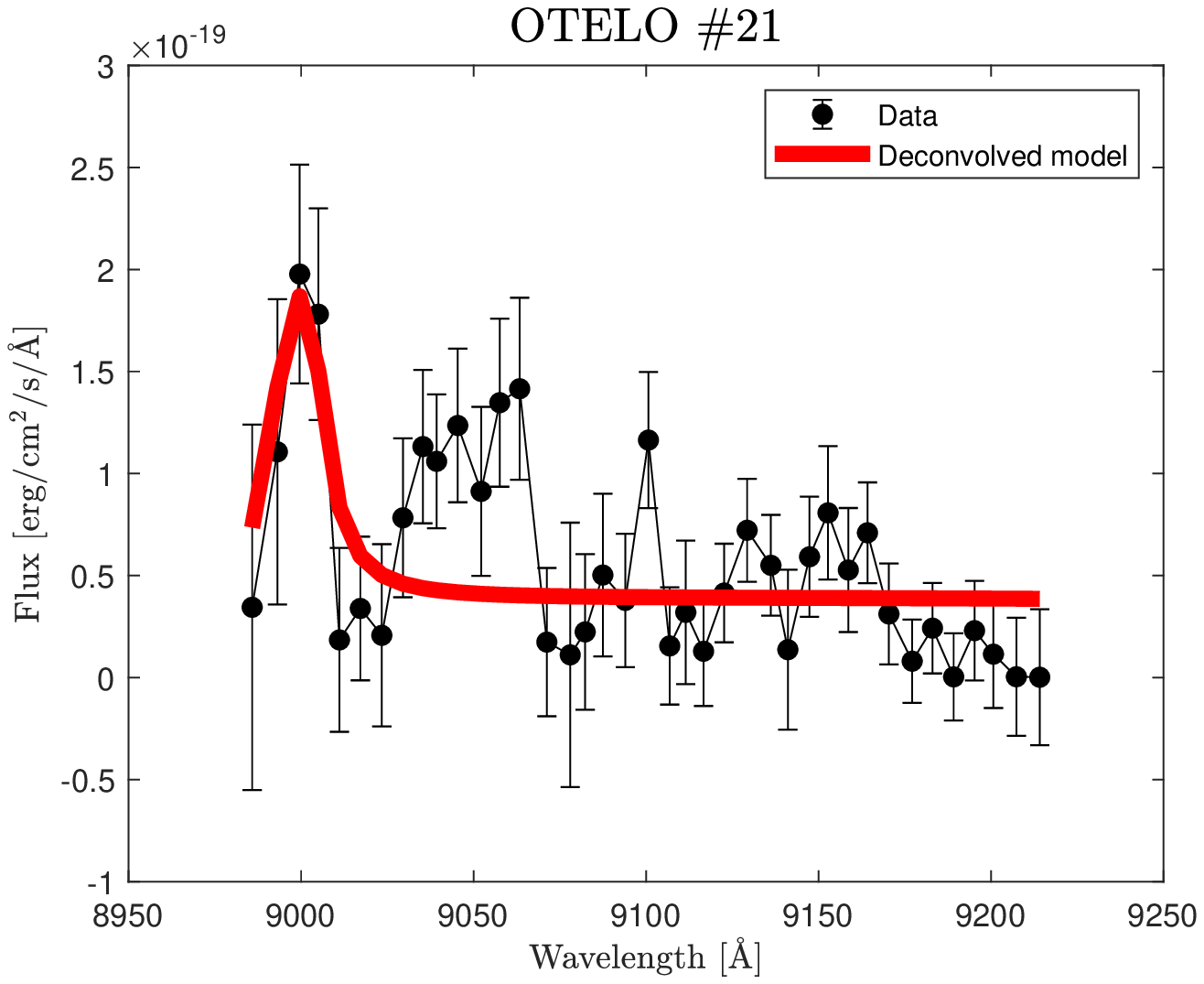}}
\subfloat{\includegraphics[width=.20\linewidth]{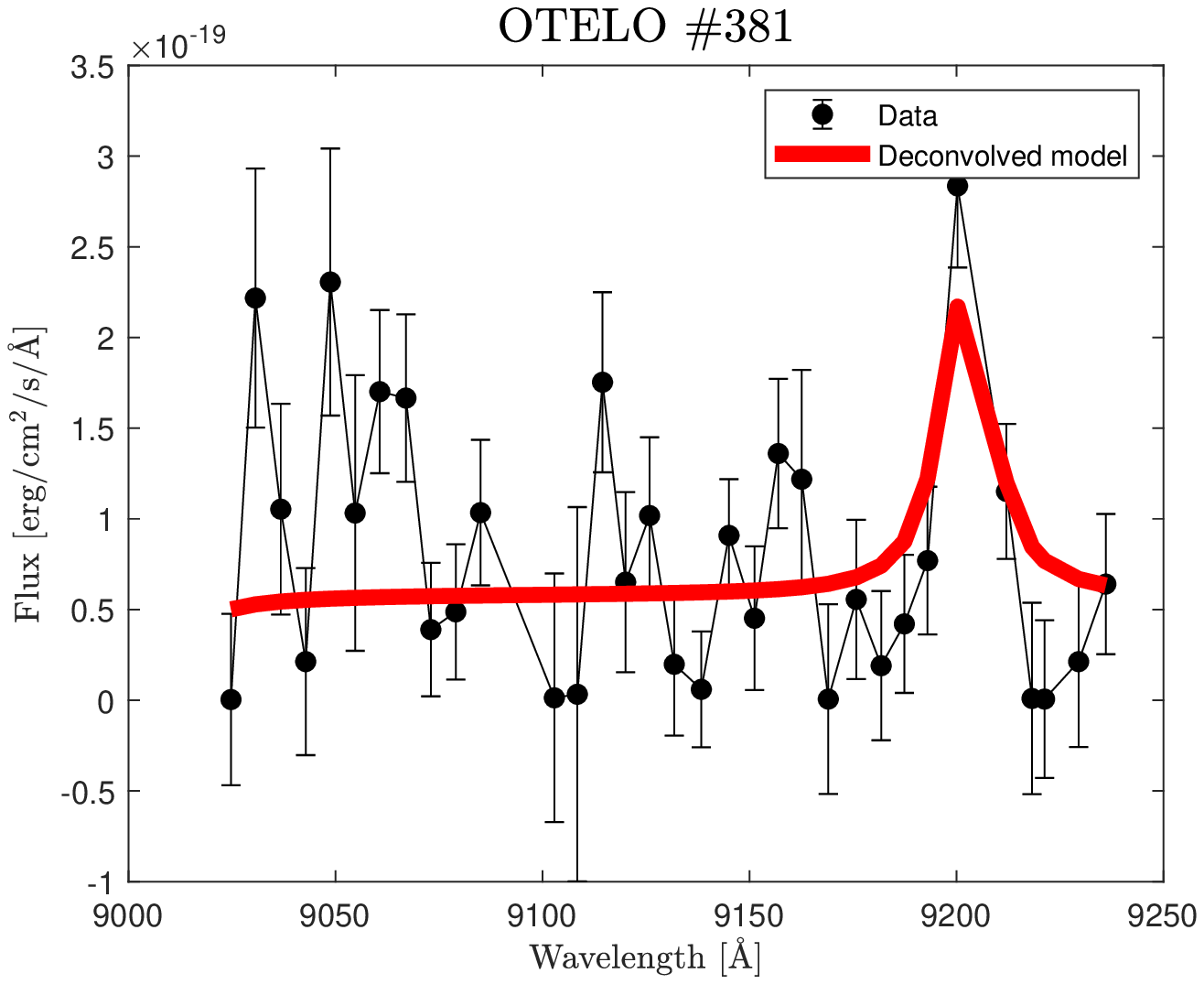}}
\subfloat{\includegraphics[width=.20\linewidth]{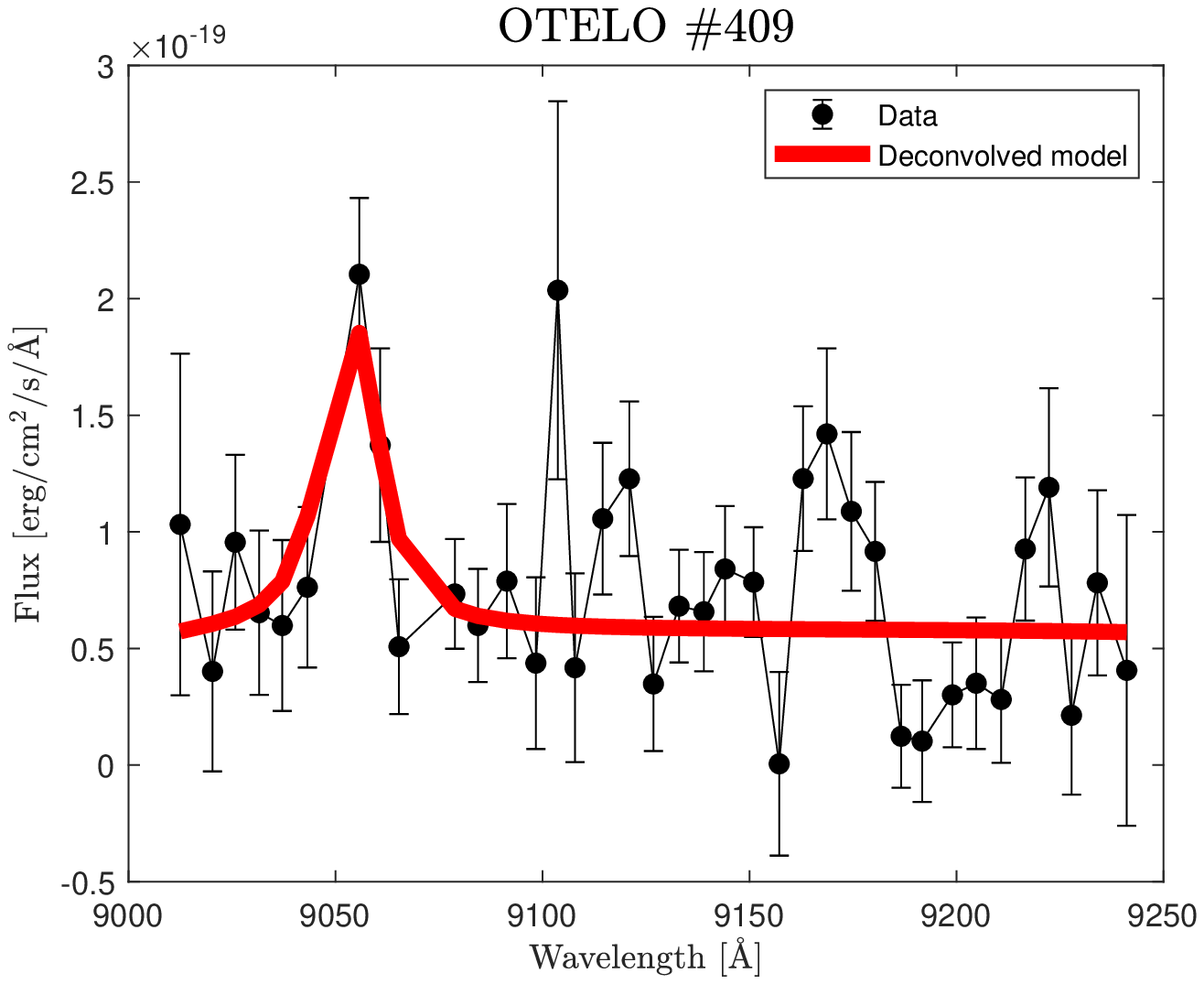}}
\subfloat{\includegraphics[width=.20\linewidth]{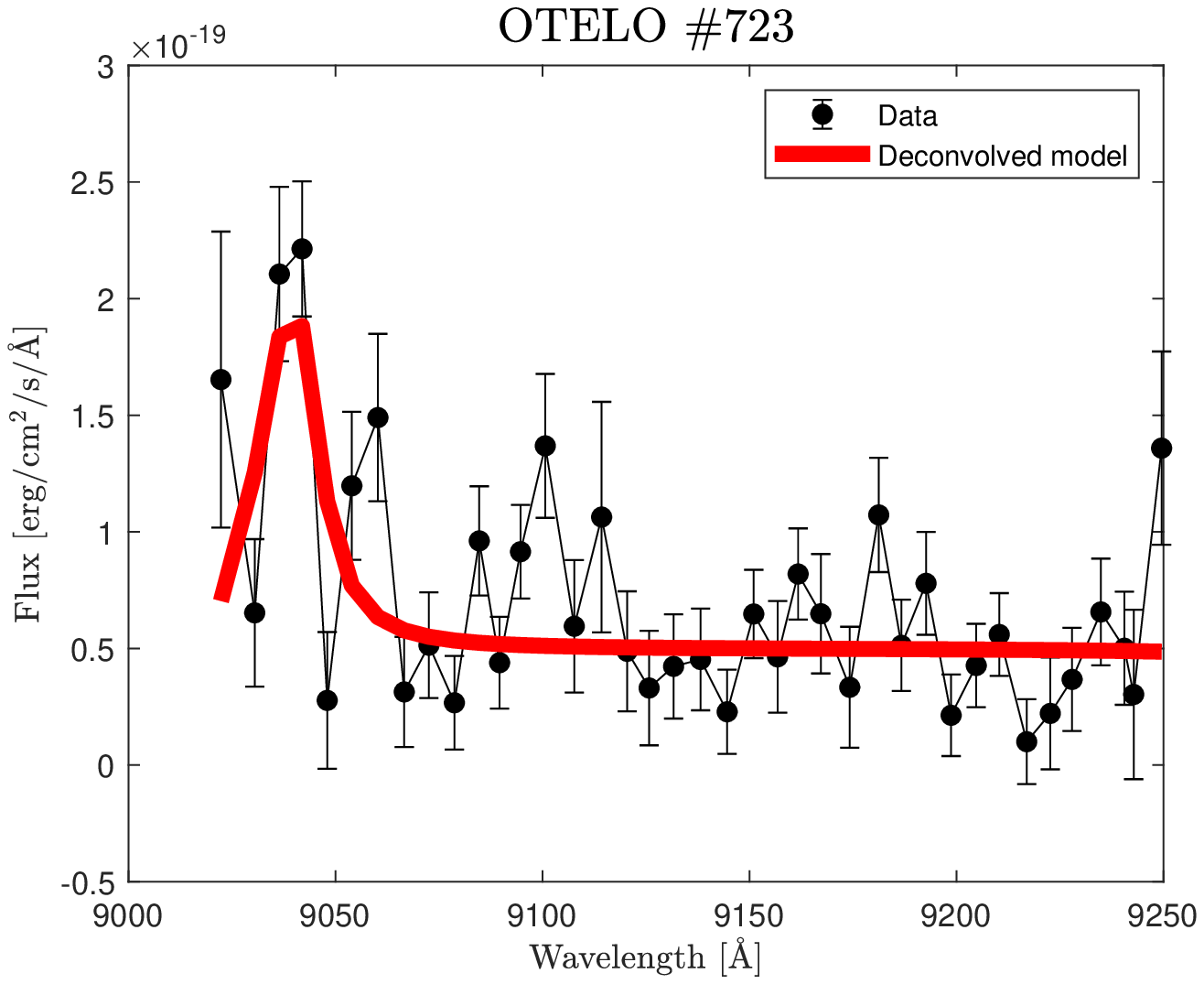}}
\
\subfloat{\includegraphics[width=.20\linewidth]{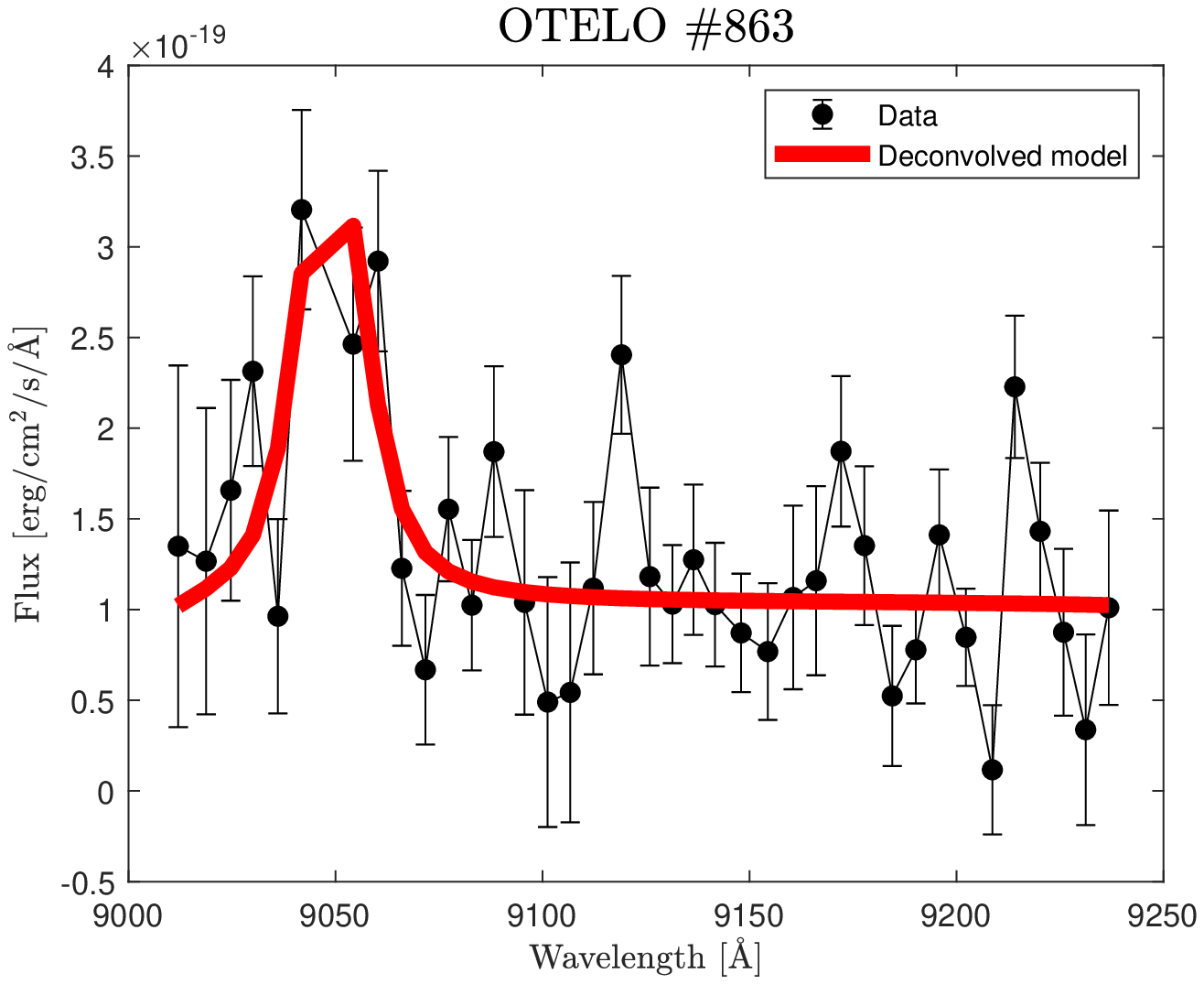}}
\subfloat{\includegraphics[width=.20\linewidth]{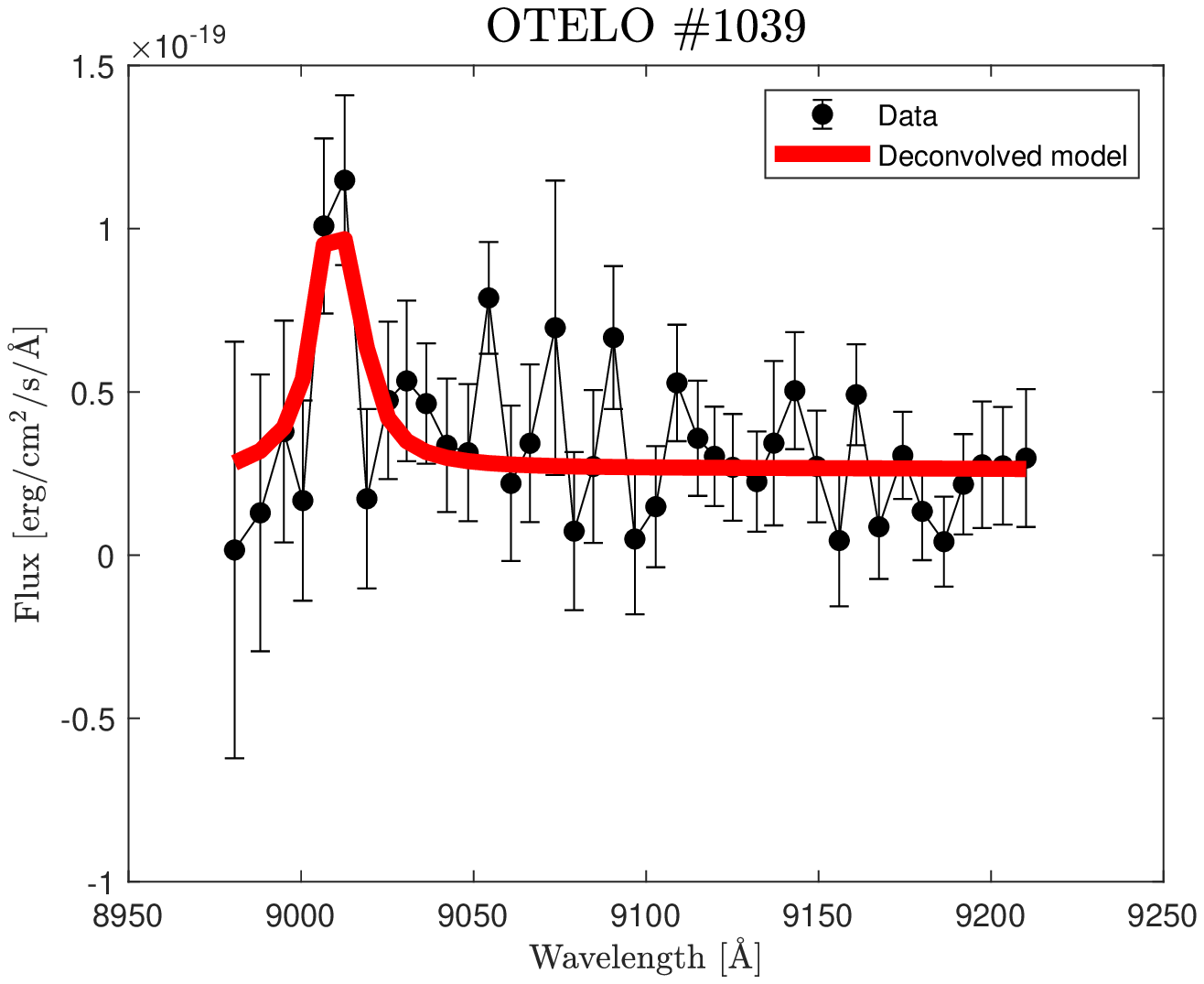}}
\subfloat{\includegraphics[width=.20\linewidth]{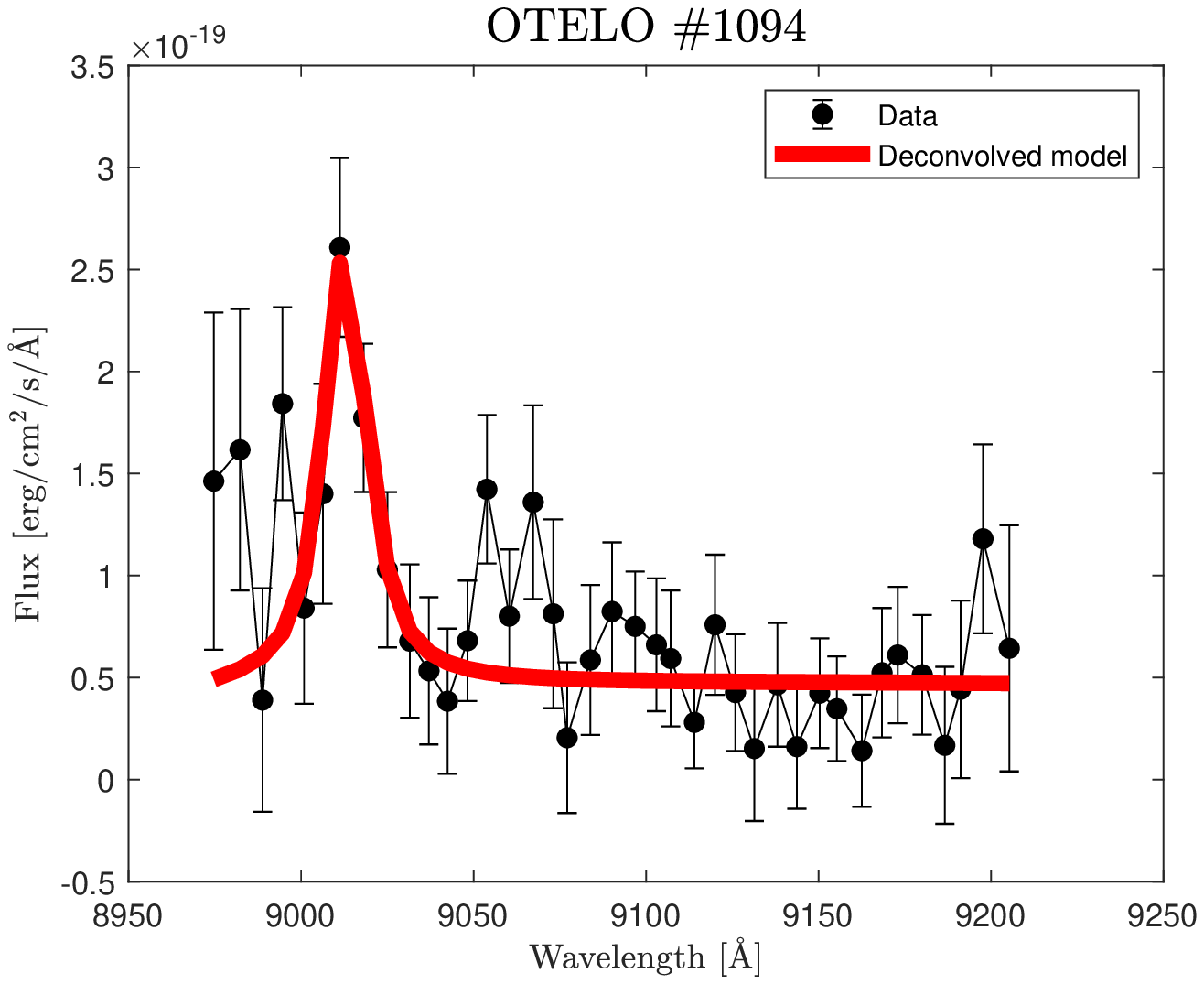}}
\subfloat{\includegraphics[width=.20\linewidth]{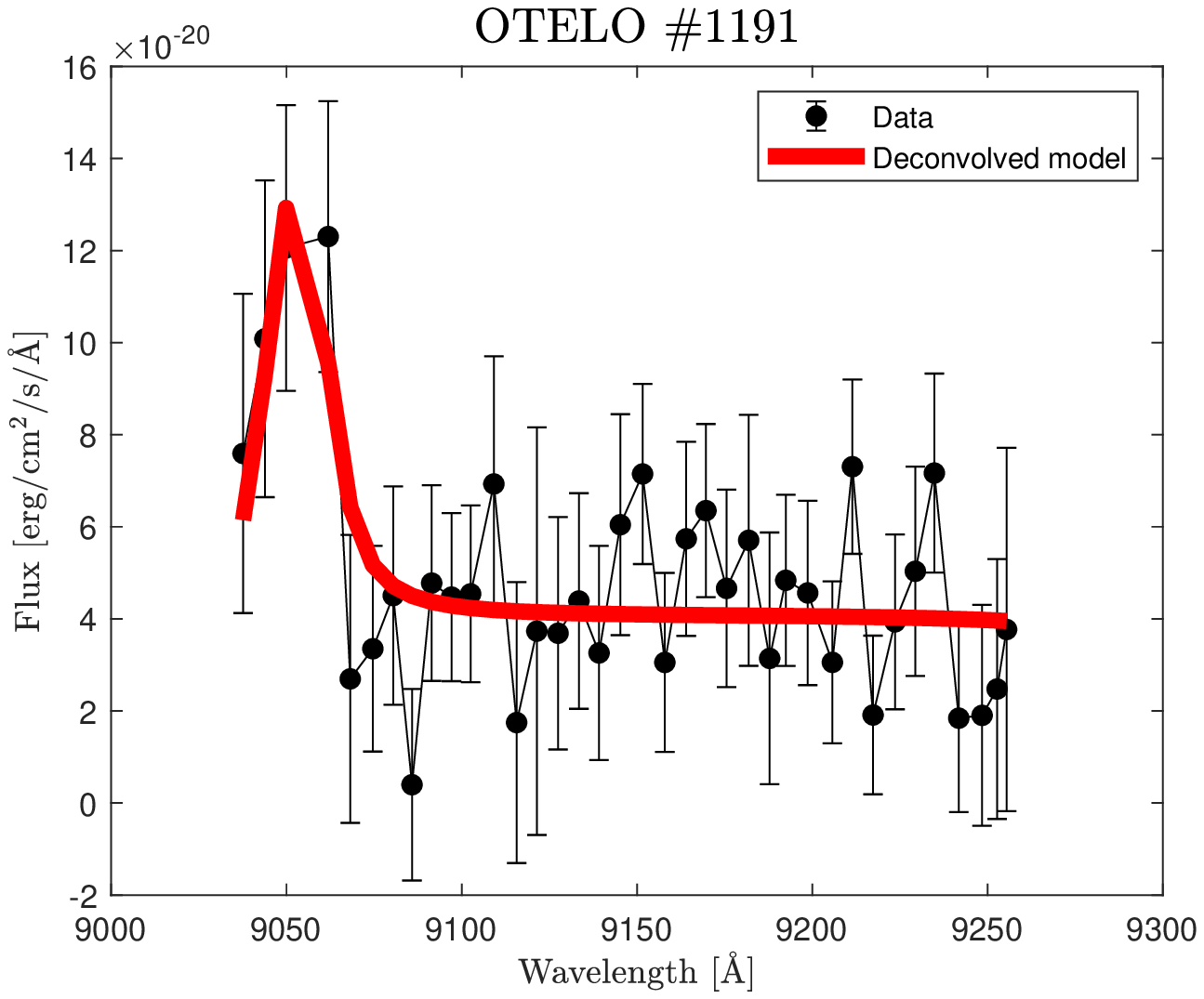}}\\
\subfloat{\includegraphics[width=.20\linewidth]{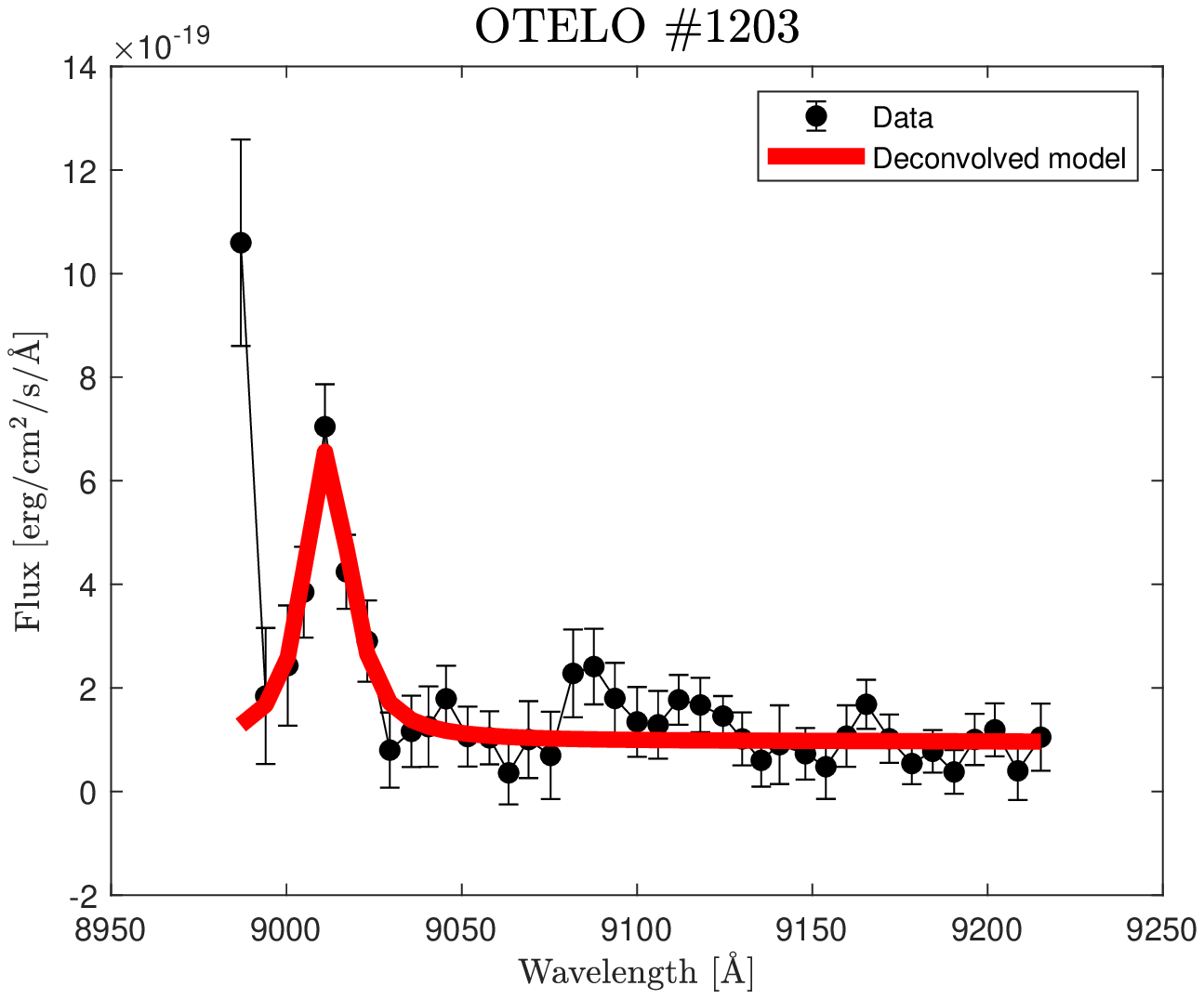}}
\subfloat{\includegraphics[width=.20\linewidth]{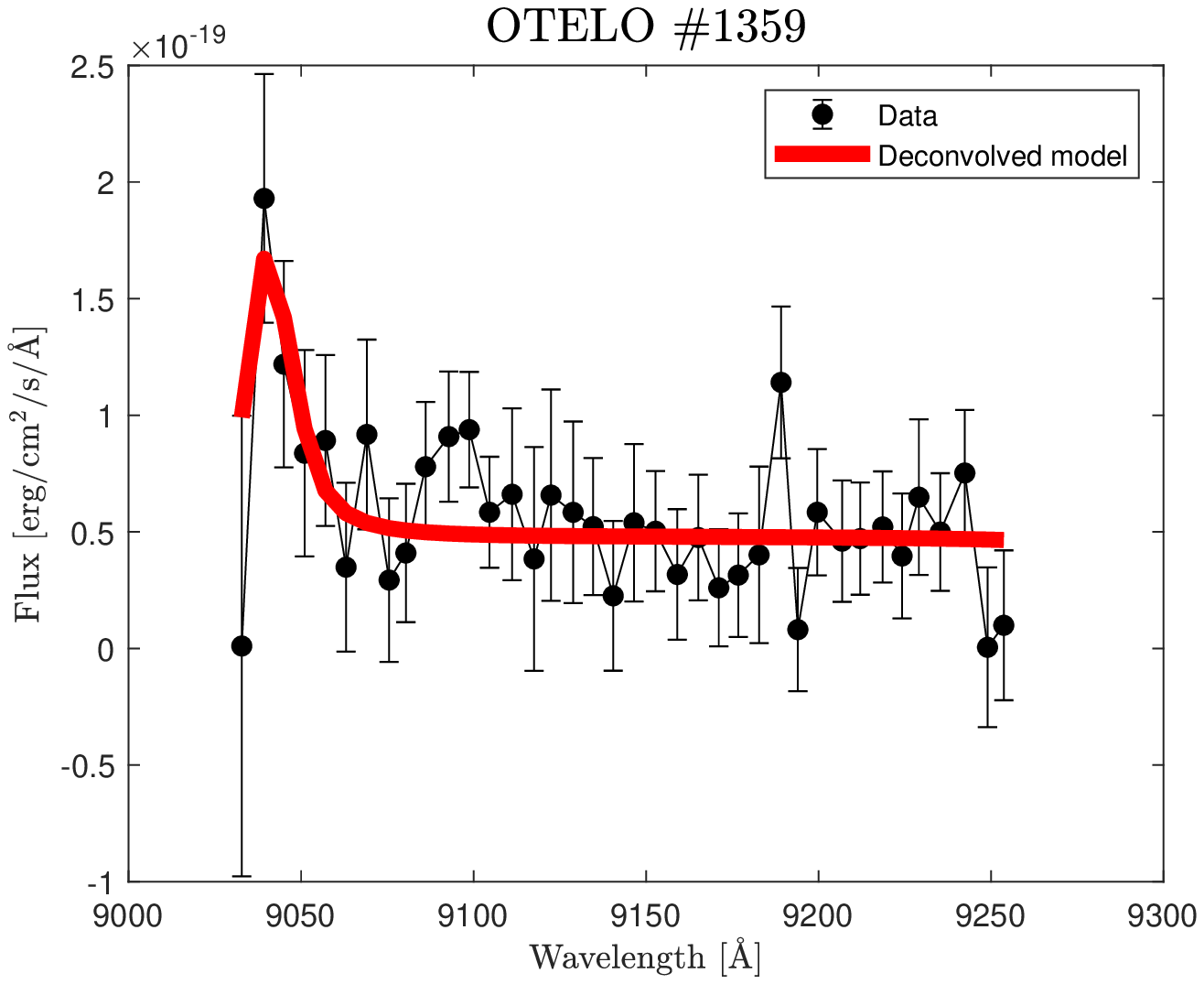}}
\subfloat{\includegraphics[width=.20\linewidth]{1494.eps}}
\subfloat{\includegraphics[width=.20\linewidth]{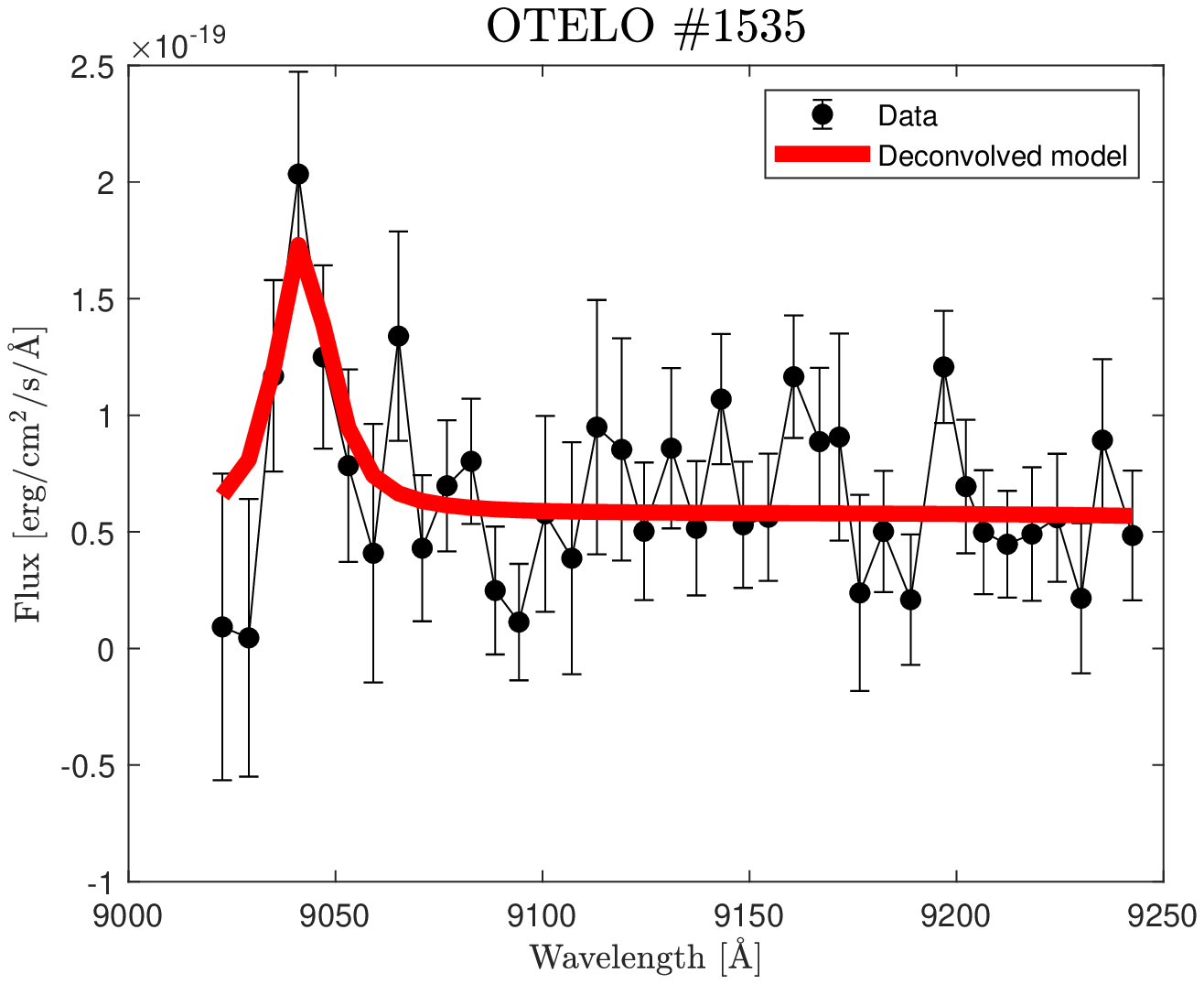}}\\
\subfloat{\includegraphics[width=.20\linewidth]{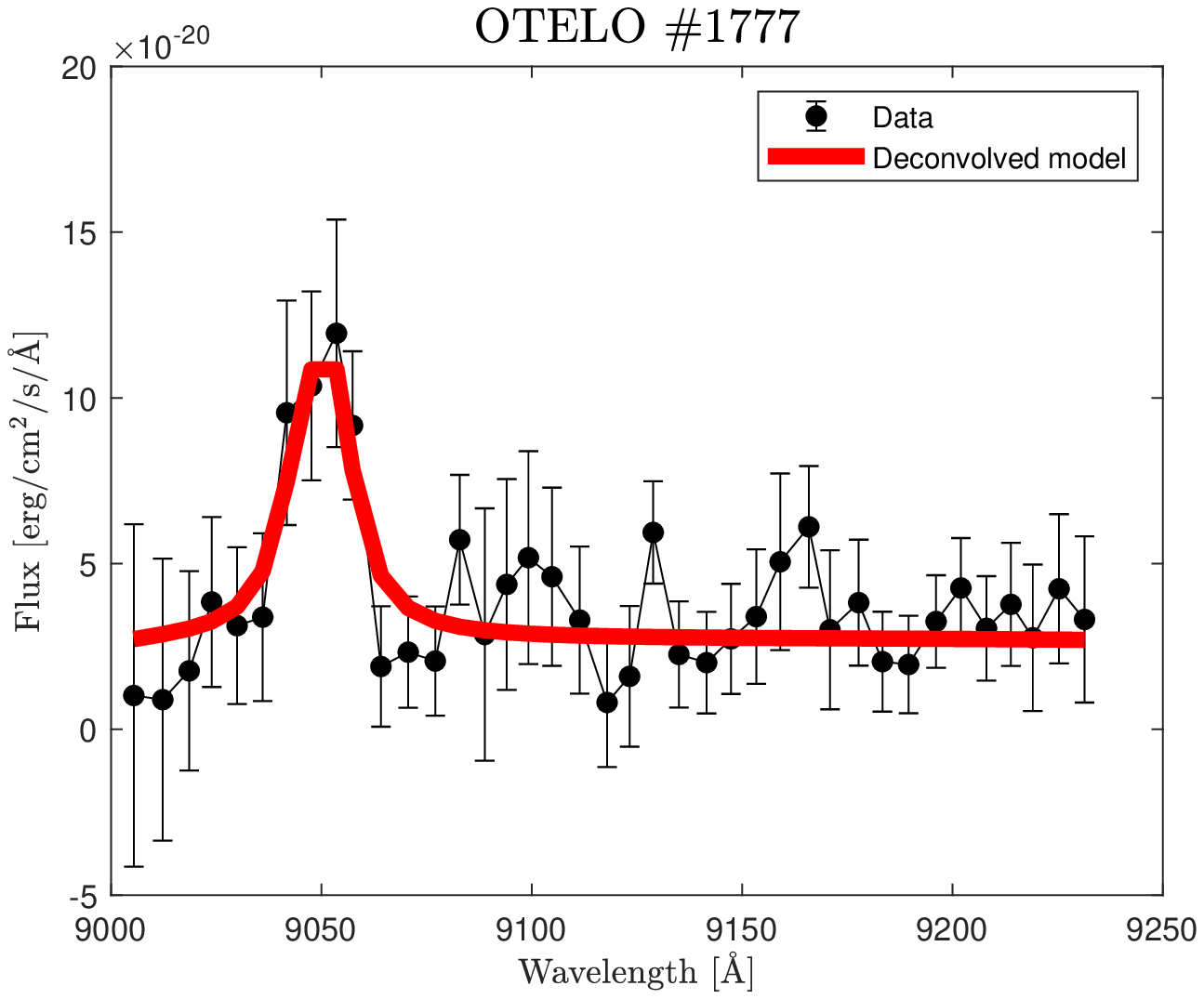}}
\subfloat{\includegraphics[width=.20\linewidth]{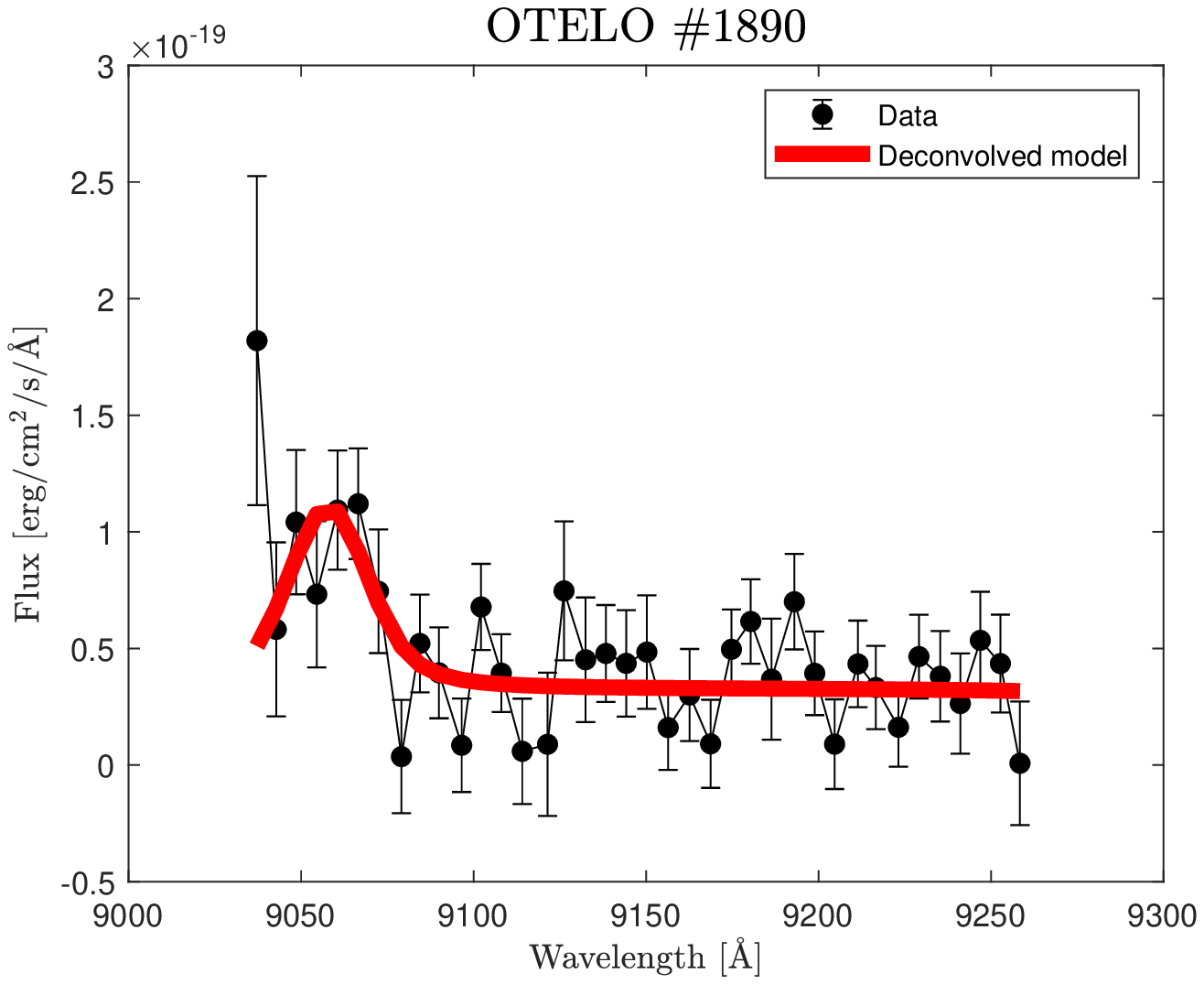}}
\subfloat{\includegraphics[width=.20\linewidth]{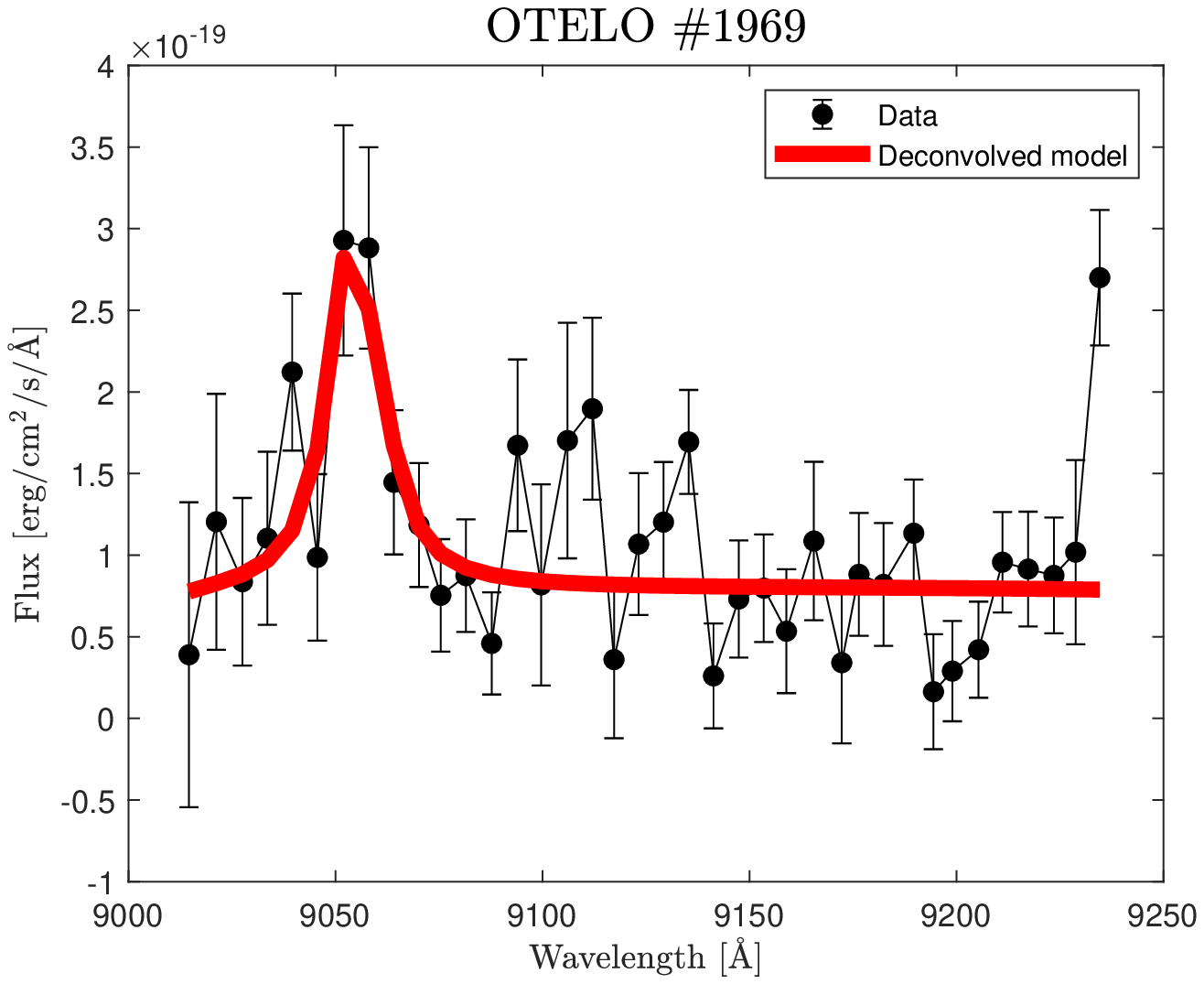}}
\subfloat{\includegraphics[width=.20\linewidth]{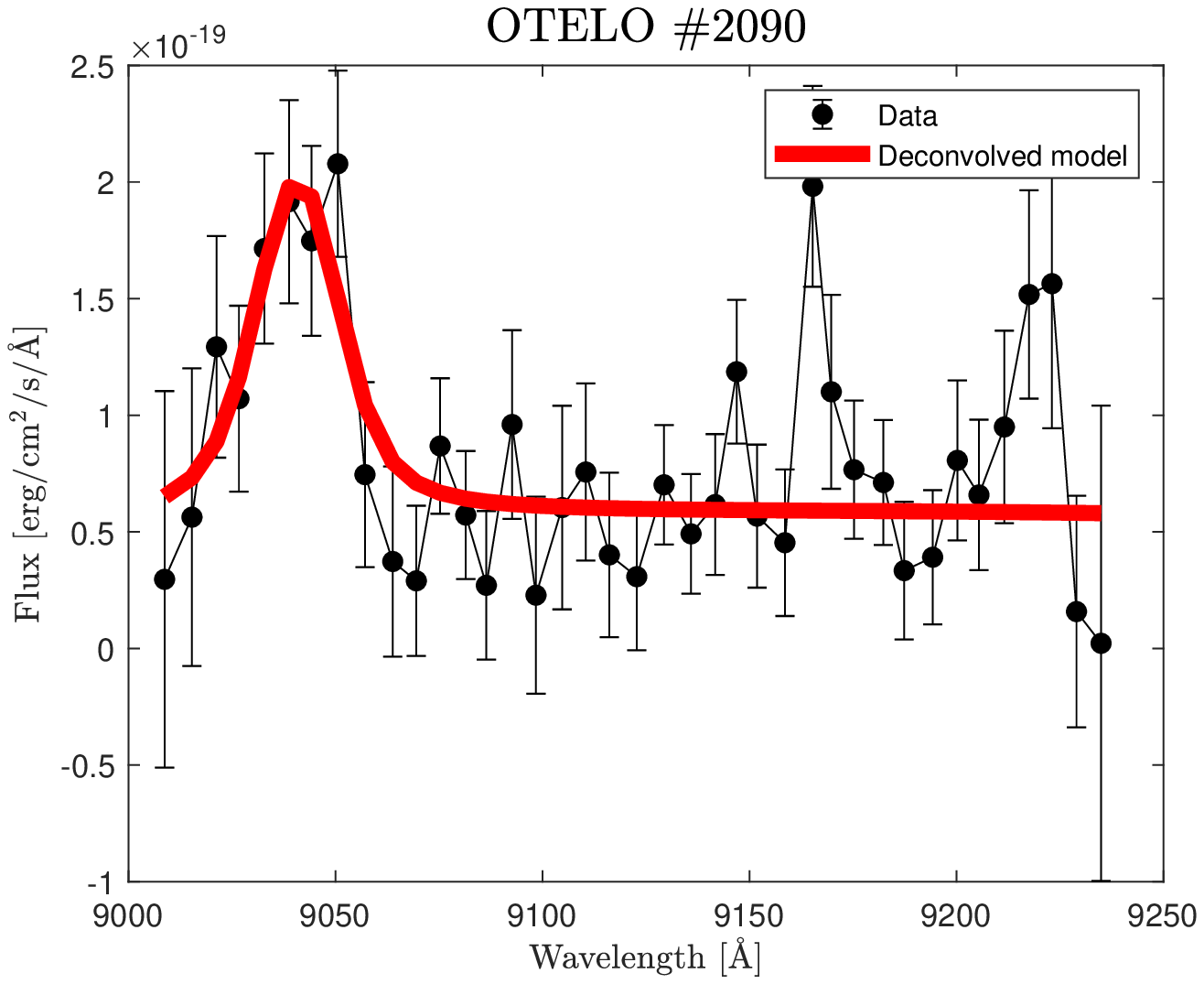}}\\
\subfloat{\includegraphics[width=.20\linewidth]{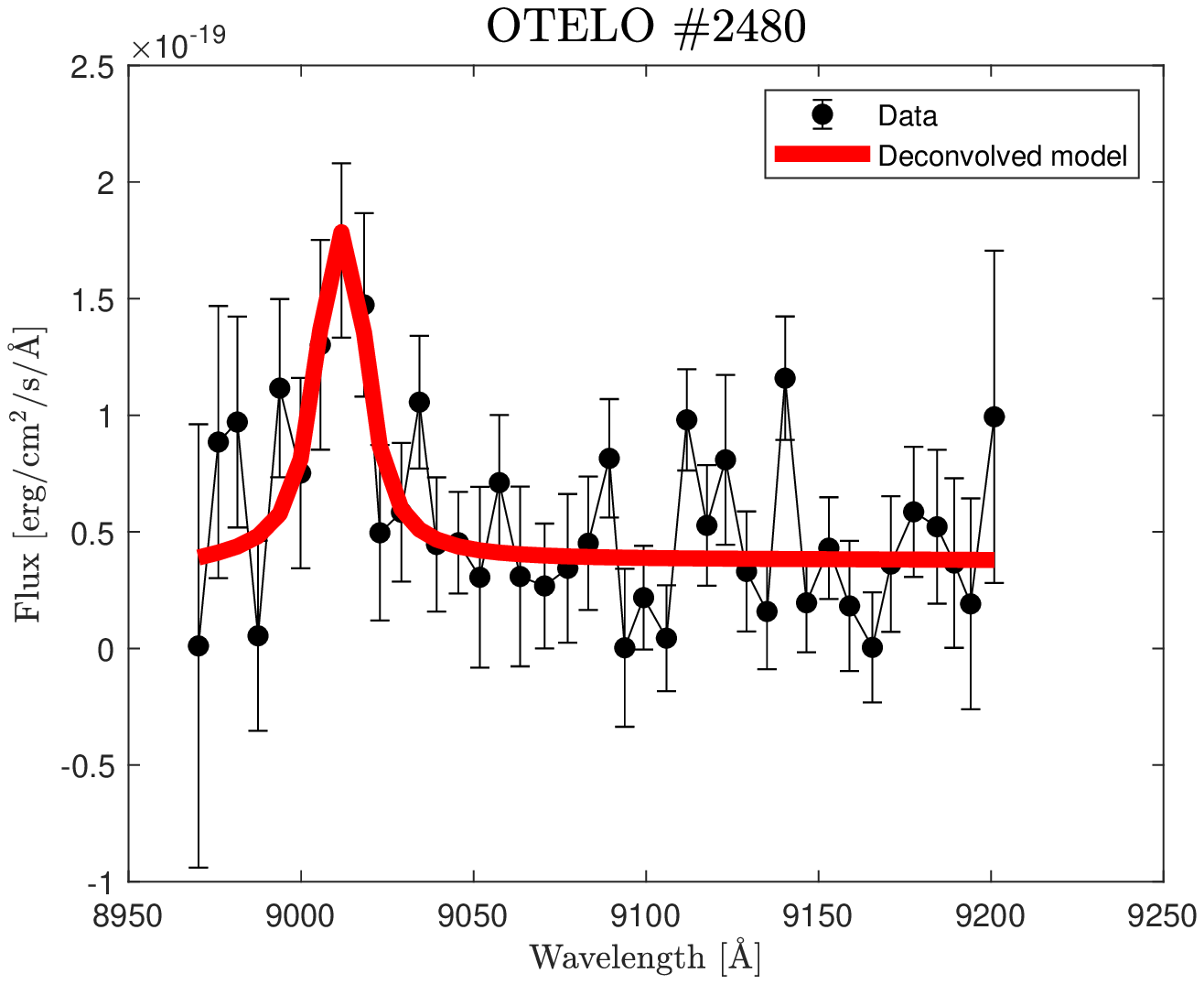}}
\subfloat{\includegraphics[width=.20\linewidth]{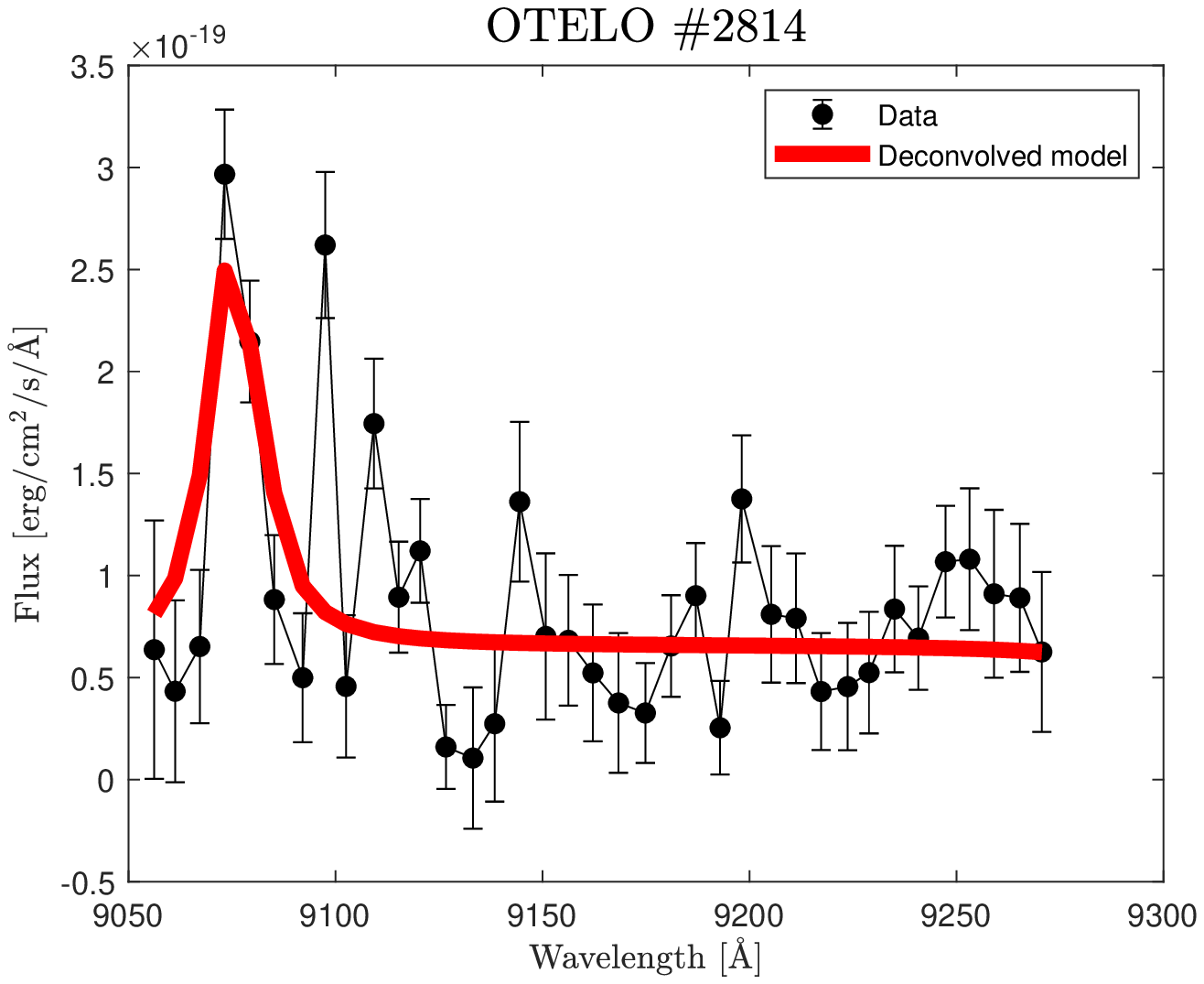}}
\subfloat{\includegraphics[width=.20\linewidth]{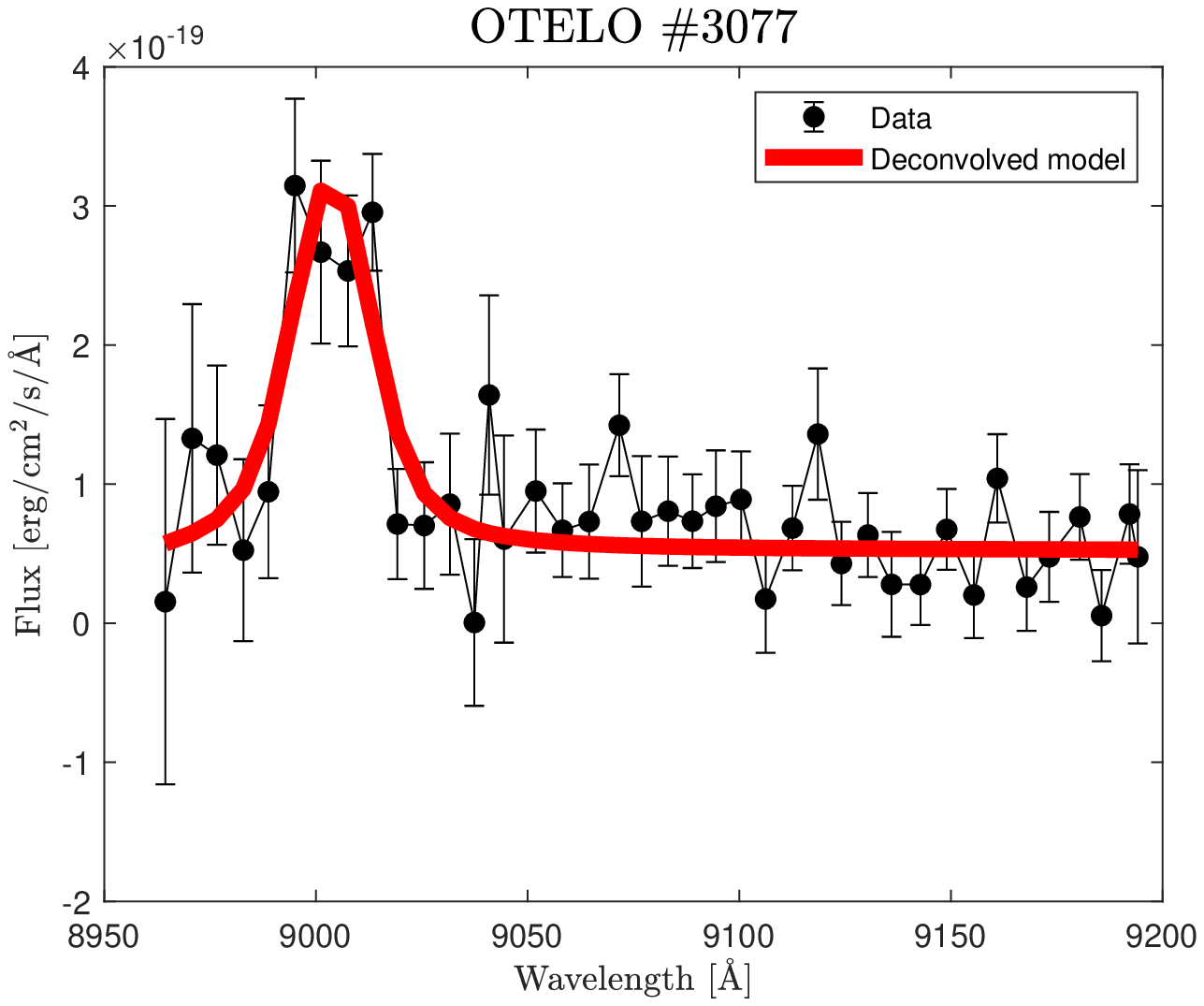}}
\subfloat{\includegraphics[width=.20\linewidth]{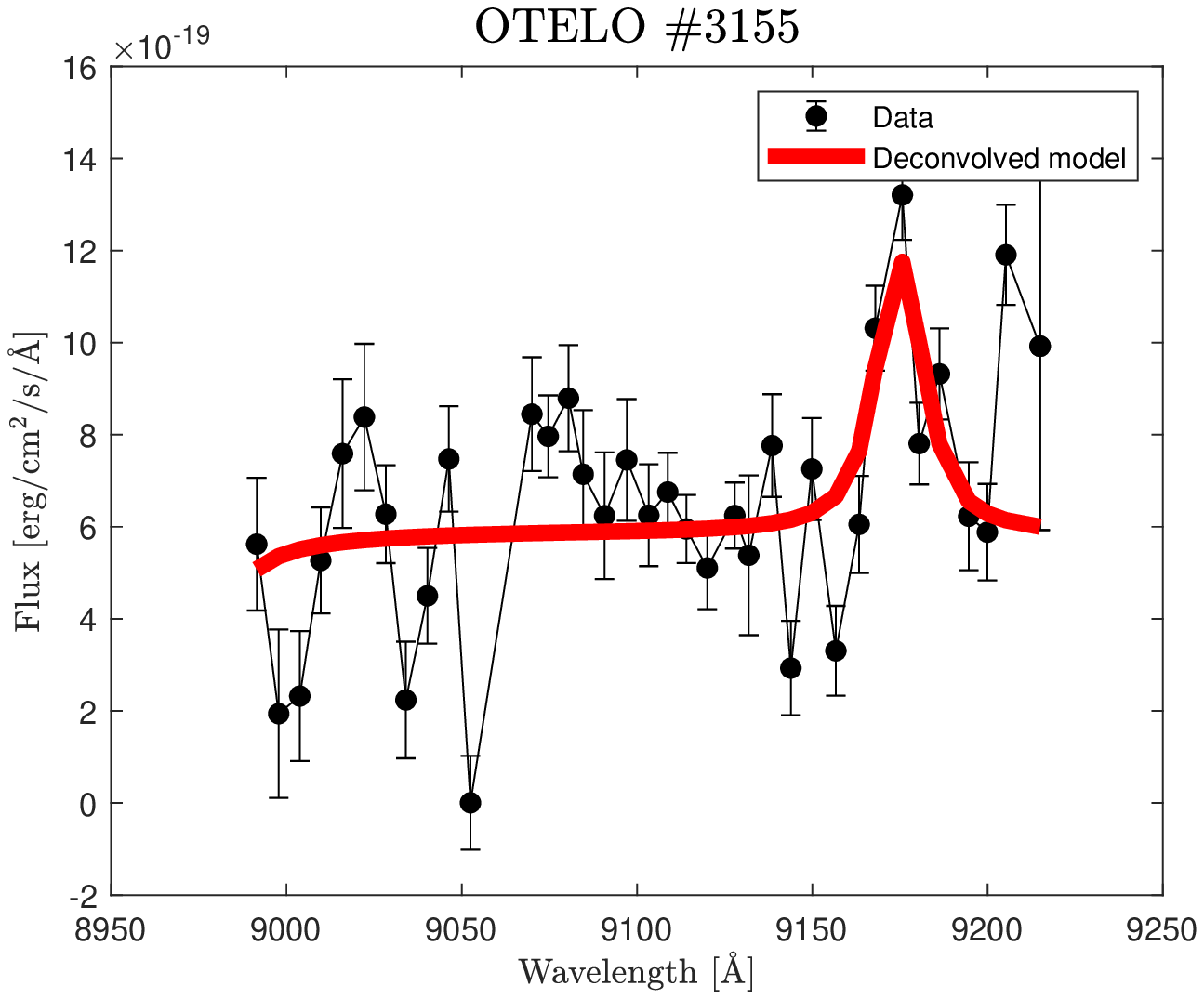}}
    \caption{Pseudo-spectra of the selected emitters. Black dots represent the measured pseudo-spectra, the red line is the best fitted deconvolved spectra.}
    \label{primera}
\end{figure*}

\begin{table*}
\caption{Selected emitters}            
\label{emisores}     
\centering
\begin{tabular}{c c c c c c c c}        
 
\hline              
OTELO ID. \# & RA & DEC & $z_{\rm\, OTELO}$ & E(B-V) & Flux \ox &  E.W. & $\log ({M_*/[{\rm M_\odot}]})$\\
 & [deg] & [deg] & & [mag] & [10$^{-17}$ erg/cm$^2$/s] & [\AA]  \\
\hline                        
00021 & 214.46764 & 52.41016 & 1.4142$\pm$0.0012 & 0.05 & 0.48$\pm$0.16 & 93$\pm$50  & 8.05$\pm$0.53\\
00381 & 214.40169 & 52.41219 & 1.4692$\pm$0.0004 & 0.16 & 0.87$\pm$0.23 & 62$\pm$40  & 8.37$\pm$0.10\\
00409 & 214.37597 & 52.41288 & 1.4287$\pm$0.0010 & 0.05 & 0.41$\pm$0.10 & 53$\pm$18  & 8.41$\pm$0.04\\
00723 & 214.40686 & 52.41580 & 1.4247$\pm$0.0010 & 0.10 & 0.62$\pm$0.12 & 71$\pm$16  & 8.97$\pm$0.10\\
00863 & 214.45145 & 52.41798 & 1.4280$\pm$0.0010 & 0.45 & 7.90$\pm$1.60 & 68$\pm$18  & 9.07$\pm$0.05\\
01039 & 214.48717 & 52.41881 & 1.4178$\pm$0.0012 & 0.05 & 0.24$\pm$0.07 & 68$\pm$29  & 7.89$\pm$0.18\\
01094 & 214.31071 & 52.42011 & 1.4184$\pm$0.0012 & 0.25 & 1.80$\pm$0.55 & 97$\pm$40  & 8.83$\pm$0.08\\
01191 & 214.41452 & 52.42083 & 1.4285$\pm$0.0012 & 0.26 & 1.10$\pm$0.33 & 65$\pm$23  & N/A\\
01203 & 214.48628 & 52.42192 & 1.4181$\pm$0.0009 & 0.05 & 1.70$\pm$0.20 & 130$\pm$33 & 8.72$\pm$0.03\\
01359 & 214.42263 & 52.42350 & 1.4261$\pm$0.0016 & 0.10 & 0.48$\pm$0.35 & 58$\pm$52  & N/A\\
01494 & 214.45005 & 52.42484 & 1.4292$\pm$0.0009 & 0.15 & 1.20$\pm$0.25 & 61$\pm$16  & 8.59$\pm$0.09\\
01535 & 214.45148 & 52.42487 & 1.4263$\pm$0.0011 & 0.05 & 0.35$\pm$0.10 & 45$\pm$17  & 8.38$\pm$0.11\\
01777 & 214.47556 & 52.42722 & 1.4277$\pm$0.0010 & 0.00 & 0.22$\pm$0.05 & 78$\pm$30  & 8.67$\pm$0.09\\
01890 & 214.43149 & 52.42841 & 1.4299$\pm$0.0013 & 0.25 & 1.10$\pm$0.46 & 85$\pm$44  & 8.64$\pm$0.15\\
01969 & 214.46984 & 52.43001 & 1.4297$\pm$0.0010 & 0.05 & 0.64$\pm$0.15 & 60$\pm$18  & 8.59$\pm$0.04\\
02090 & 214.32741 & 52.43109 & 1.4253$\pm$0.0010 & 0.00 & 0.46$\pm$0.11 & 77$\pm$23  & N/A\\
02480 & 214.29067 & 52.43492 & 1.4175$\pm$0.0010 & 0.05 & 0.46$\pm$0.18 & 90$\pm$71  & 8.25$\pm$0.19\\
02814 & 214.39153 & 52.43922 & 1.4353$\pm$0.0003 & 0.20 & 1.30$\pm$0.17 & 65$\pm$15  & 9.78$\pm$0.13\\
03077 & 214.45529 & 52.53075 & 1.4155$\pm$0.0007 & 0.05 & 1.10$\pm$0.18 & 150$\pm$36 & 8.17$\pm$0.10\\
03155 & 214.40408 & 52.52958 & 1.4619$\pm$0.0004 & 0.20 & 4.20$\pm$0.62 & 24$\pm$4   & 10.9$\pm$0.01\\
03345 & 214.32088 & 52.52815 & 1.4060$\pm$0.0005 & 0.05 & 0.81$\pm$0.15 & 140$\pm$39 & N/A\\
03633 & 214.48850 & 52.52531 & 1.4053$\pm$0.0012 & 0.05 & 0.53$\pm$0.22 & 79$\pm$41  & 7.93$\pm$0.51\\
03669 & 214.47178 & 52.52496 & 1.4123$\pm$0.0011 & 0.74 & 27.0$\pm$6.80 & 85$\pm$35  & N/A\\
03707 & 214.41281 & 52.52459 & 1.4176$\pm$0.0020 & 0.36 & 1.60$\pm$1.30 & 75$\pm$77  & N/A\\
04070 & 214.36268 & 52.52005 & 1.4190$\pm$0.0018 & 0.16 & 1.20$\pm$0.39 & 73$\pm$28  & 8.65$\pm$0.03\\
04308 & 214.34333 & 52.51710 & 1.4646$\pm$0.0008 & 0.10 & 3.60$\pm$0.62 & 18$\pm$3   & 10.6$\pm$0.01\\ 
04313 & 214.38326 & 52.51718 & 1.4283$\pm$0.0003 & 0.16 & 1.60$\pm$0.27 & 38$\pm$7   & 9.37$\pm$0.01\\
04456 & 214.39479 & 52.51559 & 1.4290$\pm$0.0010 & 0.16 & 1.10$\pm$0.25 & 55$\pm$17  & 8.02$\pm$0.11\\
04611 & 214.41167 & 52.51229 & 1.4498$\pm$0.0008 & 0.16 & 3.20$\pm$0.45 & 24$\pm$4   & 9.54$\pm$0.01\\
04814 & 214.43320 & 52.51136 & 1.4458$\pm$0.0008 & 0.16 & 1.40$\pm$0.23 & 120$\pm$29 & 8.27$\pm$0.09\\
05078 & 214.34538 & 52.47537 & 1.4310$\pm$0.0010 & 0.05 & 0.24$\pm$0.06 & 78$\pm$34  & 7.99$\pm$0.10\\
05090 & 214.41584 & 52.47524 & 1.4771$\pm$0.0003 & 0.10 & 1.50$\pm$0.14 & 89$\pm$14  & 9.57$\pm$0.03\\
05260 & 214.31066 & 52.47391 & 1.4258$\pm$0.0011 & 0.30 & 1.30$\pm$0.39 & 33$\pm$14  & 8.72$\pm$0.04\\
05957 & 214.30764 & 52.46609 & 1.4737$\pm$0.0012 & 0.16 & 0.88$\pm$0.36 & 28$\pm$13  & 9.26$\pm$0.01\\
06196 & 214.47412 & 52.46364 & 1.4484$\pm$0.0009 & 0.00 & 0.22$\pm$0.08 & 45$\pm$19  & 9.24$\pm$0.06\\
06398 & 214.37361 & 52.46292 & 1.4580$\pm$0.0010 & 0.16 & 0.29$\pm$0.11 & 45$\pm$20  & 8.71$\pm$0.04\\
06445 & 214.47302 & 52.46081 & 1.4410$\pm$0.0007 & 0.45 & 1.40$\pm$0.26 & 180$\pm$46 & N/A\\
06758 & 214.40821 & 52.45696 & 1.4439$\pm$0.0010 & 0.00 & 0.21$\pm$0.08 & 59$\pm$28  & N/A\\
06844 & 214.30997 & 52.45600 & 1.4267$\pm$0.0013 & 0.16 & 0.44$\pm$0.13 & 55$\pm$23  & N/A\\
06959 & 214.46646 & 52.45496 & 1.4291$\pm$0.0012 & 0.00 & 0.19$\pm$0.06 & 63$\pm$26  & 8.86$\pm$0.12\\
07183 & 214.46175 & 52.45150 & 1.4359$\pm$0.0011 & 0.45 & 3.90$\pm$1.70 & 10$\pm$5   & 9.93$\pm$0.01\\
07431 & 214.36263 & 52.44858 & 1.4601$\pm$0.0003 & 0.05 & 2.60$\pm$0.19 & 130$\pm$13 & 9.05$\pm$0.01\\
07623 & 214.30638 & 52.44692 & 1.4301$\pm$0.0010 & 0.00 & 0.42$\pm$0.14 & 50$\pm$20  & N/A\\
08276 & 214.41274 & 52.44036 & 1.4414$\pm$0.0007 & 0.30 & 0.92$\pm$0.27 & 65$\pm$29  & N/A\\
08313 & 214.35177 & 52.50993 & 1.4307$\pm$0.0012 & 0.66 & 17.0$\pm$4.50 & 23$\pm$7   & 10.3$\pm$0.01\\
08333 & 214.43944 & 52.51062 & 1.4264$\pm$0.0013 & 0.16 & 0.79$\pm$0.27 & 70$\pm$36  & 8.48$\pm$0.03\\
08406 & 214.41518 & 52.50931 & 1.4756$\pm$0.0003 & 0.15 & 0.63$\pm$0.12 & 67$\pm$17  & 8.97$\pm$0.05\\
08453 & 214.46035 & 52.49464 & 1.4307$\pm$0.0012 & 0.05 & 0.32$\pm$0.10 & 55$\pm$24  & 8.6$\pm$ 0.04\\
08532 & 214.32581 & 52.50783 & 1.4255$\pm$0.0004 & 0.16 & 12.0$\pm$0.51 & 185$\pm$16 & 9.41$\pm$0.01\\
08533 & 214.33172 & 52.50768 & 1.4174$\pm$0.0018 & 0.00 & 0.25$\pm$0.12 & 44$\pm$25  & 8.35$\pm$0.07\\
09285 & 214.42861 & 52.50071 & 1.4286$\pm$0.0013 & 0.05 & 0.40$\pm$0.13 & 50$\pm$20  & N/A\\
09393 & 214.34782 & 52.49960 & 1.4277$\pm$0.0011 & 0.10 & 0.80$\pm$0.23 & 68$\pm$25  & 9.21$\pm$0.03\\
09419 & 214.44289 & 52.49953 & 1.4274$\pm$0.0011 & 0.05 & 0.31$\pm$0.10 & 57$\pm$22  & 8.24$\pm$0.09\\
09532 & 214.34223 & 52.49744 & 1.4606$\pm$0.0004 & 0.16 & 6.80$\pm$0.41 & 120$\pm$14 & 9.22$\pm$0.01\\
09992 & 214.41634 & 52.47749 & 1.4763$\pm$0.0008 & 0.15 & 1.70$\pm$0.27 & 30$\pm$5   & 9.35$\pm$0.01\\
10114 & 214.48834 & 52.49102 & 1.4172$\pm$0.0013 & 0.05 & 0.87$\pm$0.20 & 76$\pm$22  & 8.78$\pm$0.02\\
10272 & 214.40064 & 52.48954 & 1.4742$\pm$0.0014 & 0.16 & 0.53$\pm$0.27 & 25$\pm$15  & 8.83$\pm$0.01\\
10628 & 214.36998 & 52.48522 & 1.4827$\pm$0.0014 & 0.16 & 0.40$\pm$0.17 & 34$\pm$16  & 8.65$\pm$0.02\\
10808 & 214.29183 & 52.48187 & 1.4146$\pm$0.0013 & 0.05 & 0.57$\pm$0.16 & 100$\pm$45 & 8.38$\pm$0.24\\
10814 & 214.36935 & 52.48267 & 1.4306$\pm$0.0018 & 0.15 & 0.61$\pm$0.37 & 18$\pm$11  & 9.31$\pm$0.01\\

\hline                                   
\end{tabular}
\end{table*}
In Fig. \ref{primera}, we represent, for all the emitters, the pseudo-spectrum flux in erg/cm$^2$/s/\AA\, (black dots and black lines) and the deconvolved model (red line and same units), as a function of the wavelength in \AA.\\

\begin{figure*}\continuedfloat
    \centering
\subfloat{\includegraphics[width=.20\linewidth]{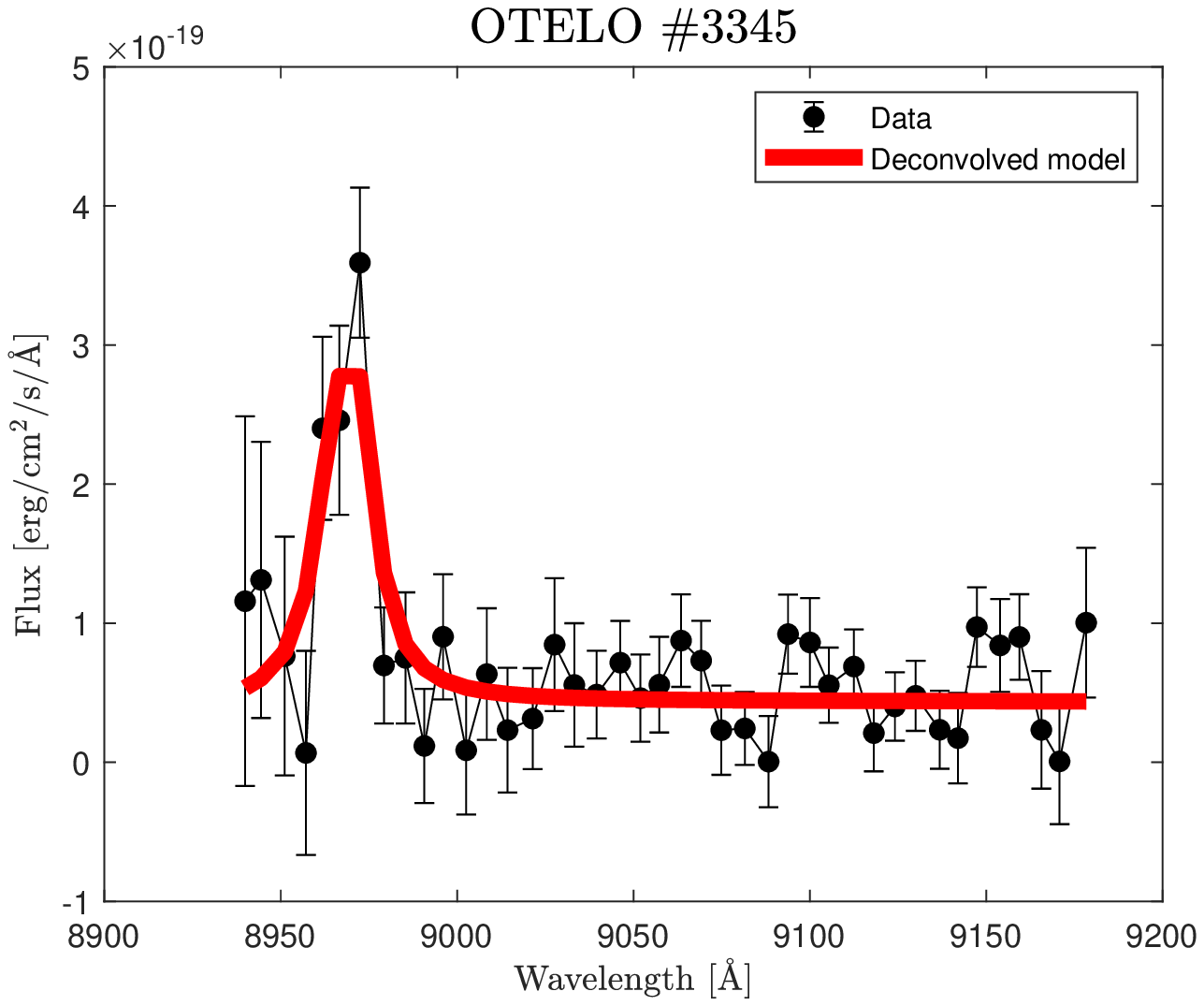}}
\subfloat{\includegraphics[width=.20\linewidth]{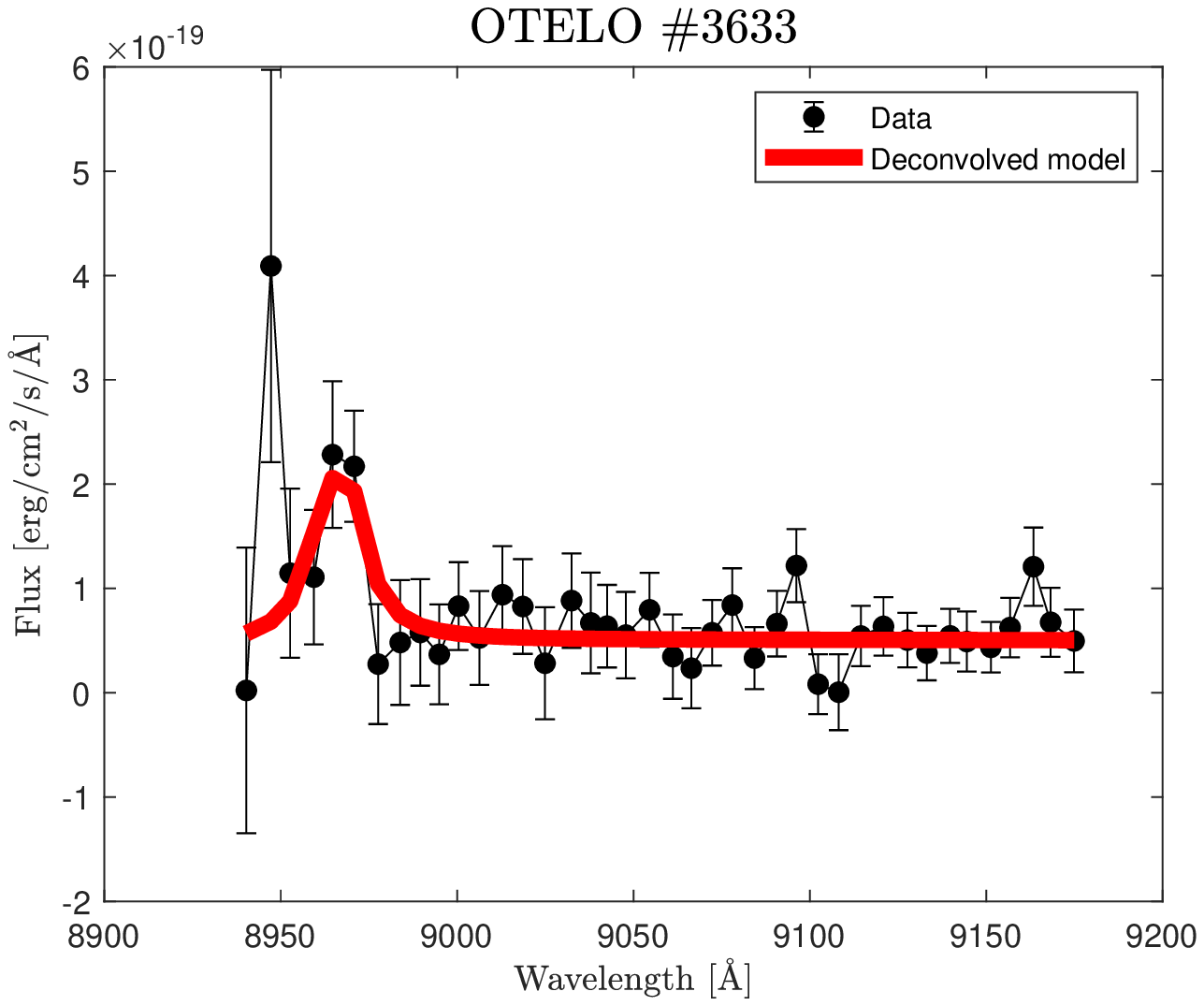}}
\subfloat{\includegraphics[width=.20\linewidth]{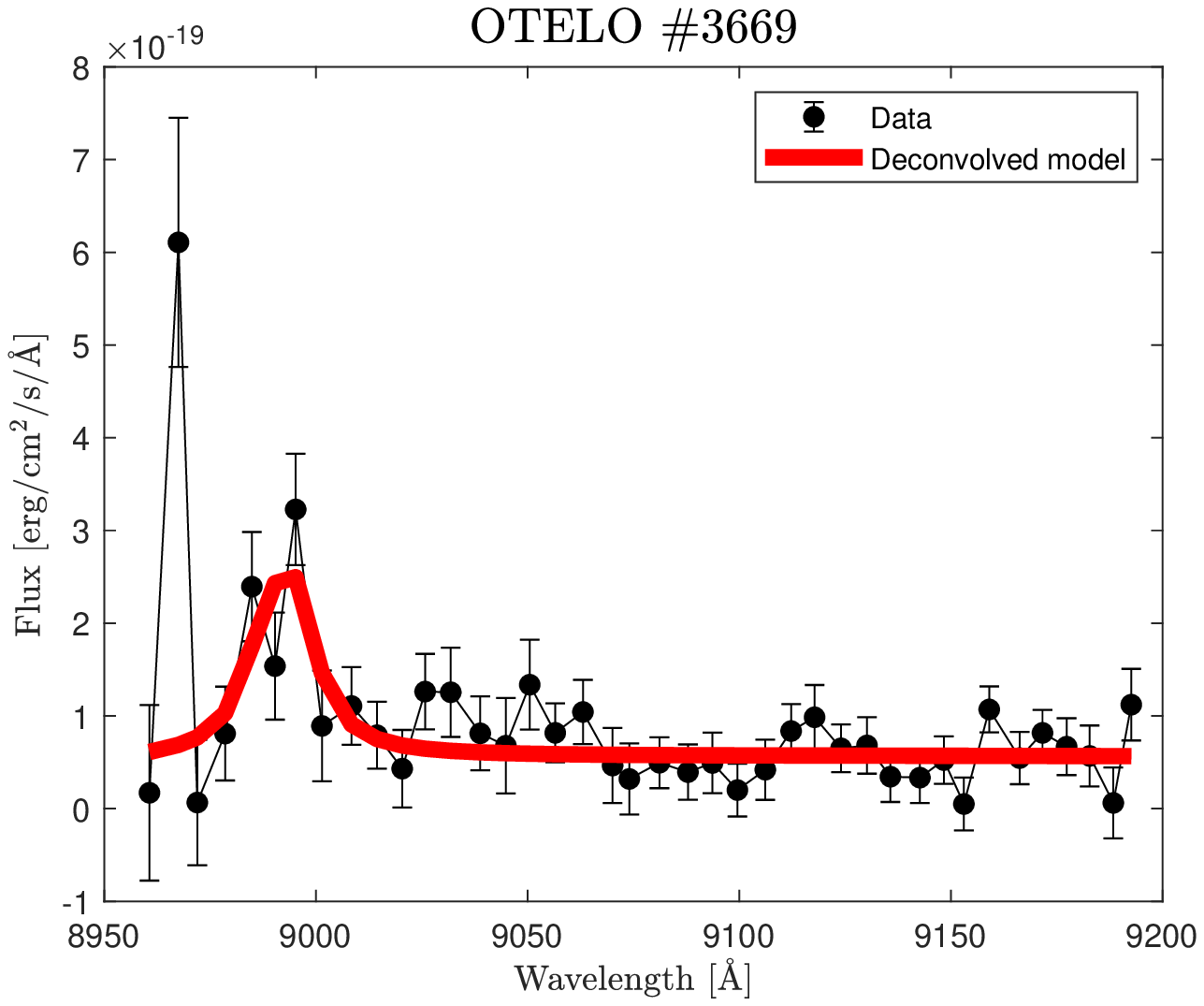}}
\subfloat{\includegraphics[width=.20\linewidth]{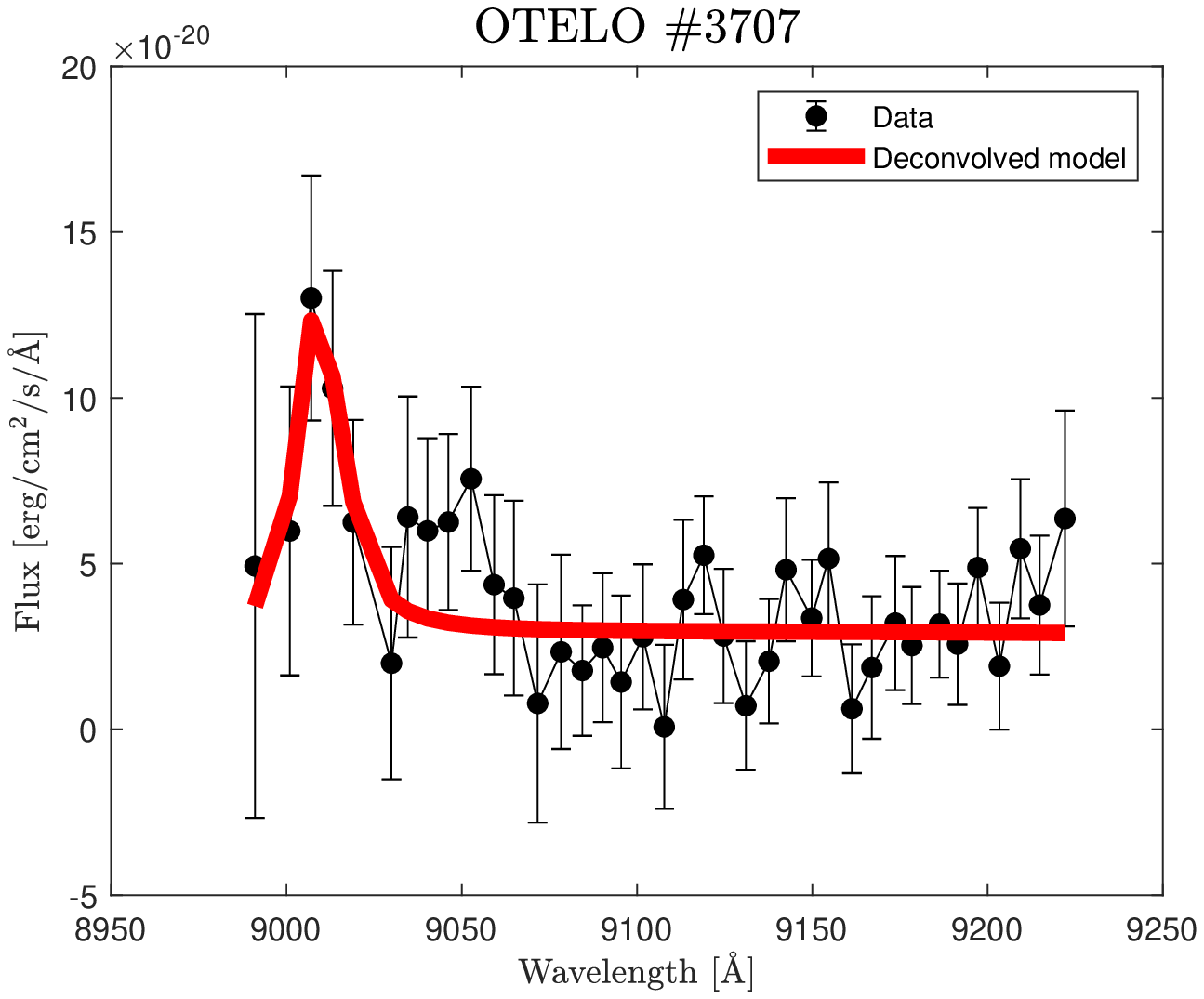}}\\
\subfloat{\includegraphics[width=.20\linewidth]{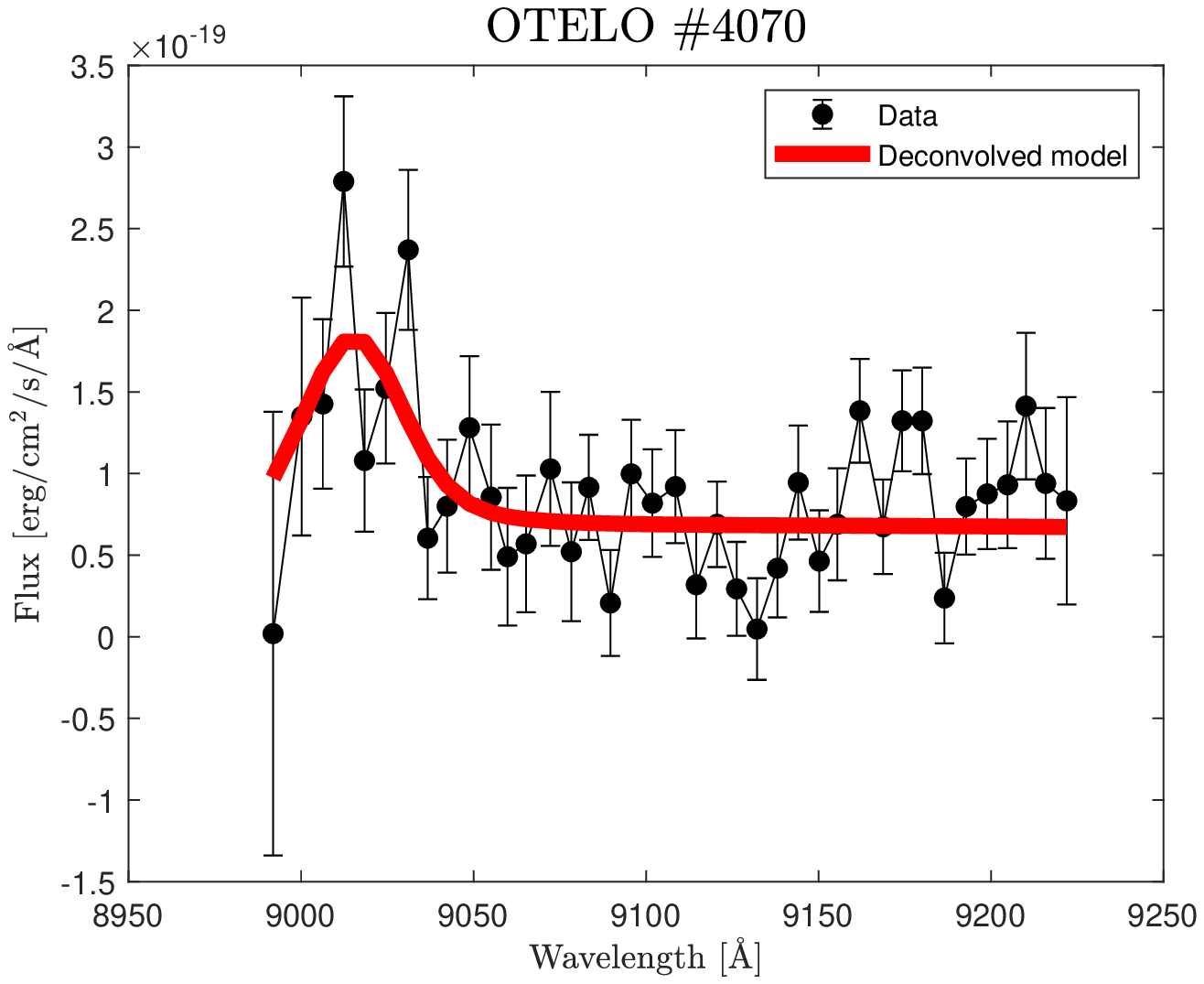}}
\subfloat{\includegraphics[width=.20\linewidth]{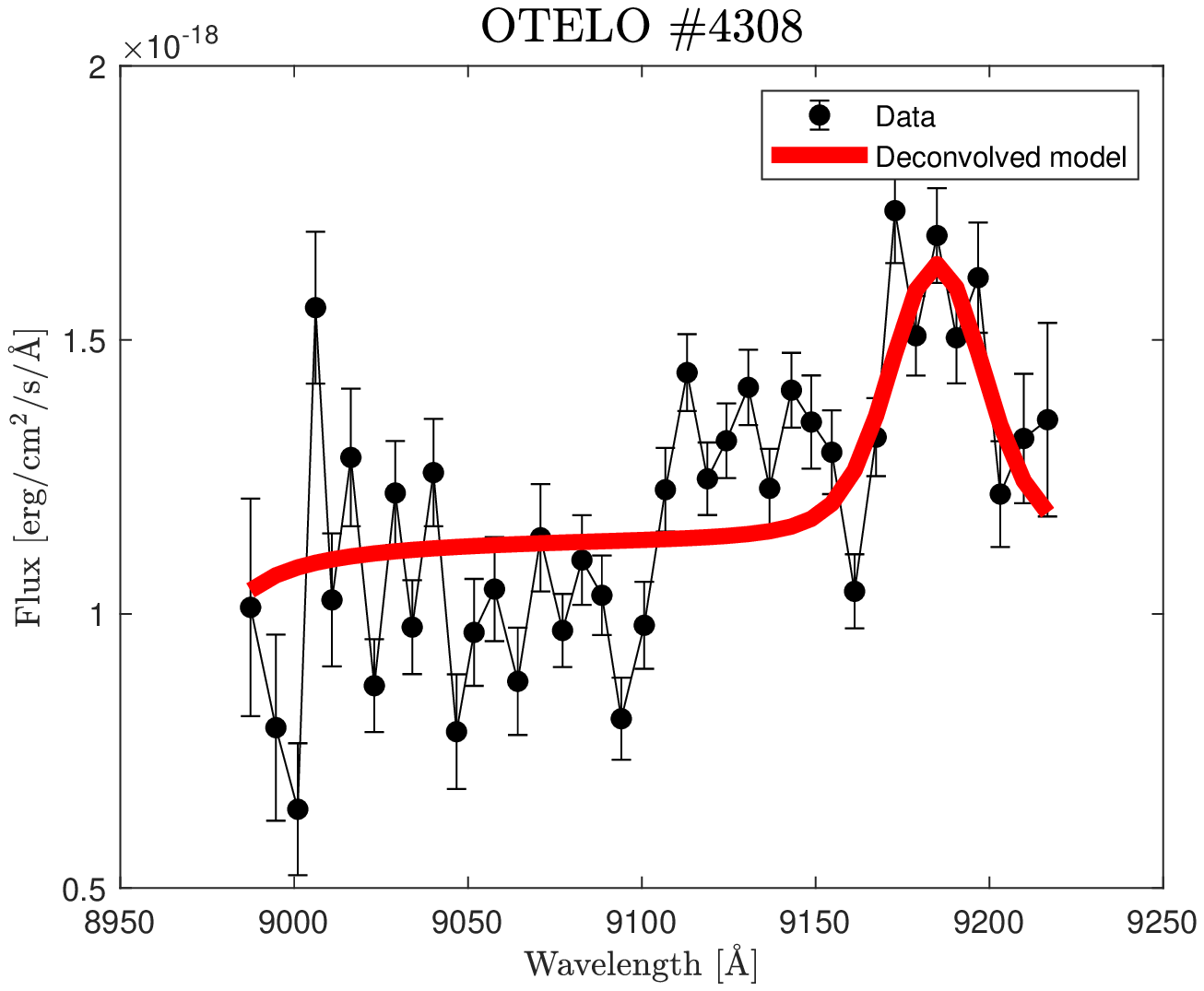}}
\subfloat{\includegraphics[width=.20\linewidth]{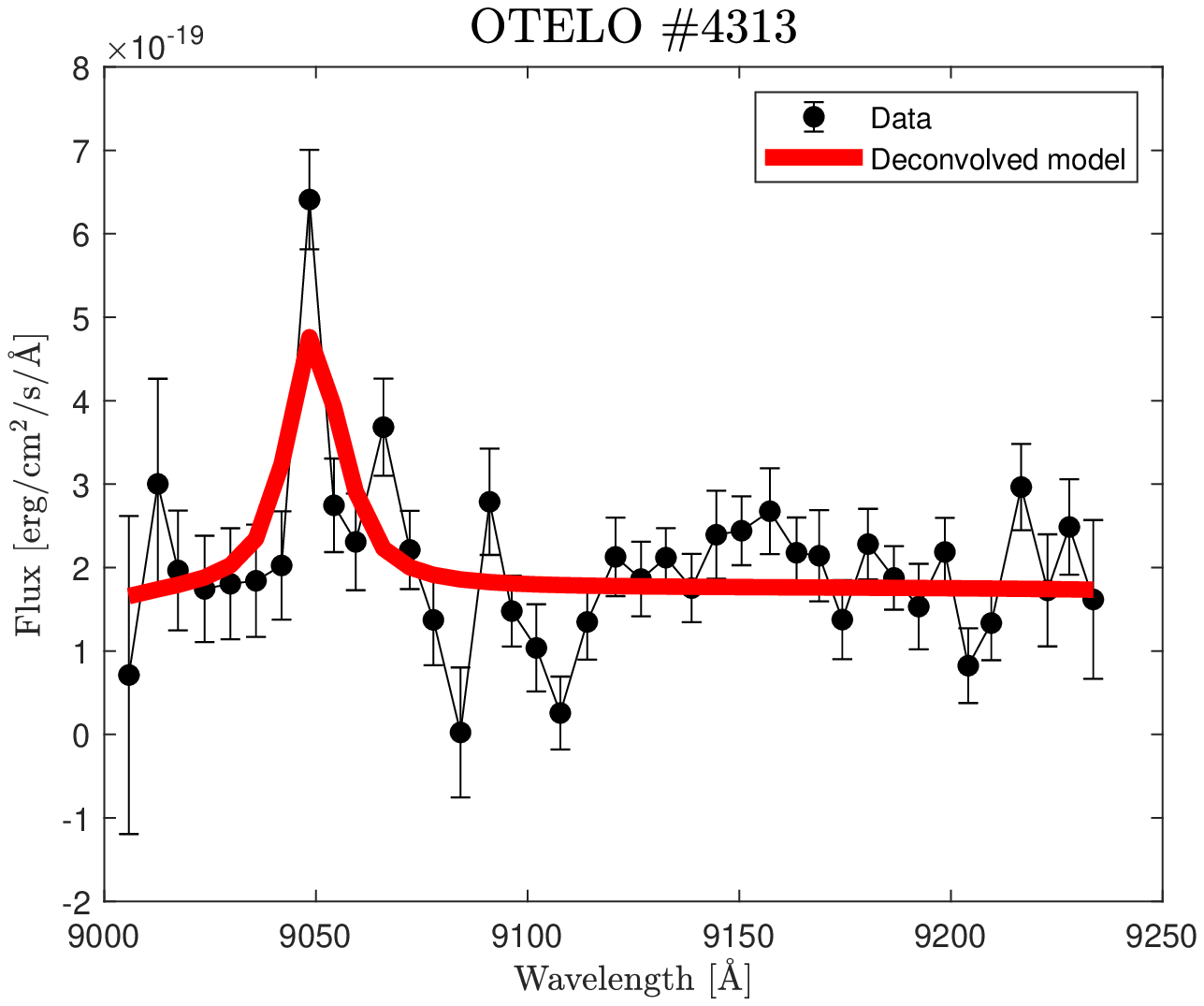}}
\subfloat{\includegraphics[width=.20\linewidth]{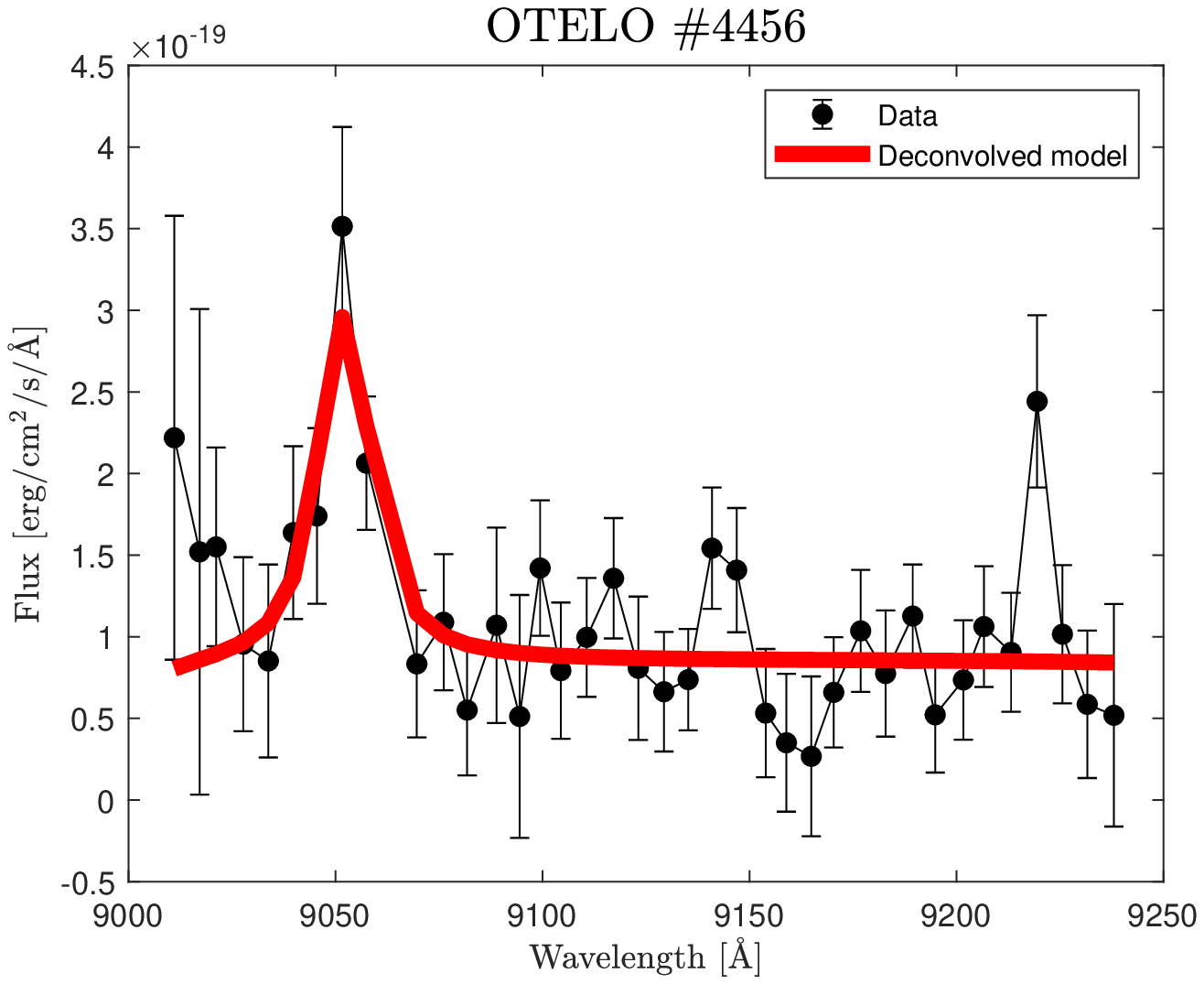}}\\
\subfloat{\includegraphics[width=.20\linewidth]{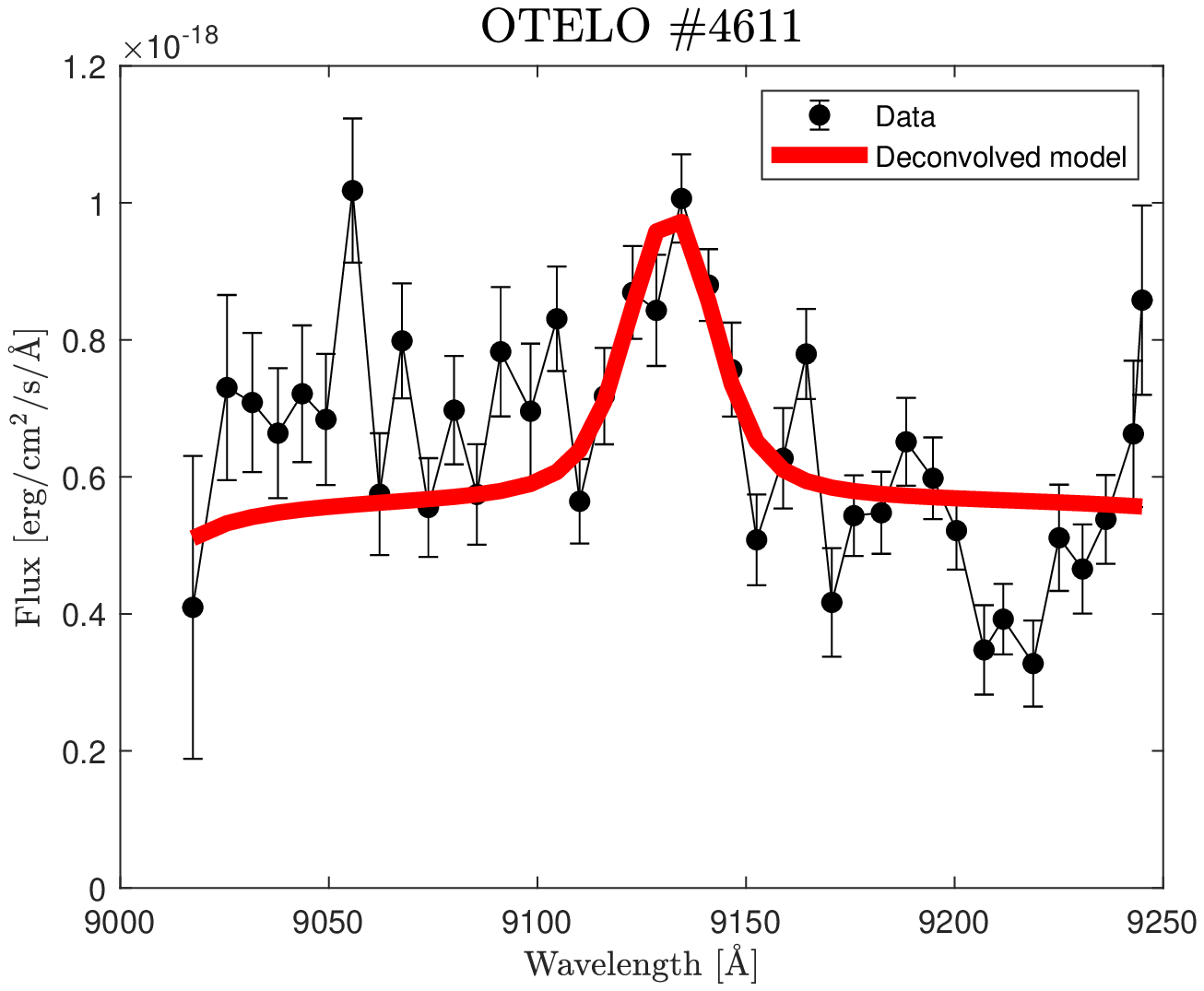}}
\subfloat{\includegraphics[width=.20\linewidth]{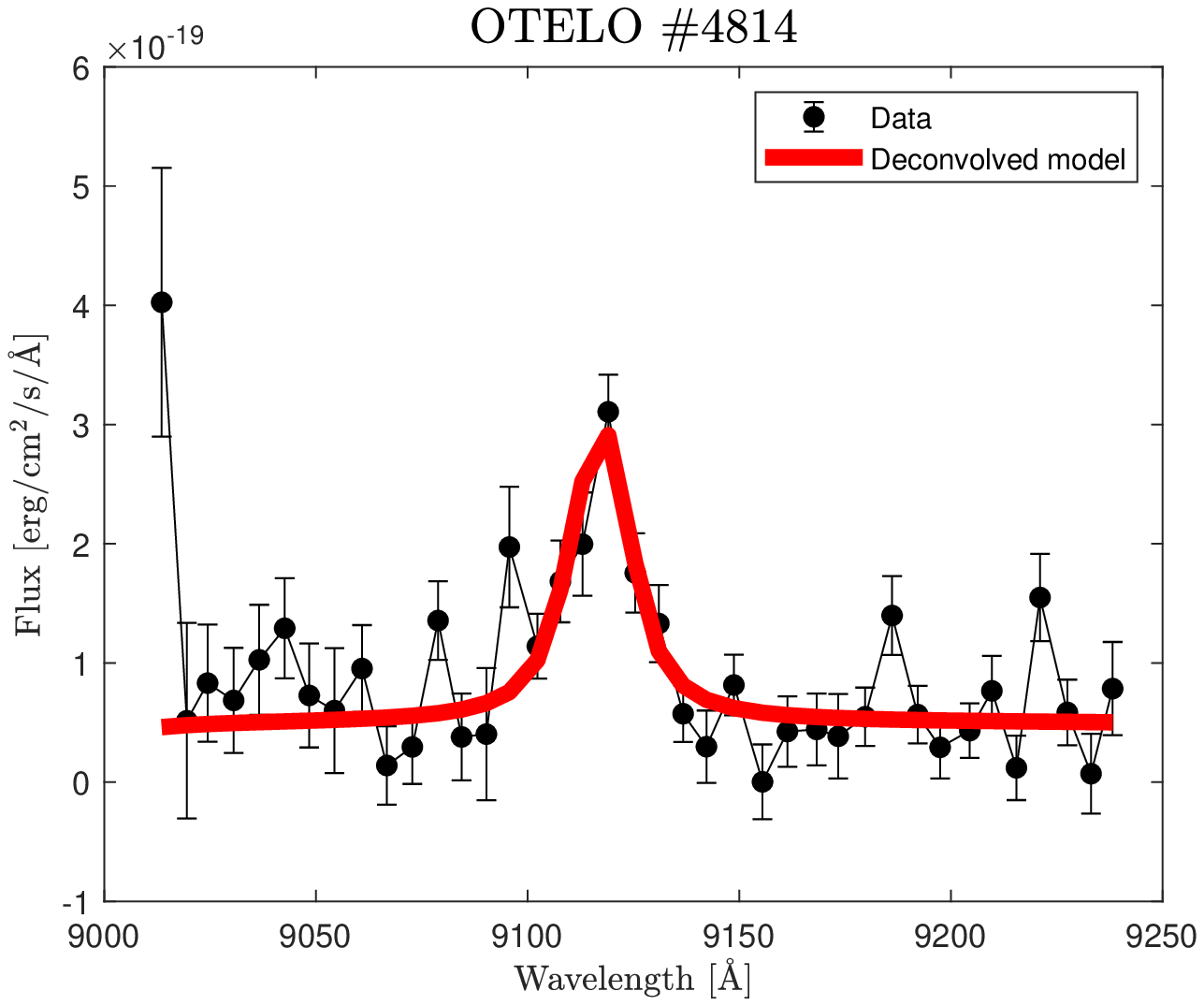}}
\subfloat{\includegraphics[width=.20\linewidth]{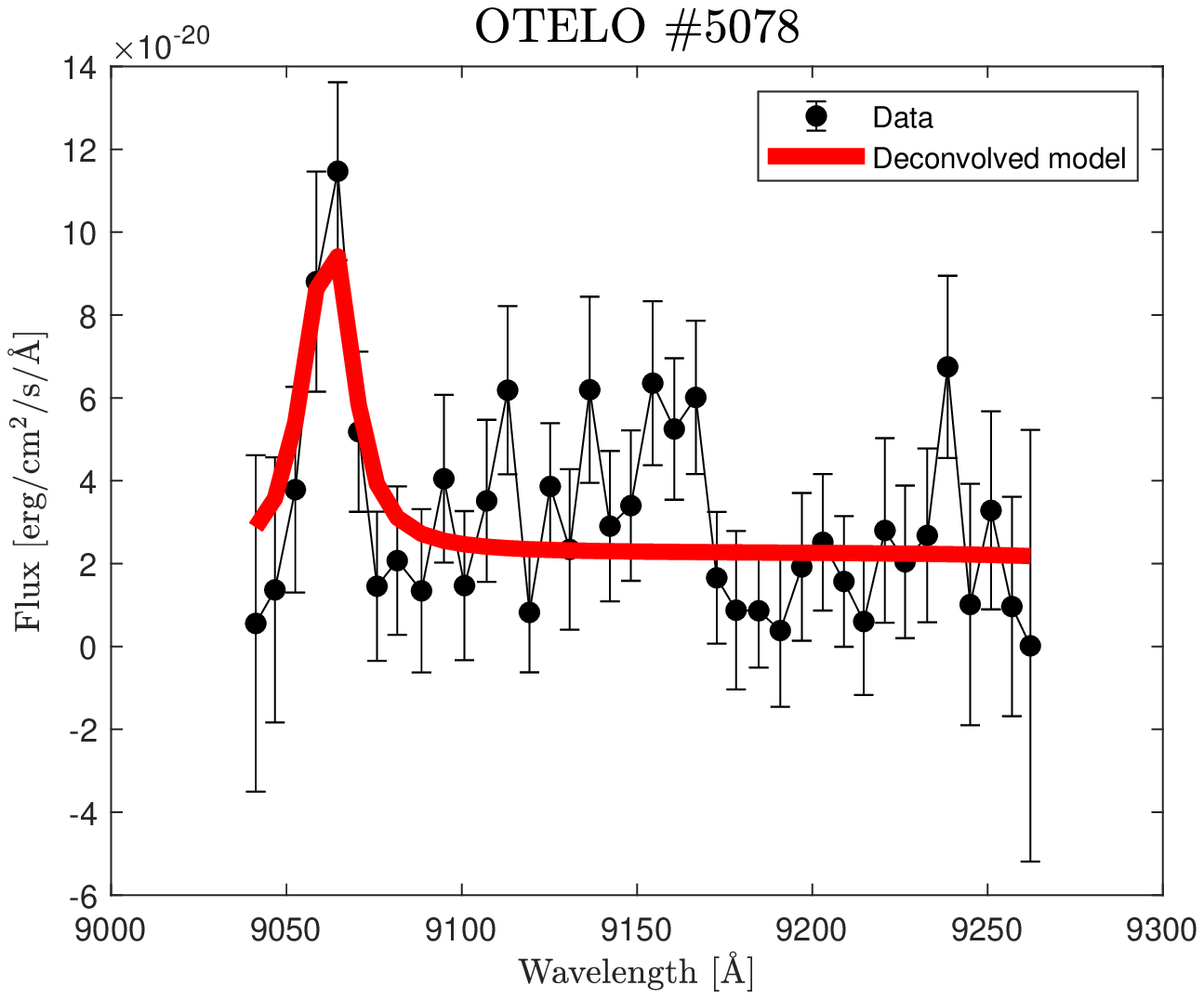}}
\subfloat{\includegraphics[width=.20\linewidth]{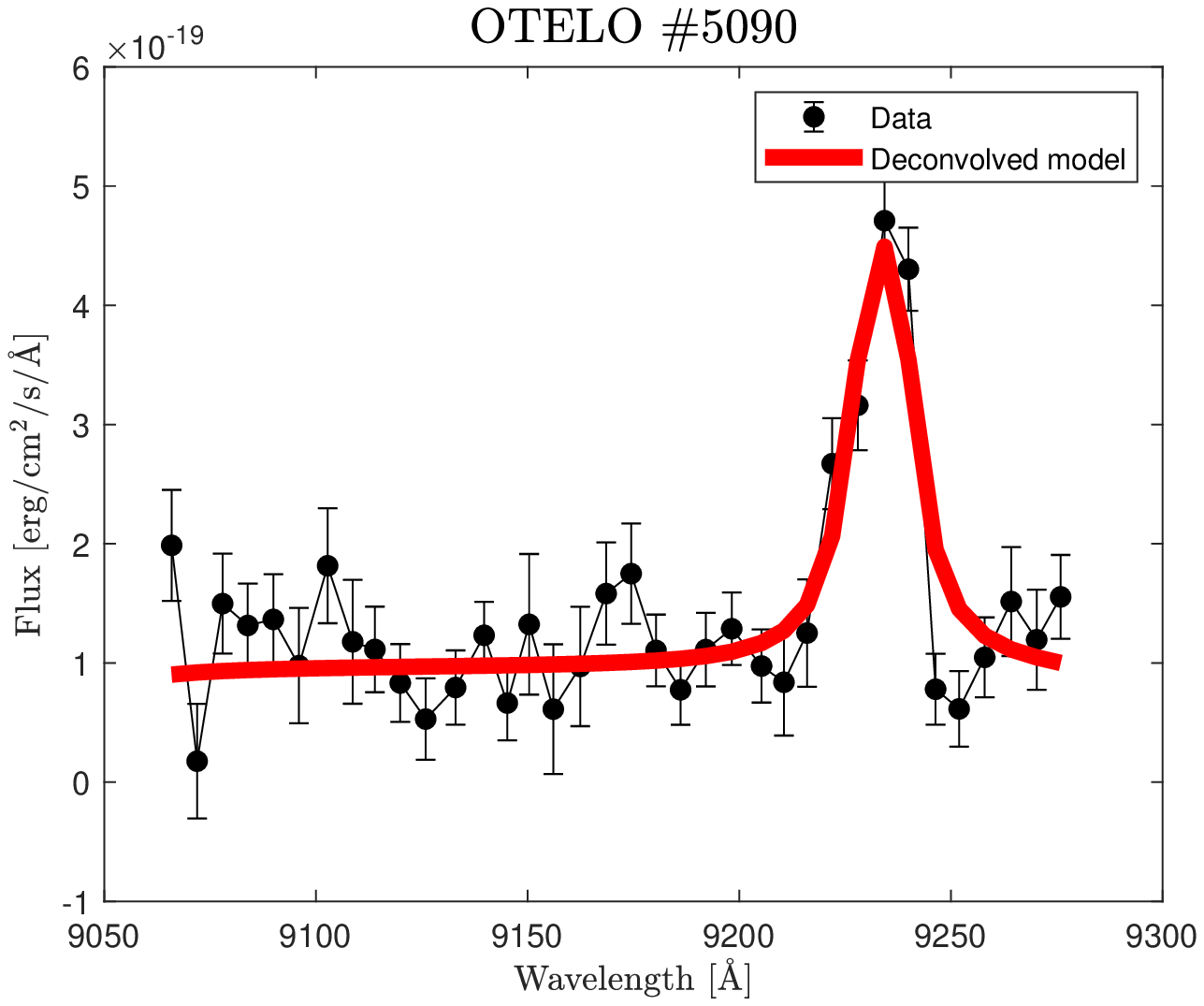}}\\
\subfloat{\includegraphics[width=.20\linewidth]{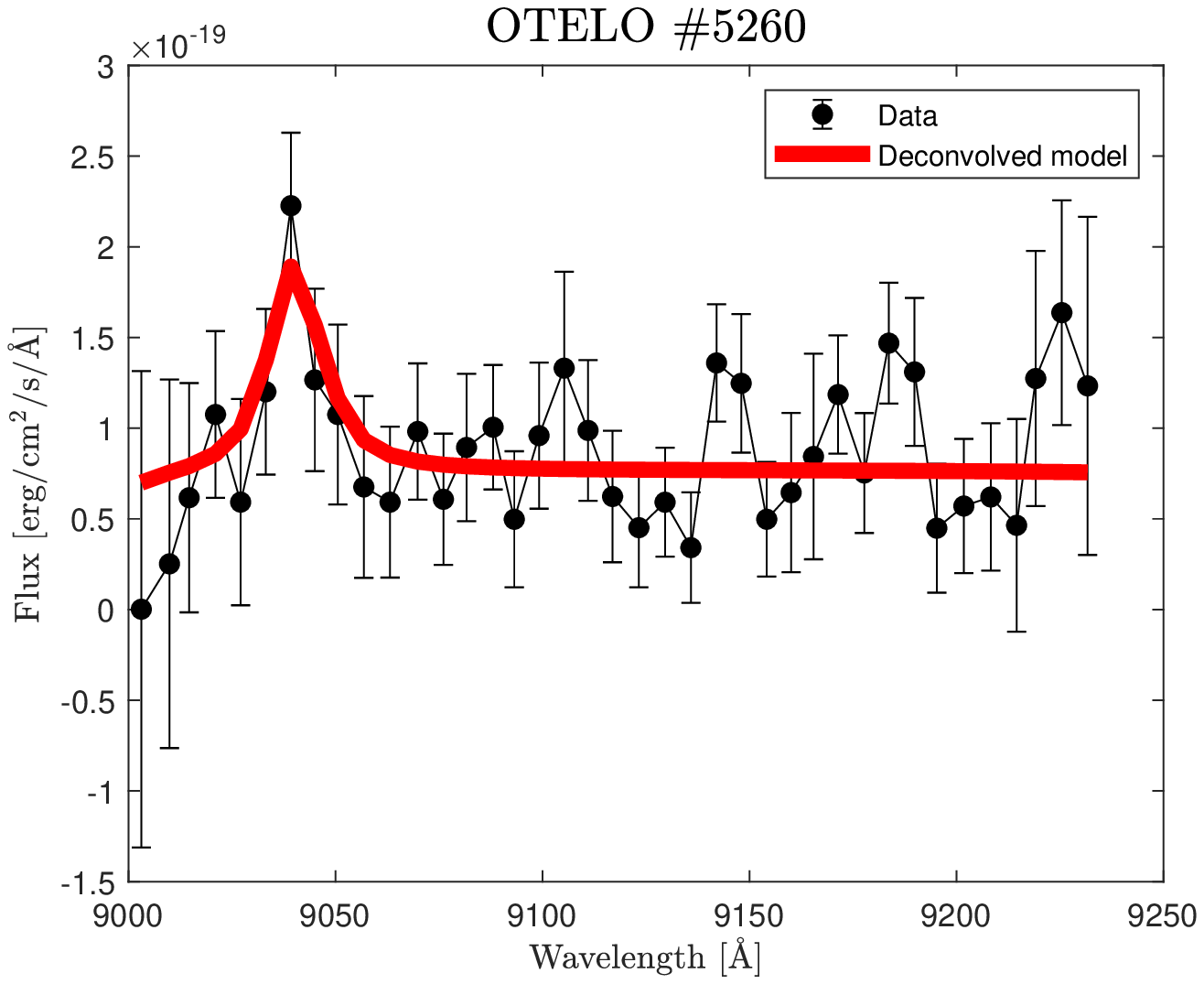}}
\subfloat{\includegraphics[width=.20\linewidth]{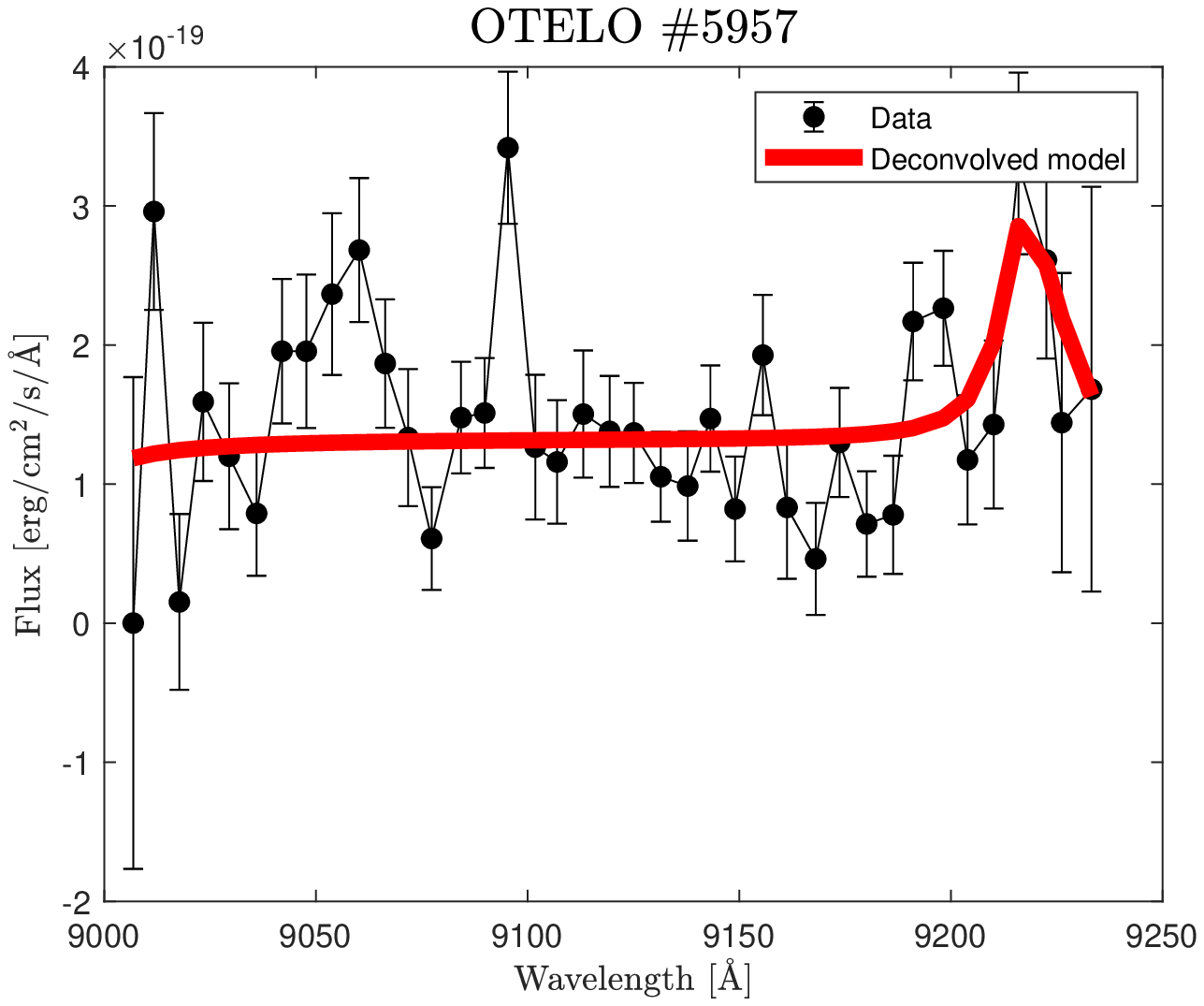}}
\subfloat{\includegraphics[width=.20\linewidth]{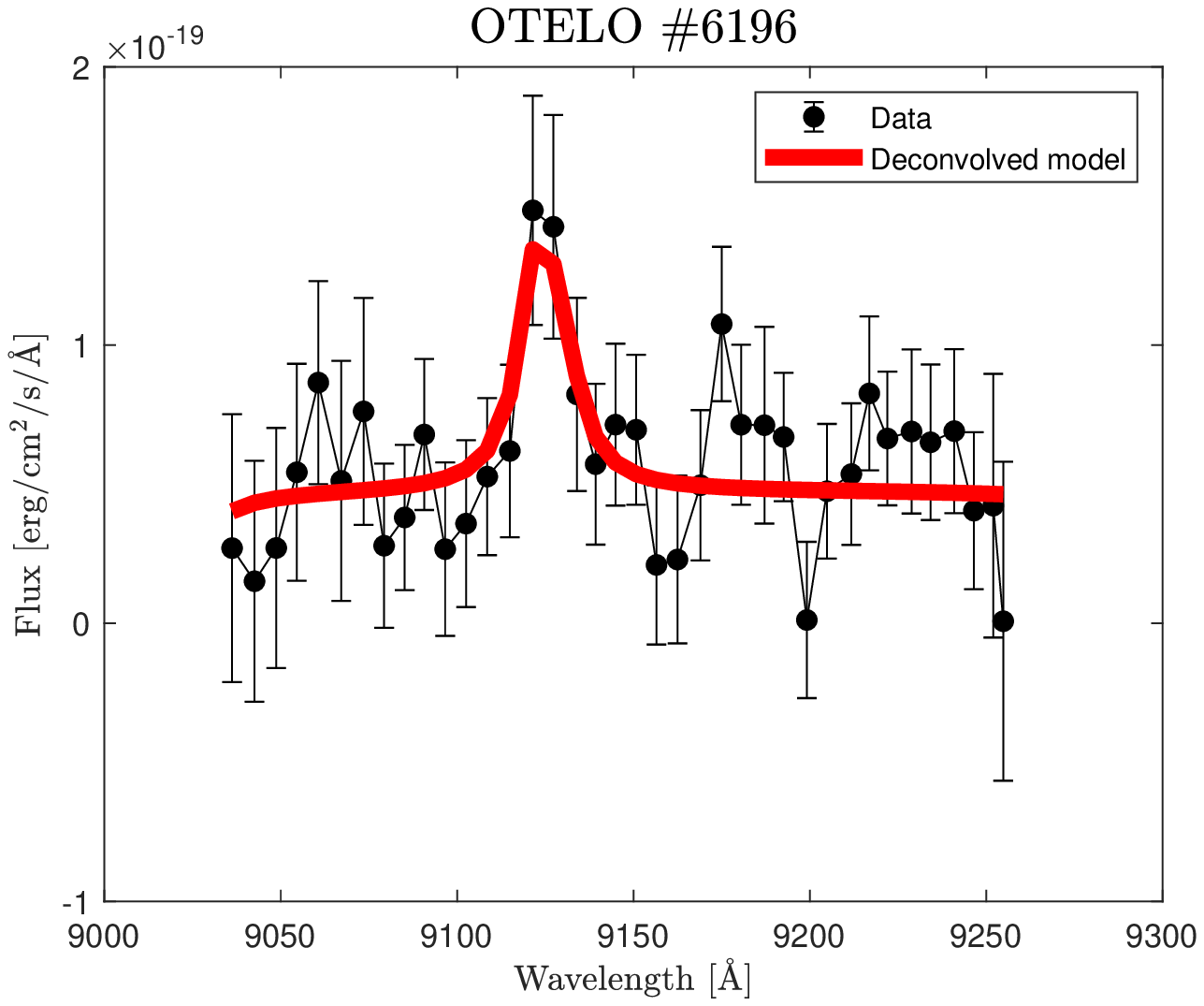}}
\subfloat{\includegraphics[width=.20\linewidth]{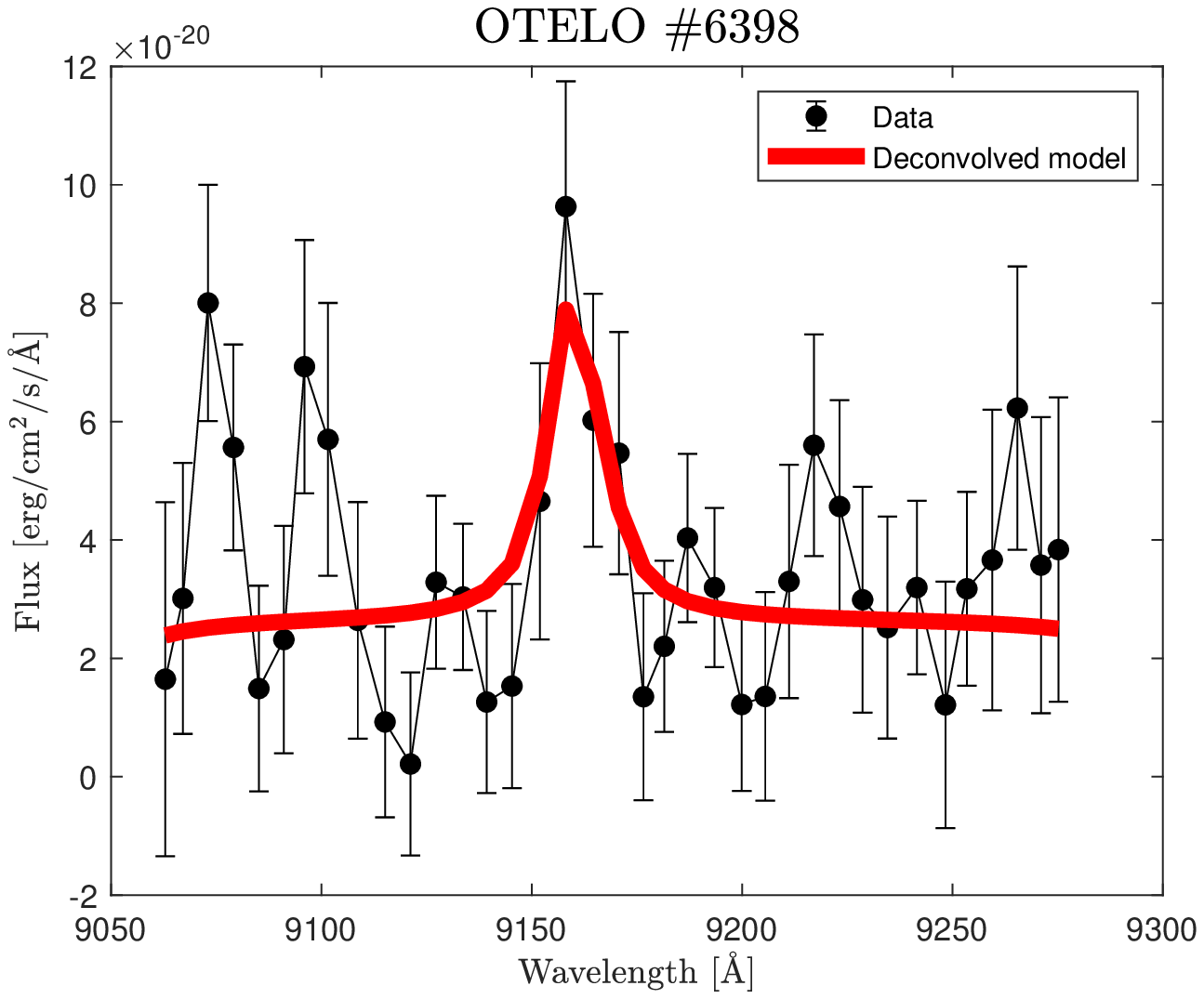}}\\
\subfloat{\includegraphics[width=.20\linewidth]{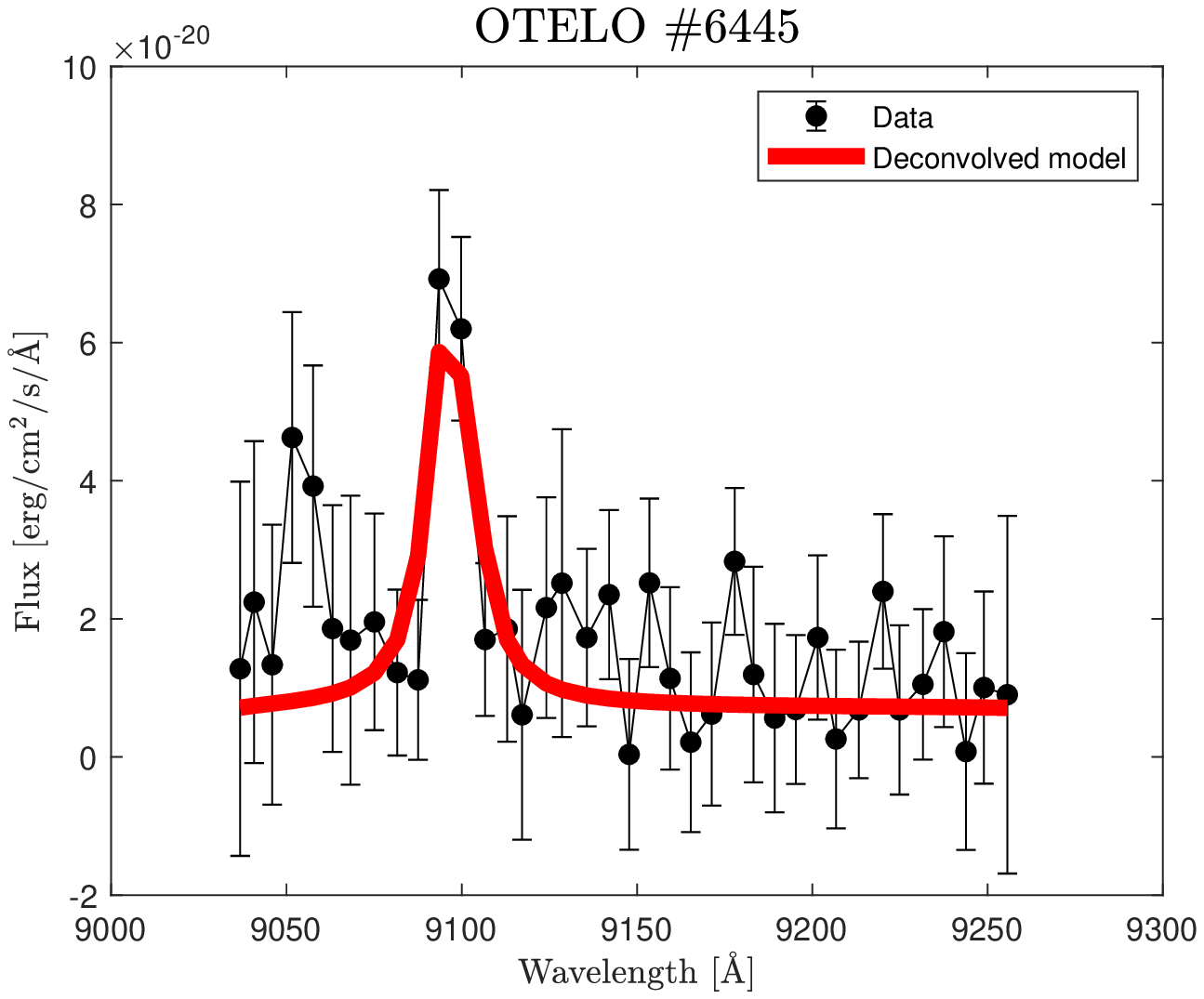}}
\subfloat{\includegraphics[width=.20\linewidth]{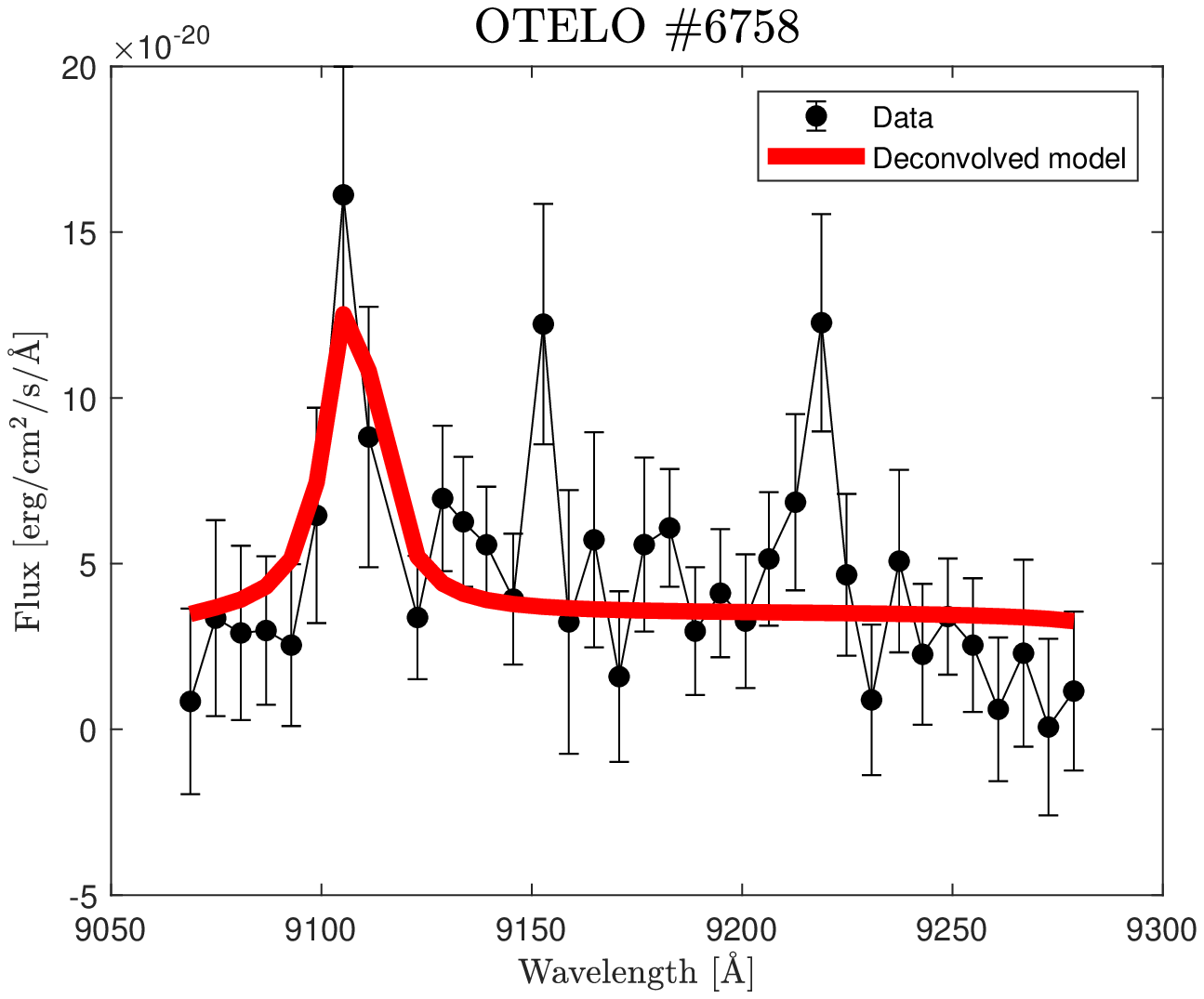}}
\subfloat{\includegraphics[width=.20\linewidth]{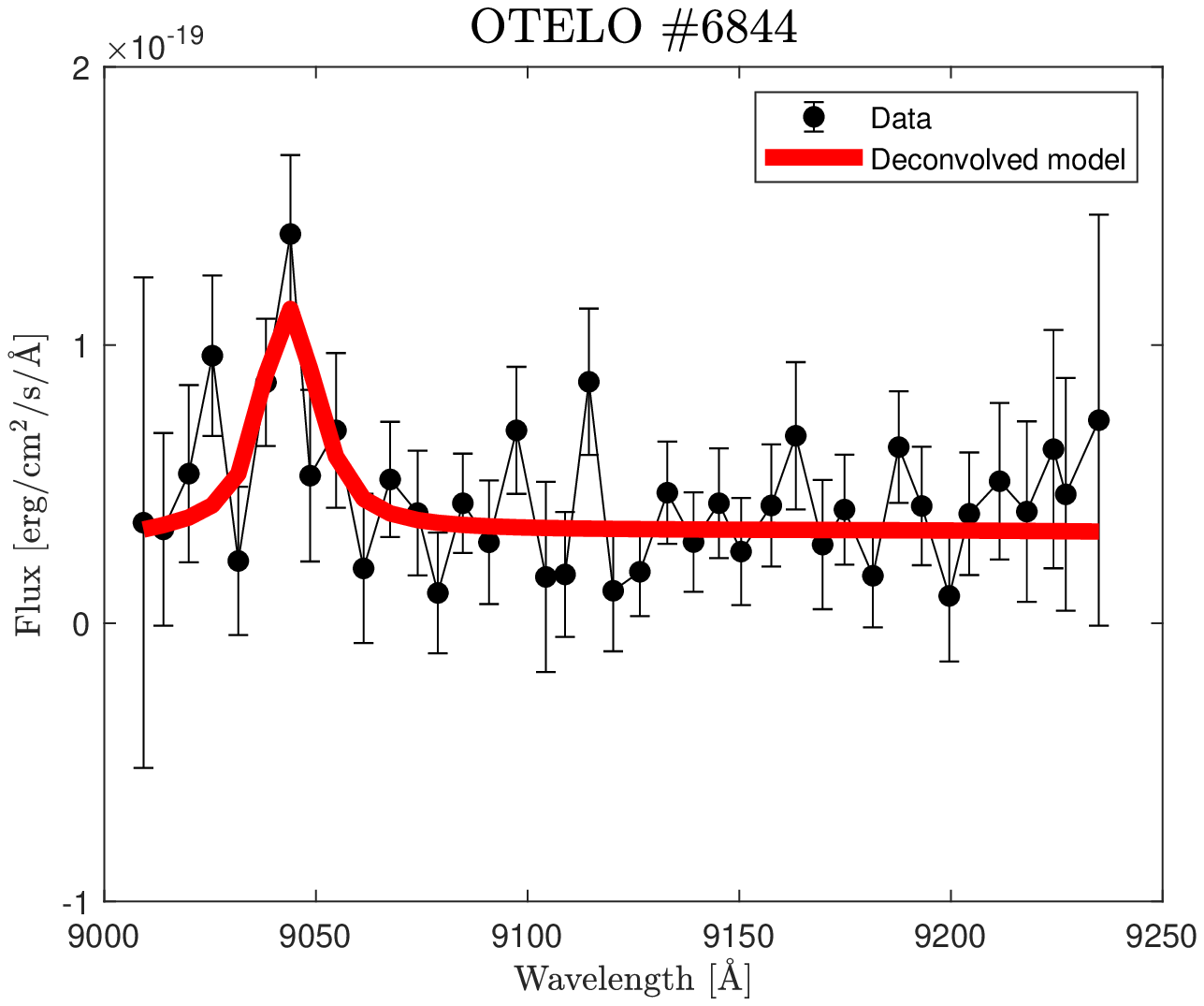}}
\subfloat{\includegraphics[width=.20\linewidth]{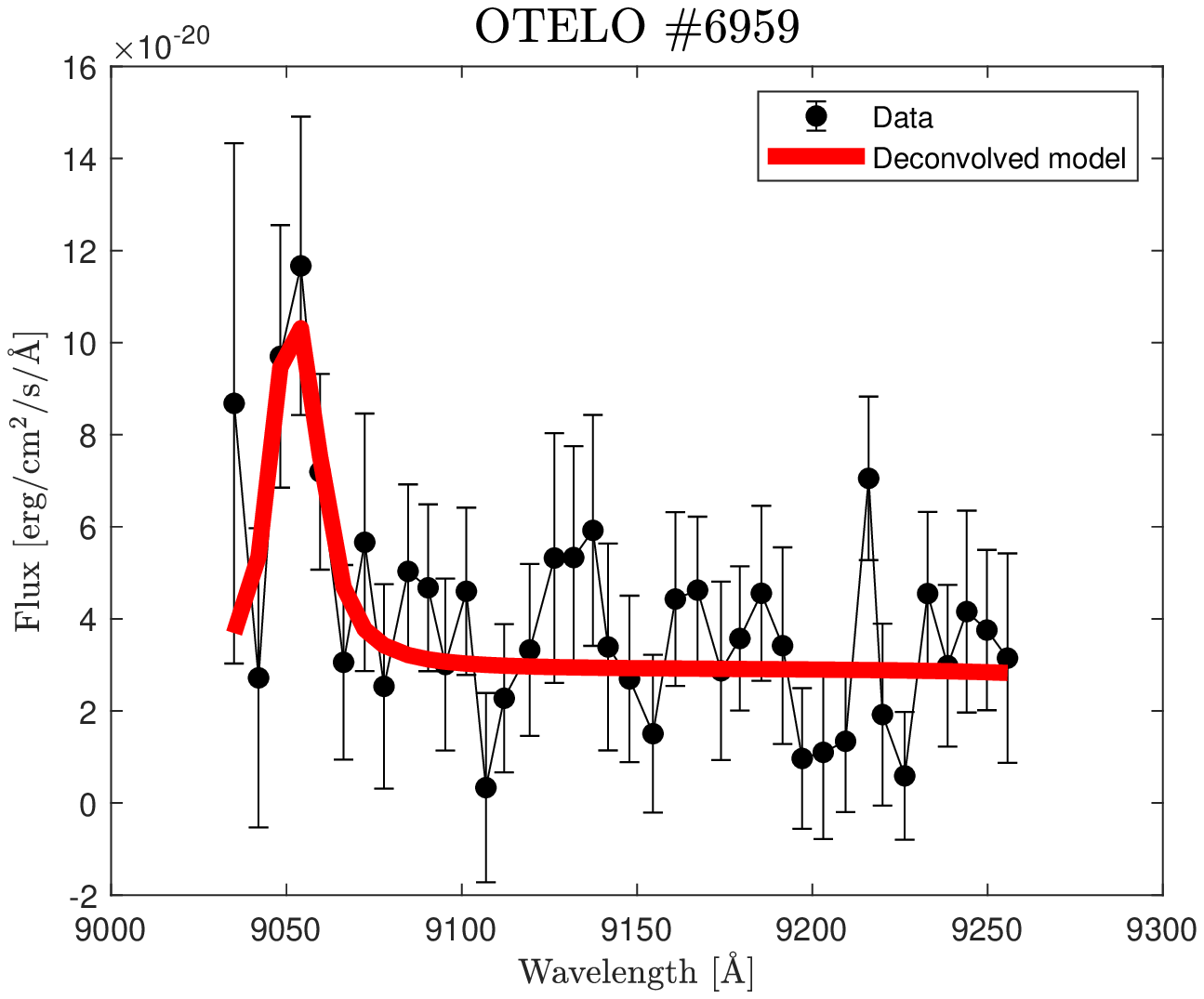}}
   \caption{Pseudo-spectra of the selected emitters. Black dots represent the measured pseudo-spectra, the red line is the best fitted deconvolved spectra.}
\end{figure*}
\begin{figure*}\continuedfloat
    \centering
\subfloat{\includegraphics[width=.20\linewidth]{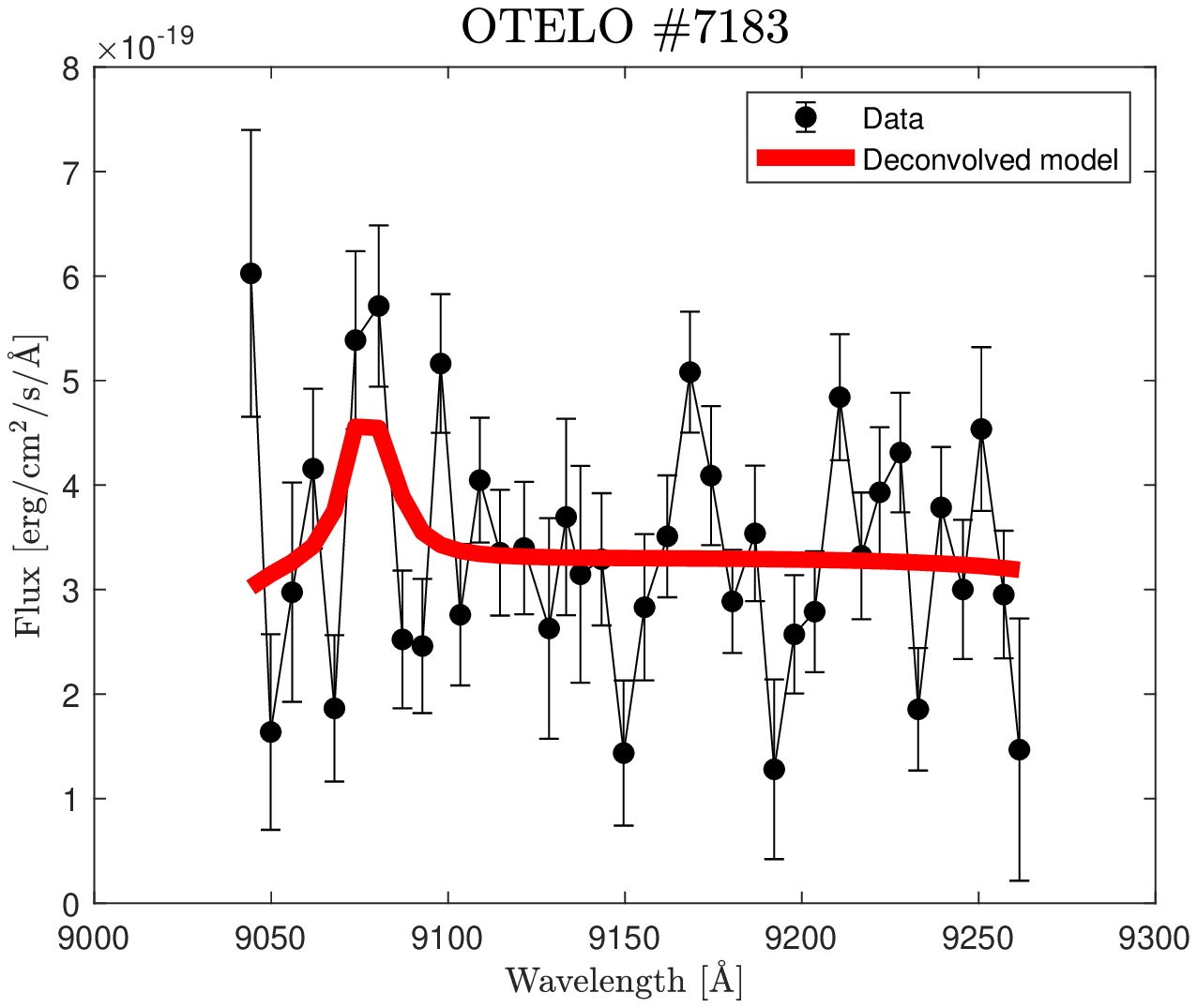}}
\subfloat{\includegraphics[width=.20\linewidth]{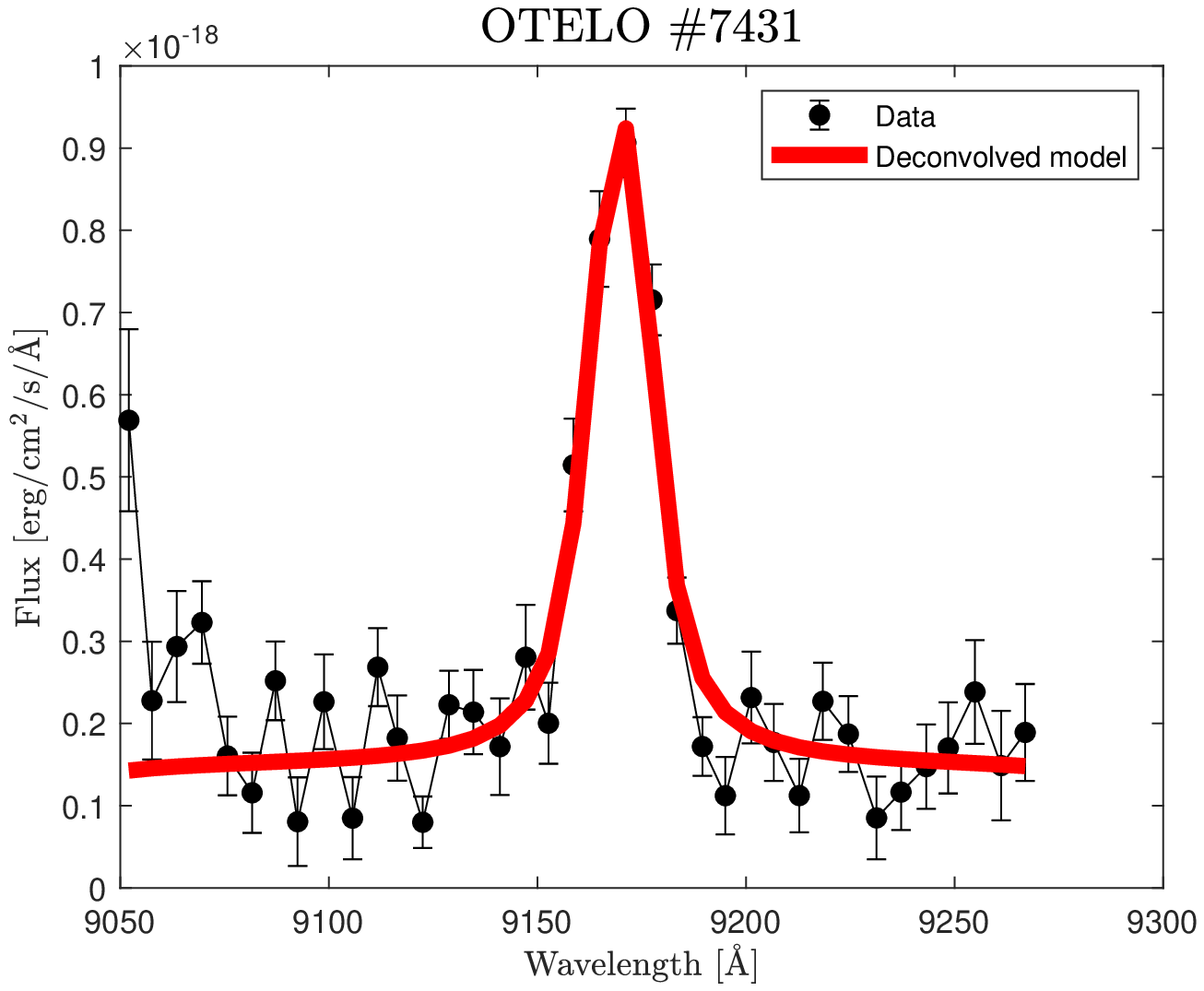}}
\subfloat{\includegraphics[width=.20\linewidth]{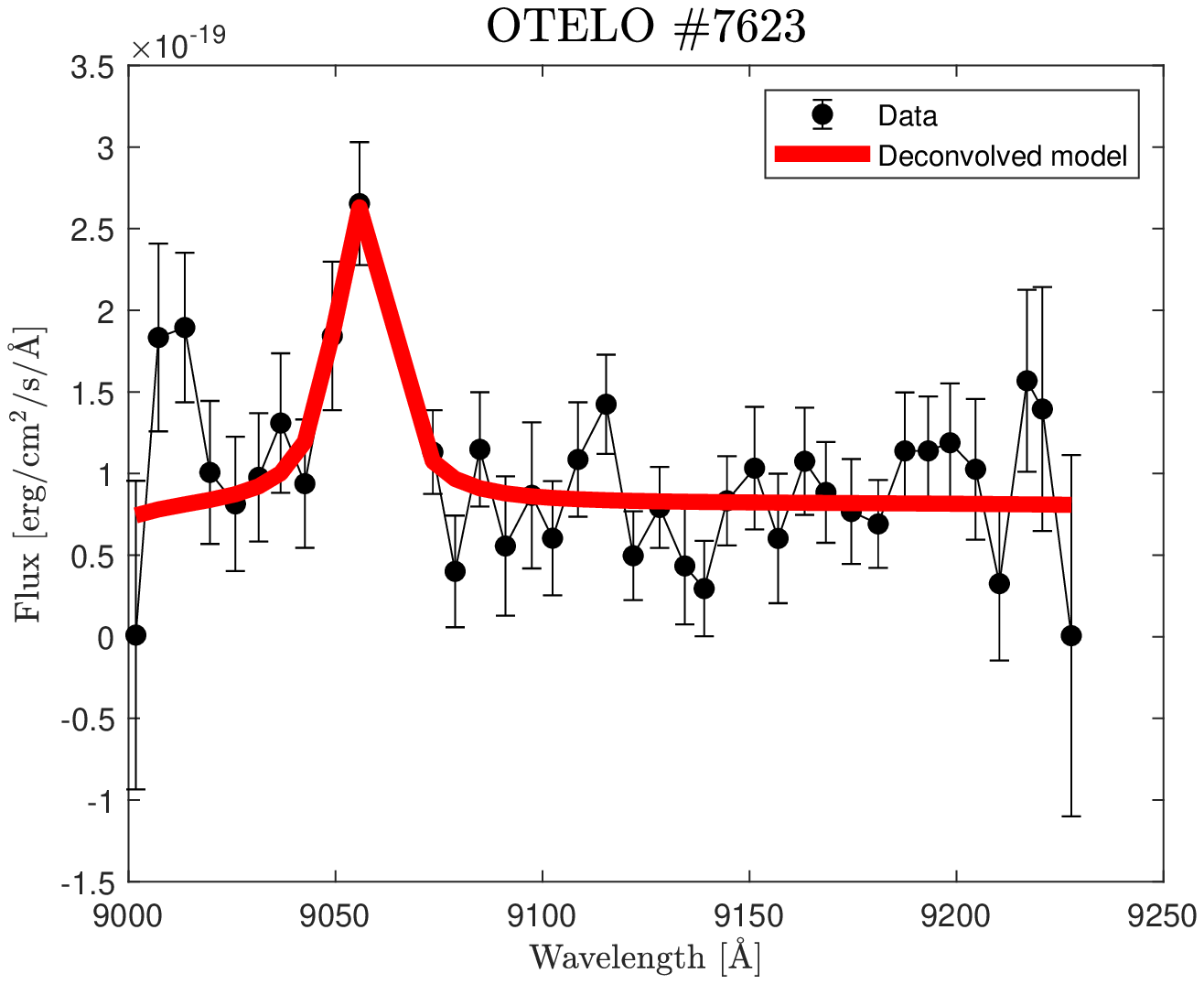}}
\subfloat{\includegraphics[width=.20\linewidth]{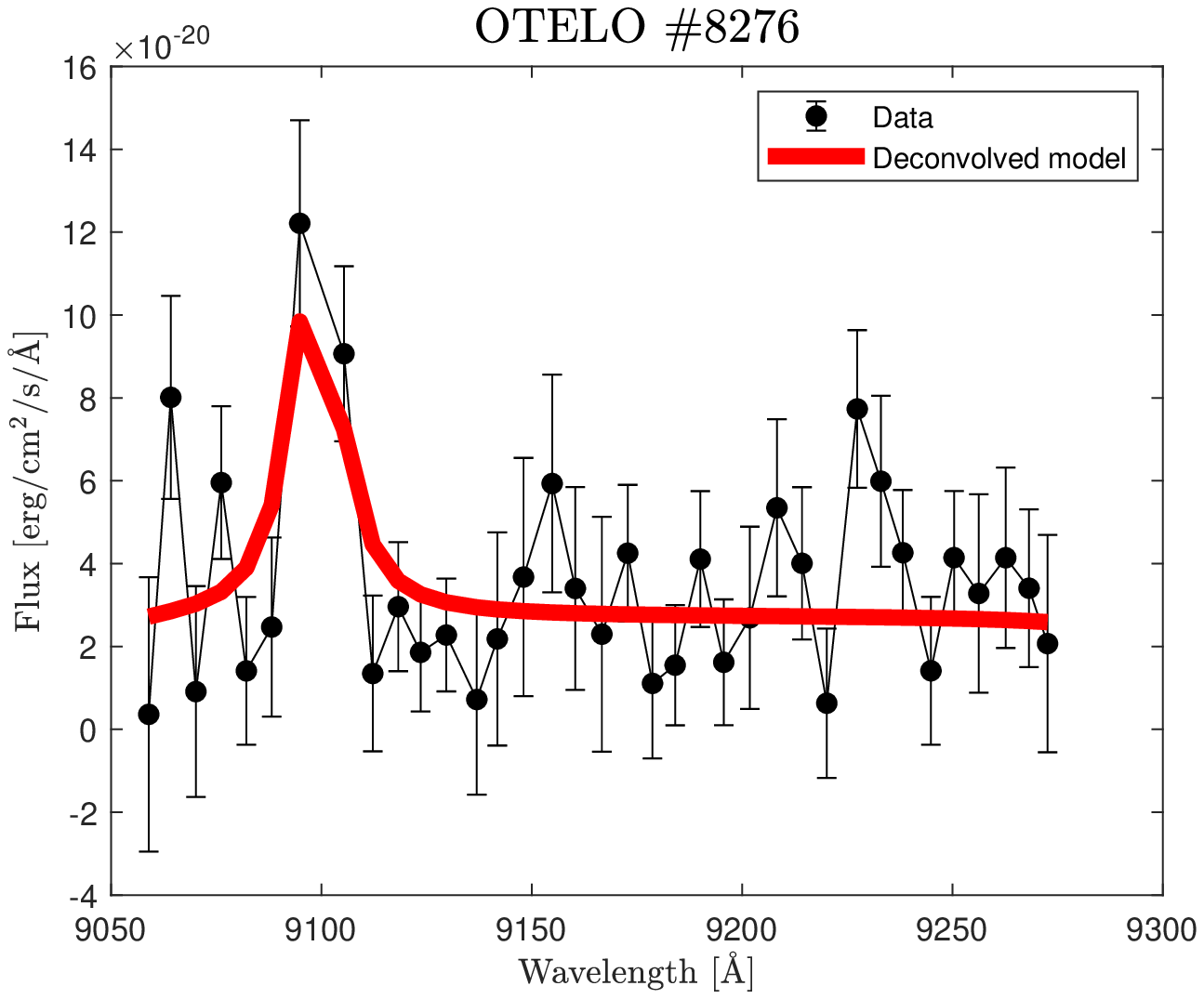}}\\
\subfloat{\includegraphics[width=.20\linewidth]{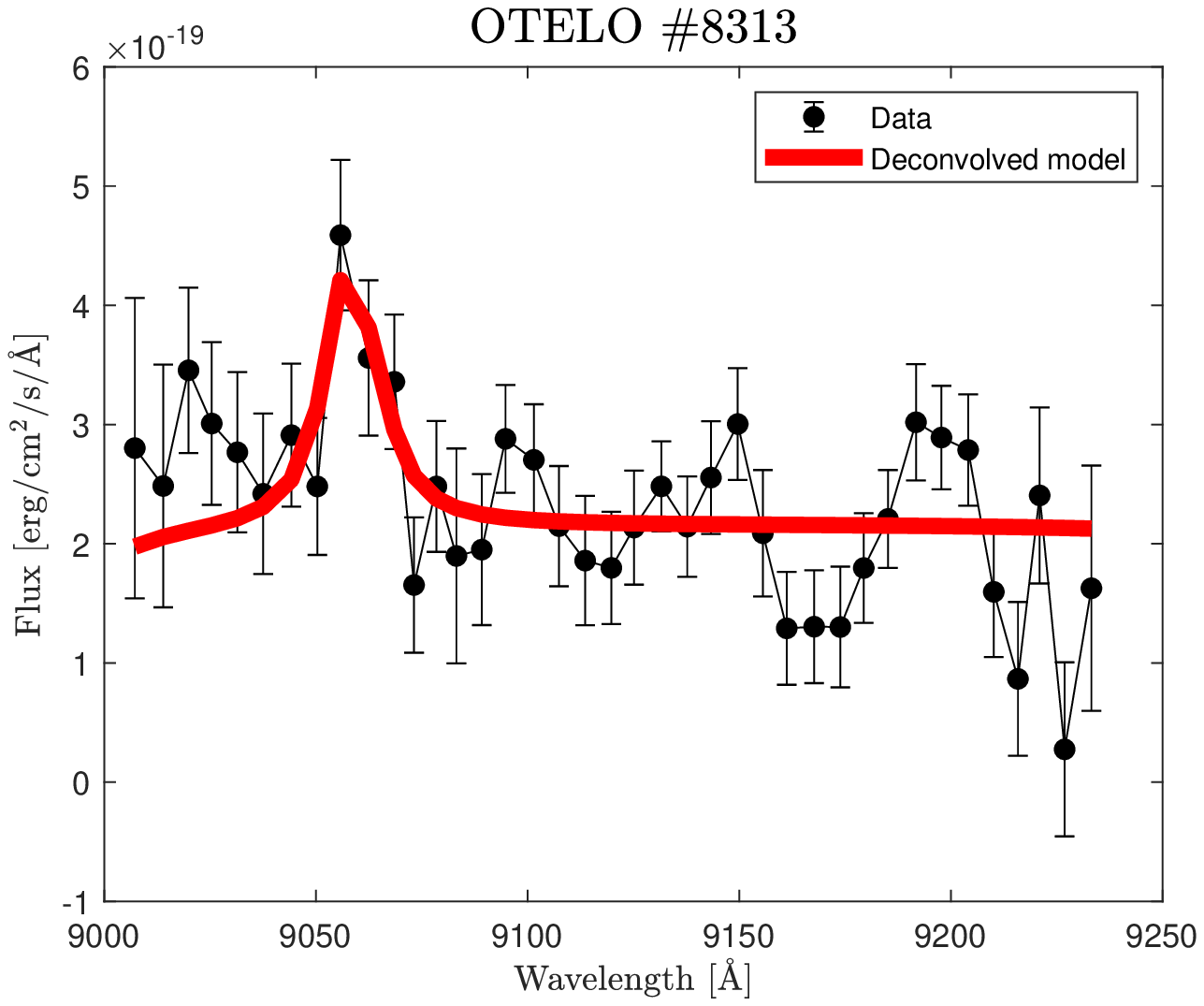}}
\subfloat{\includegraphics[width=.20\linewidth]{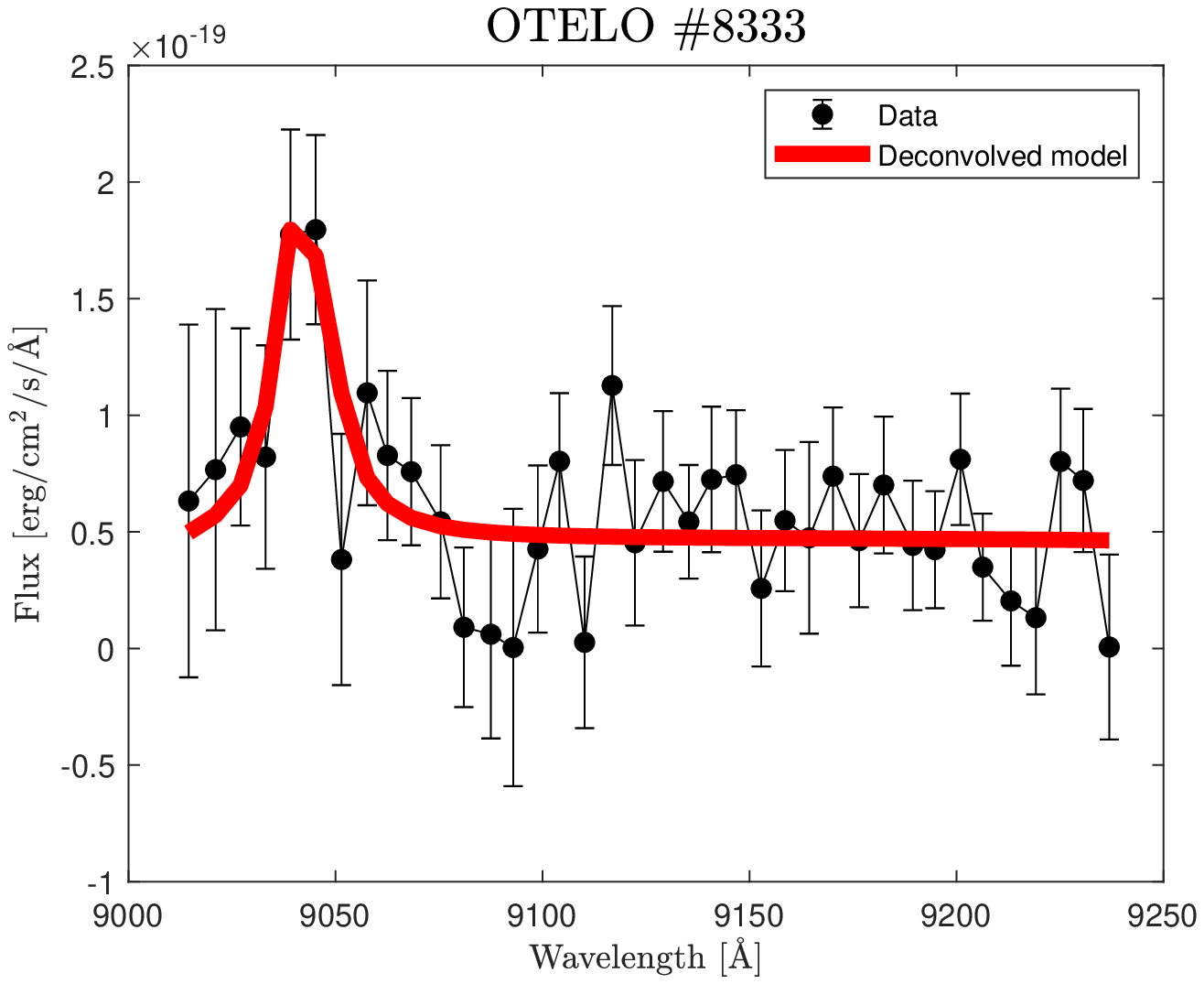}}
\subfloat{\includegraphics[width=.20\linewidth]{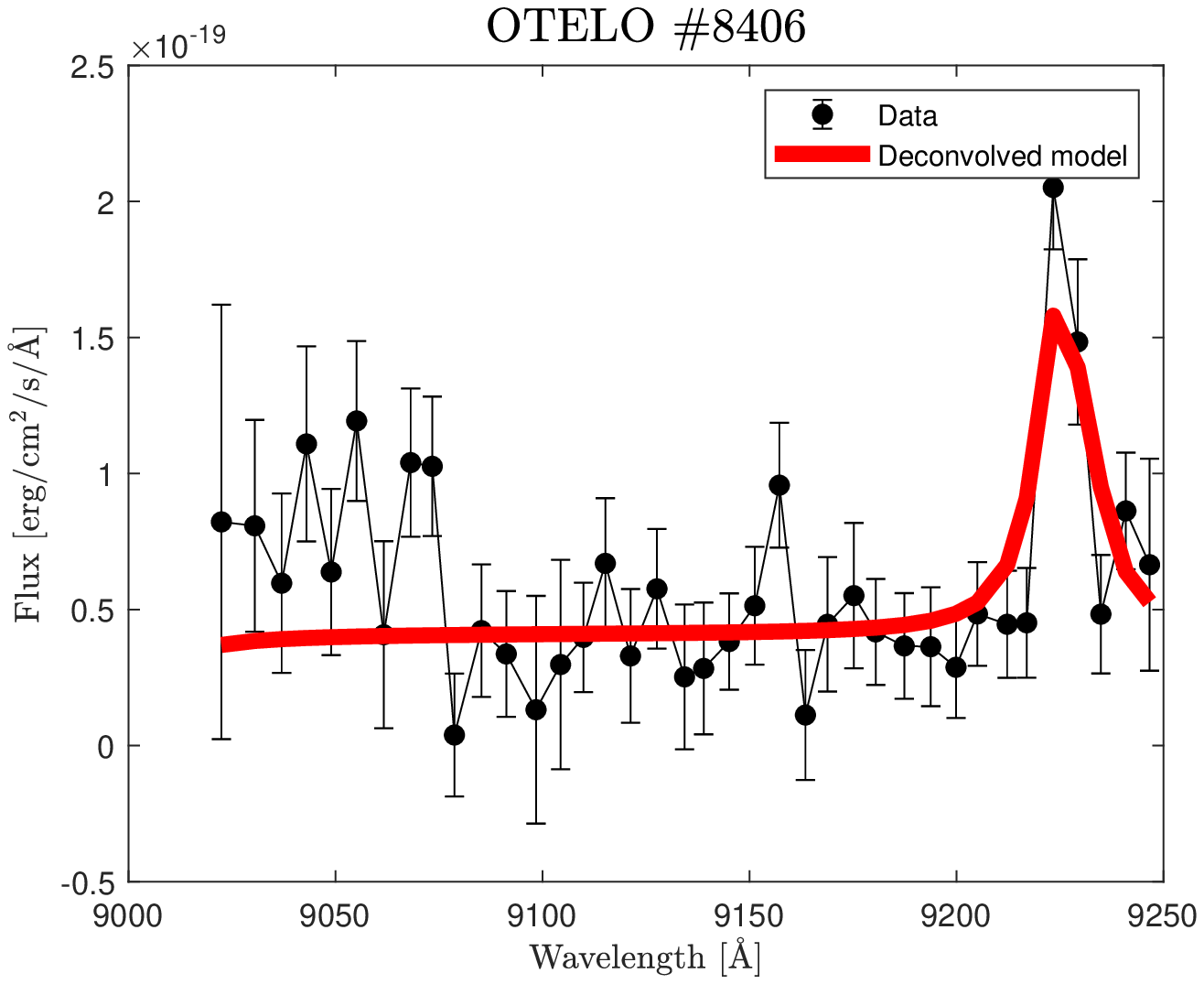}}
\subfloat{\includegraphics[width=.20\linewidth]{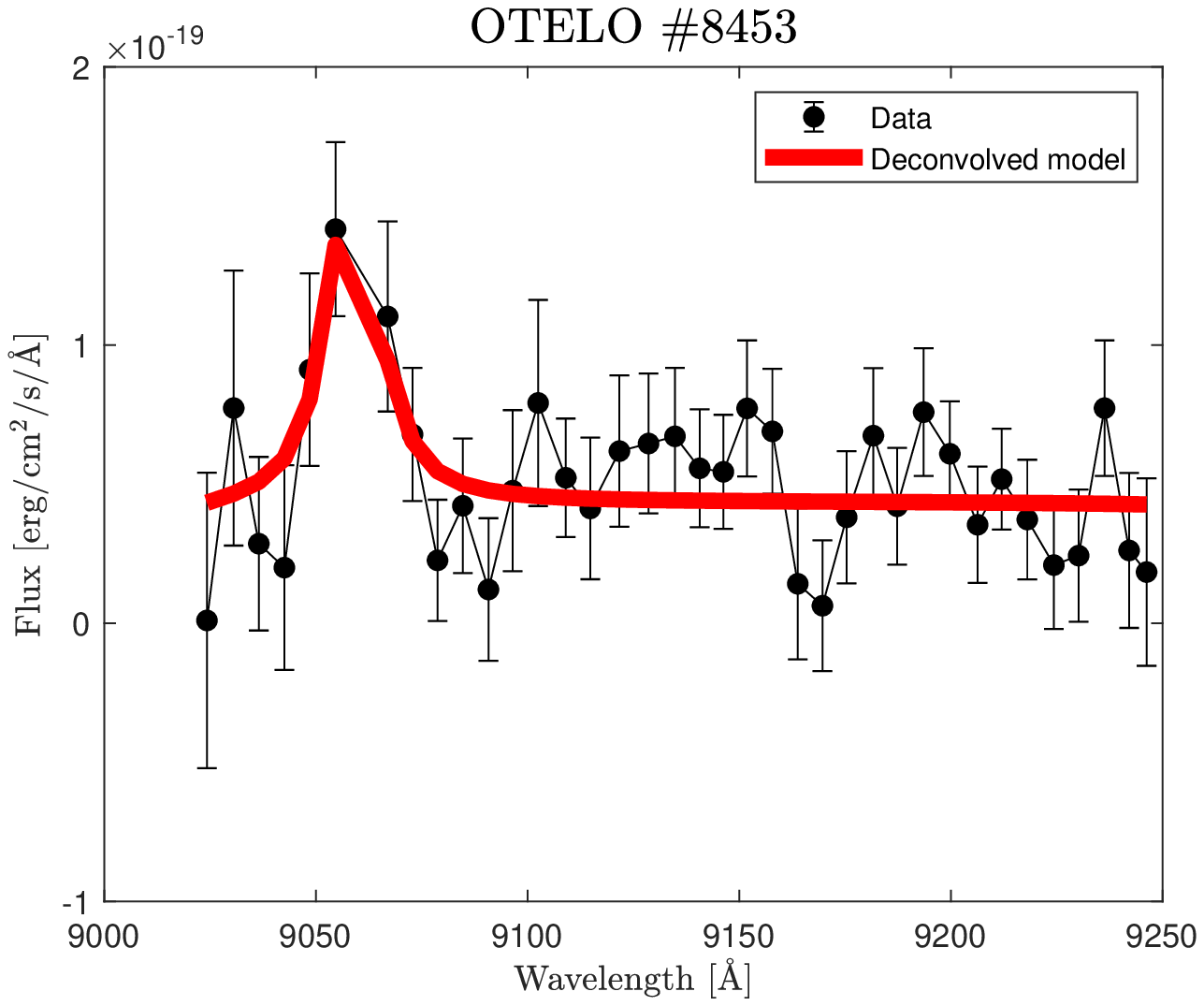}}
\\
\subfloat{\includegraphics[width=.20\linewidth]{8532.eps}}
\subfloat{\includegraphics[width=.20\linewidth]{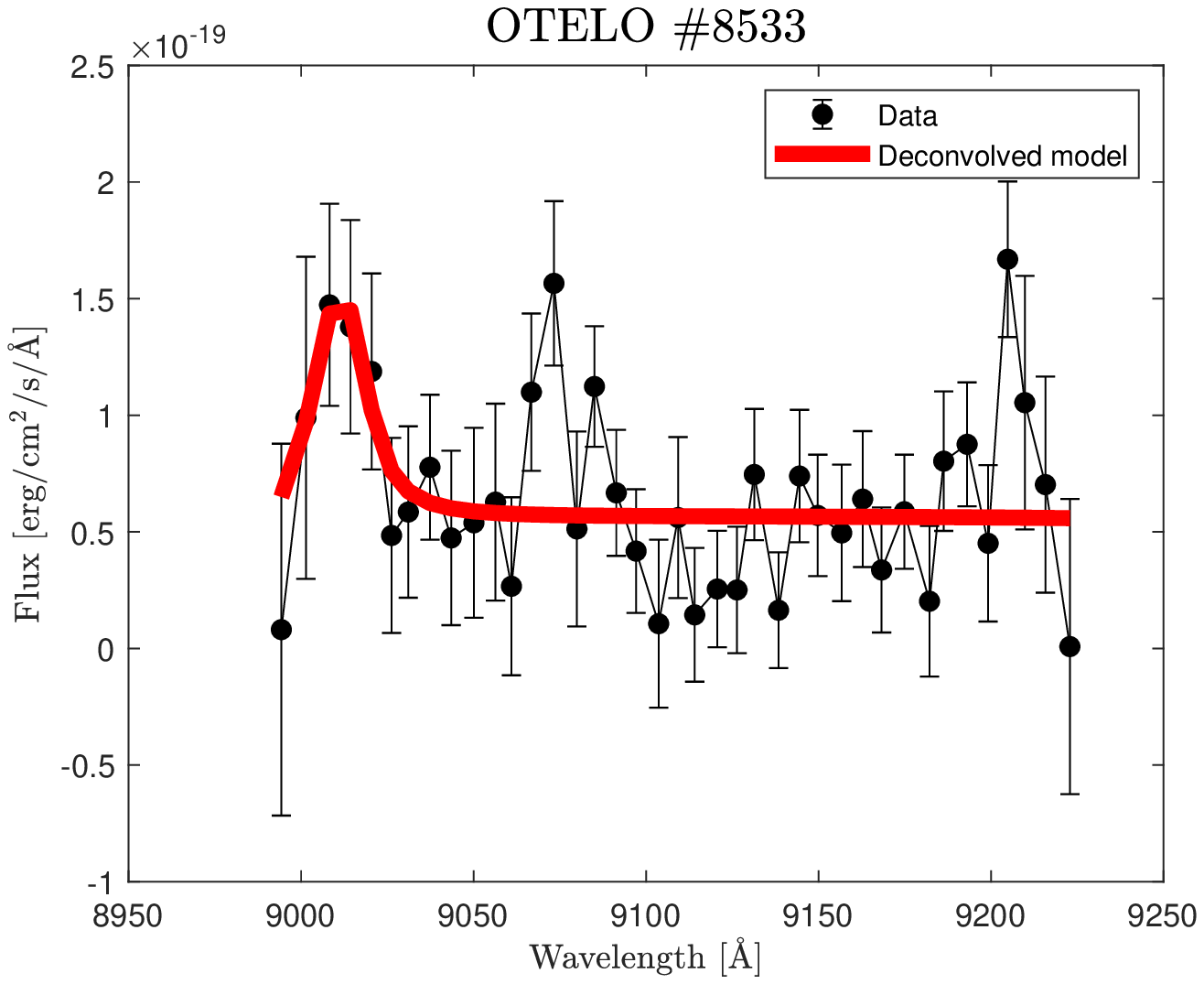}}
\subfloat{\includegraphics[width=.20\linewidth]{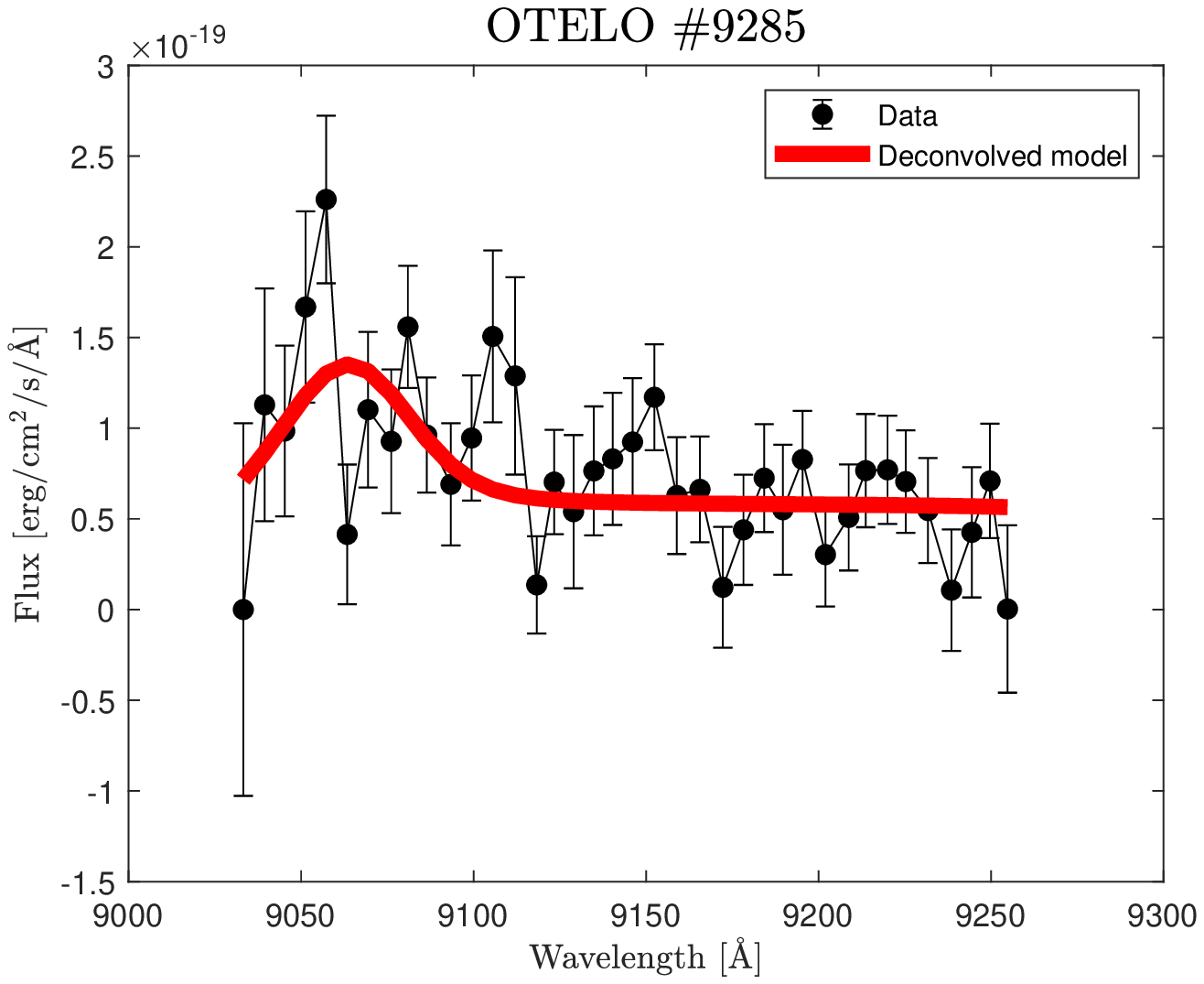}}
\subfloat{\includegraphics[width=.20\linewidth]{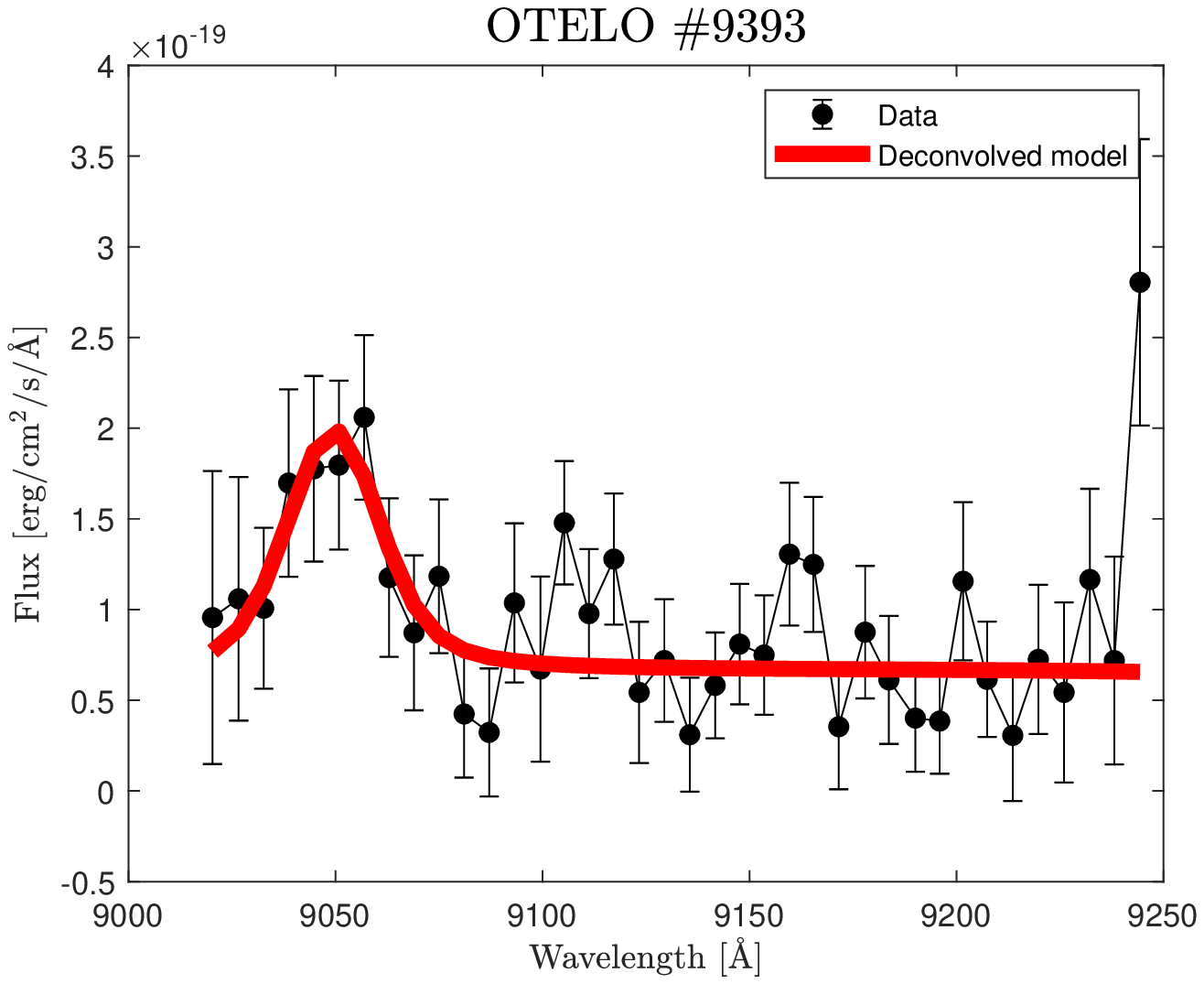}}
\\
\subfloat{\includegraphics[width=.20\linewidth]{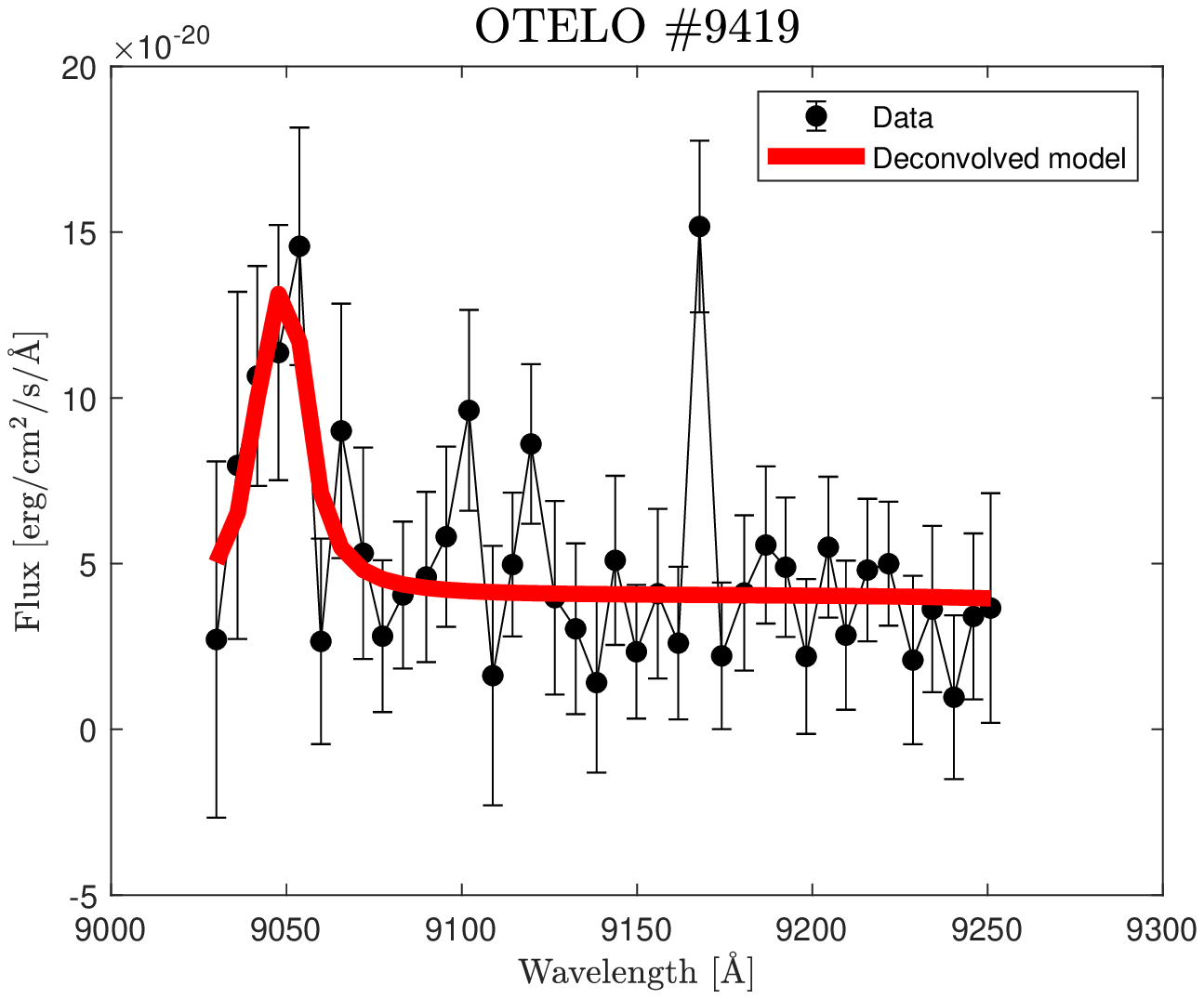}}
\subfloat{\includegraphics[width=.20\linewidth]{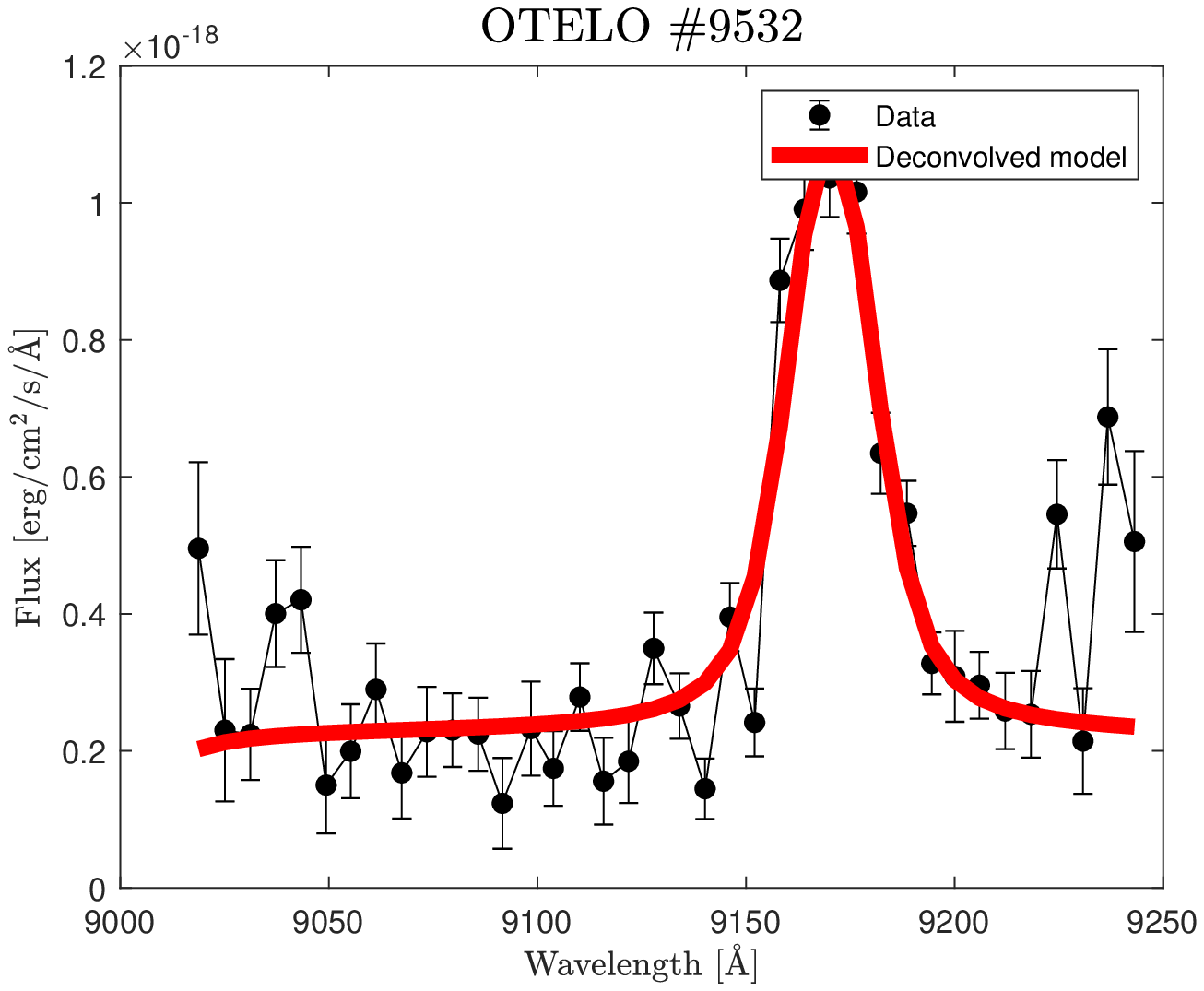}}
\subfloat{\includegraphics[width=.20\linewidth]{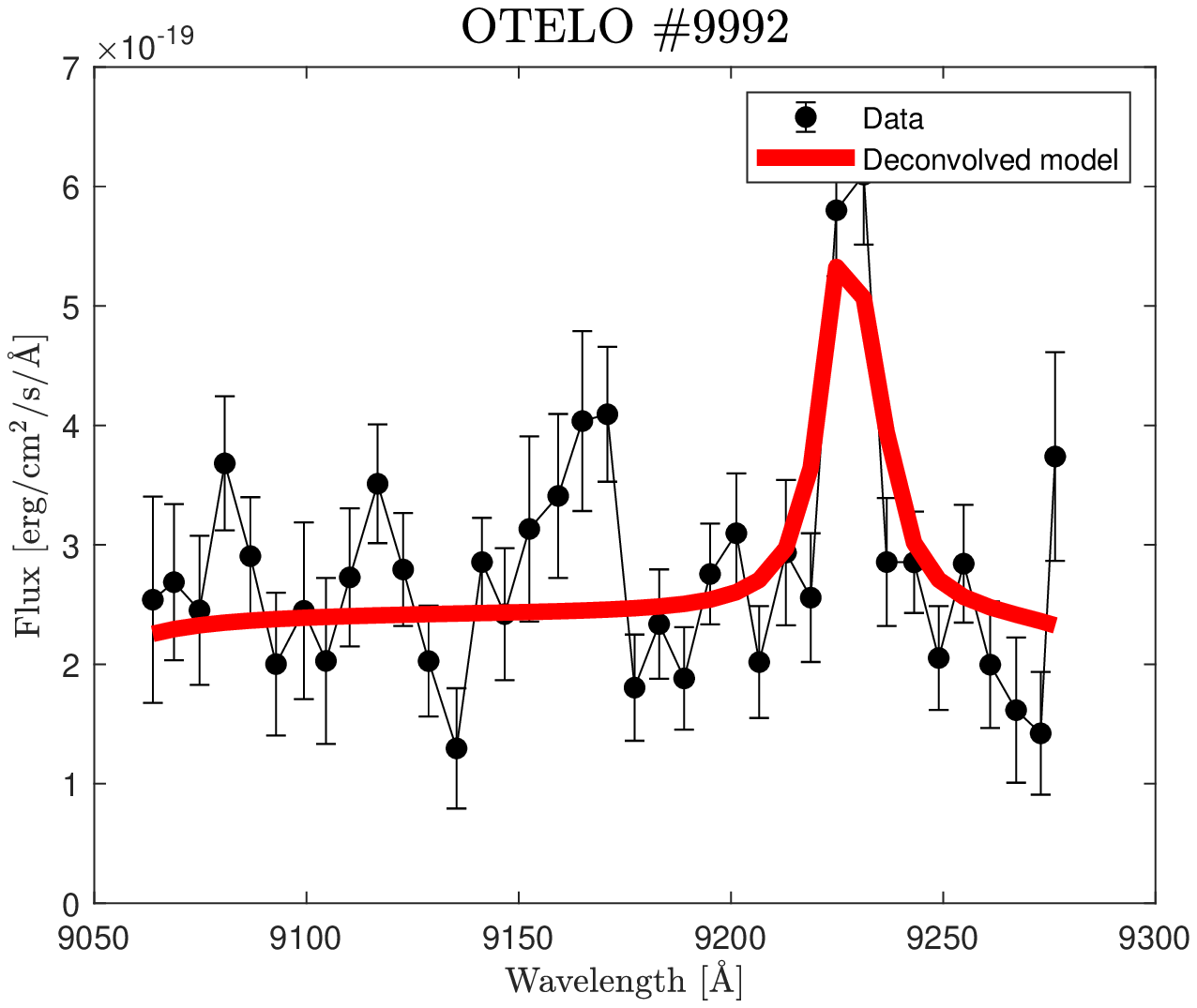}}
\subfloat{\includegraphics[width=.20\linewidth]{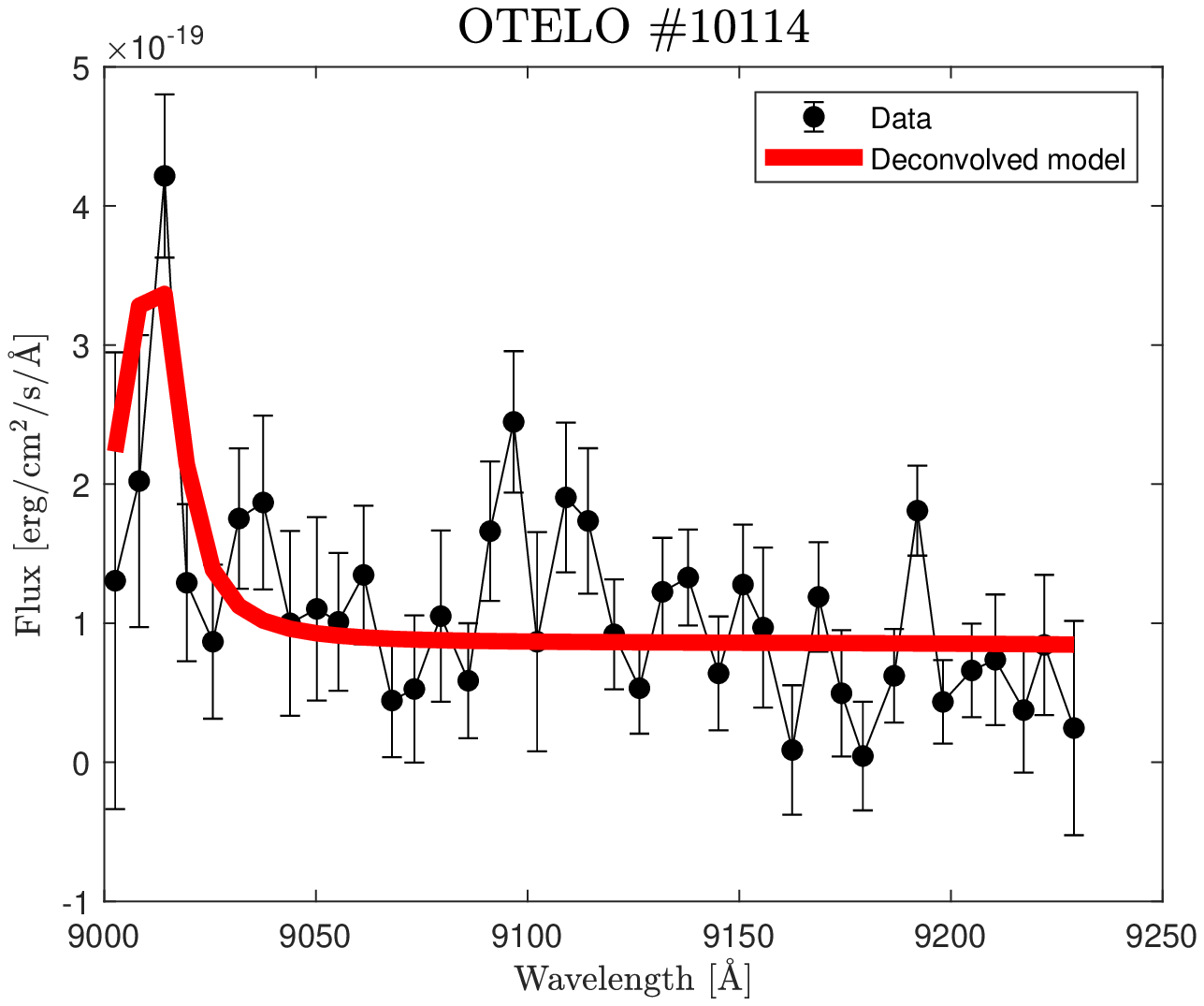}}\\
\subfloat{\includegraphics[width=.20\linewidth]{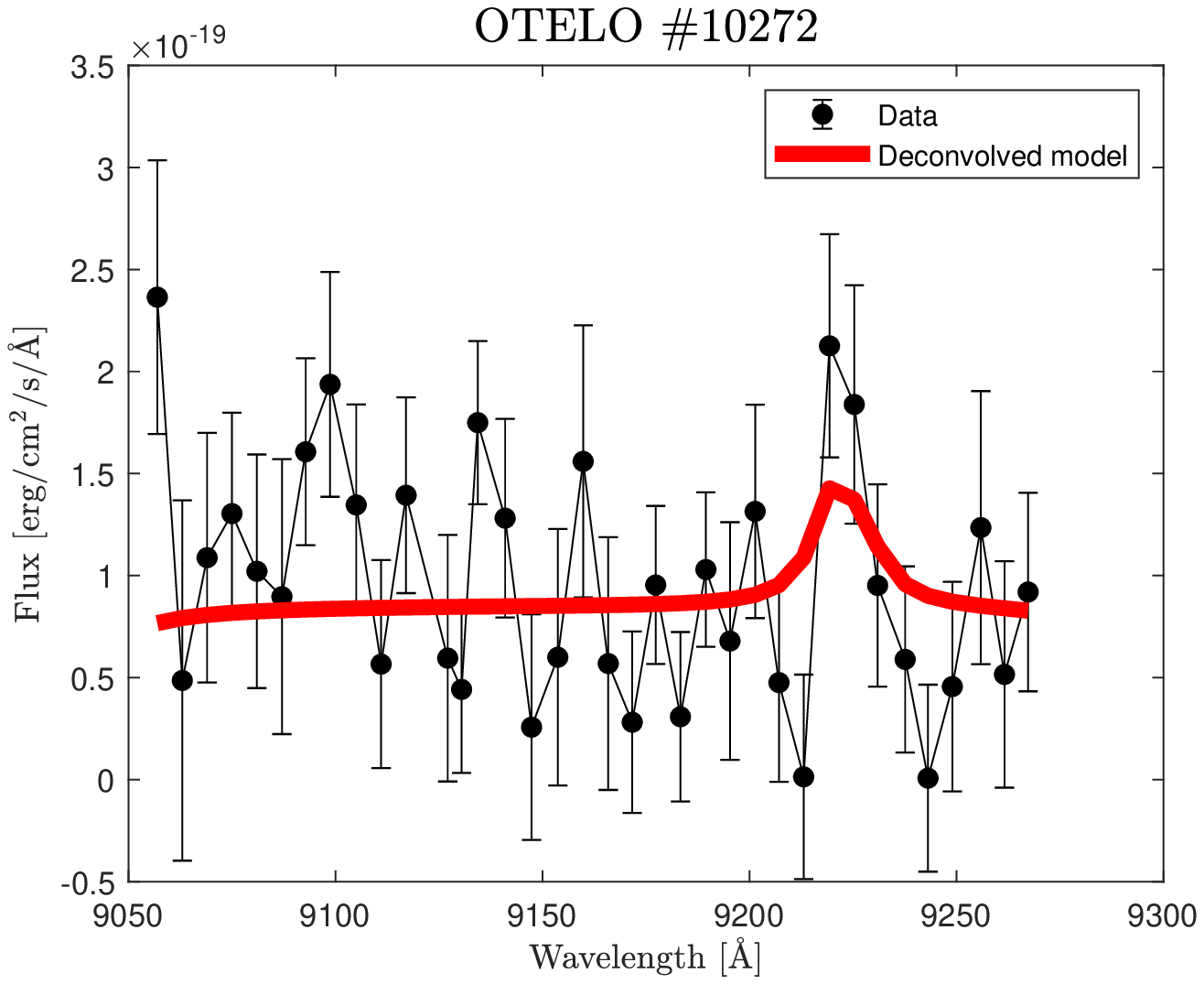}}
\subfloat{\includegraphics[width=.20\linewidth]{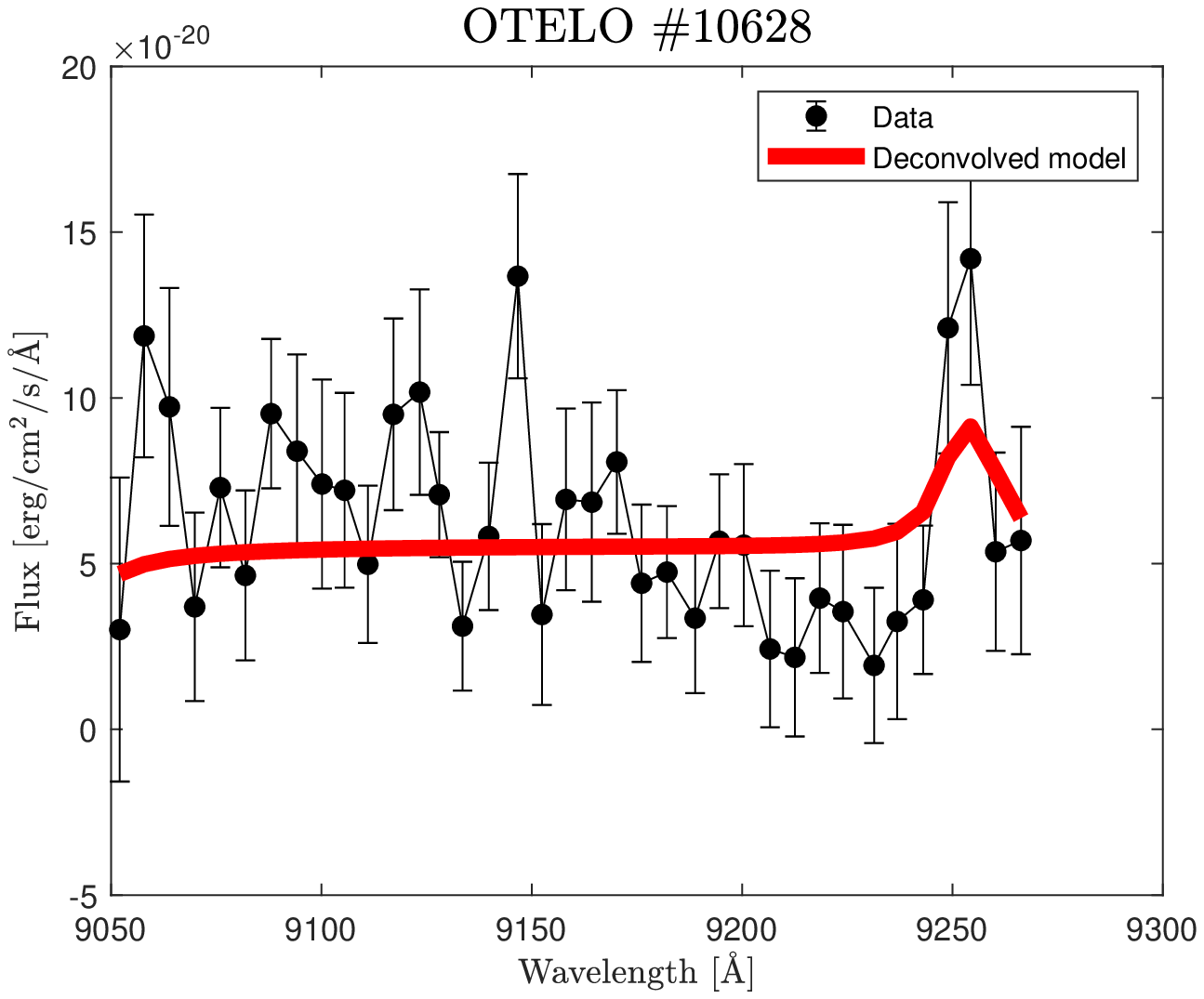}}
\subfloat{\includegraphics[width=.20\linewidth]{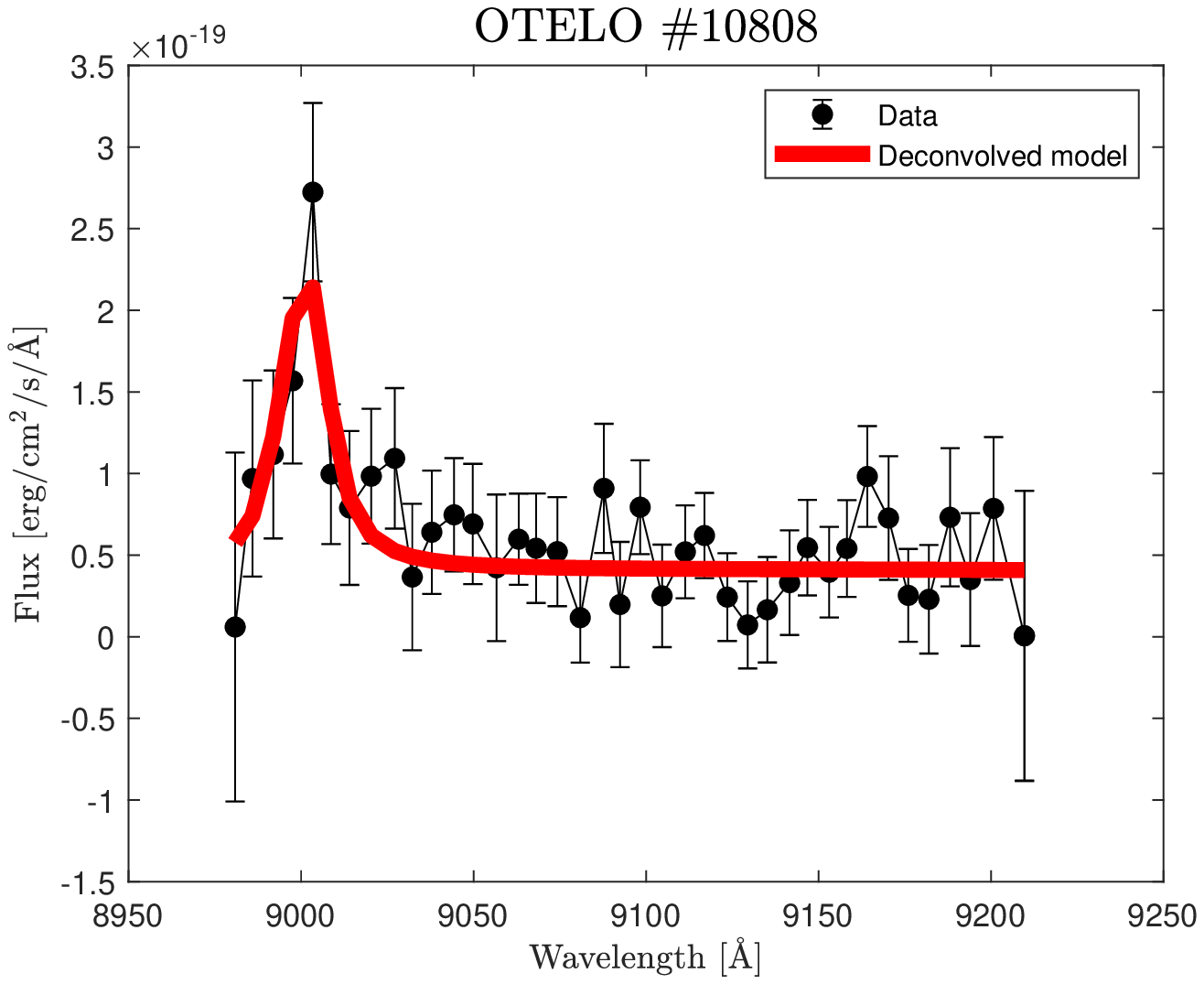}}
\subfloat{\includegraphics[width=.20\linewidth]{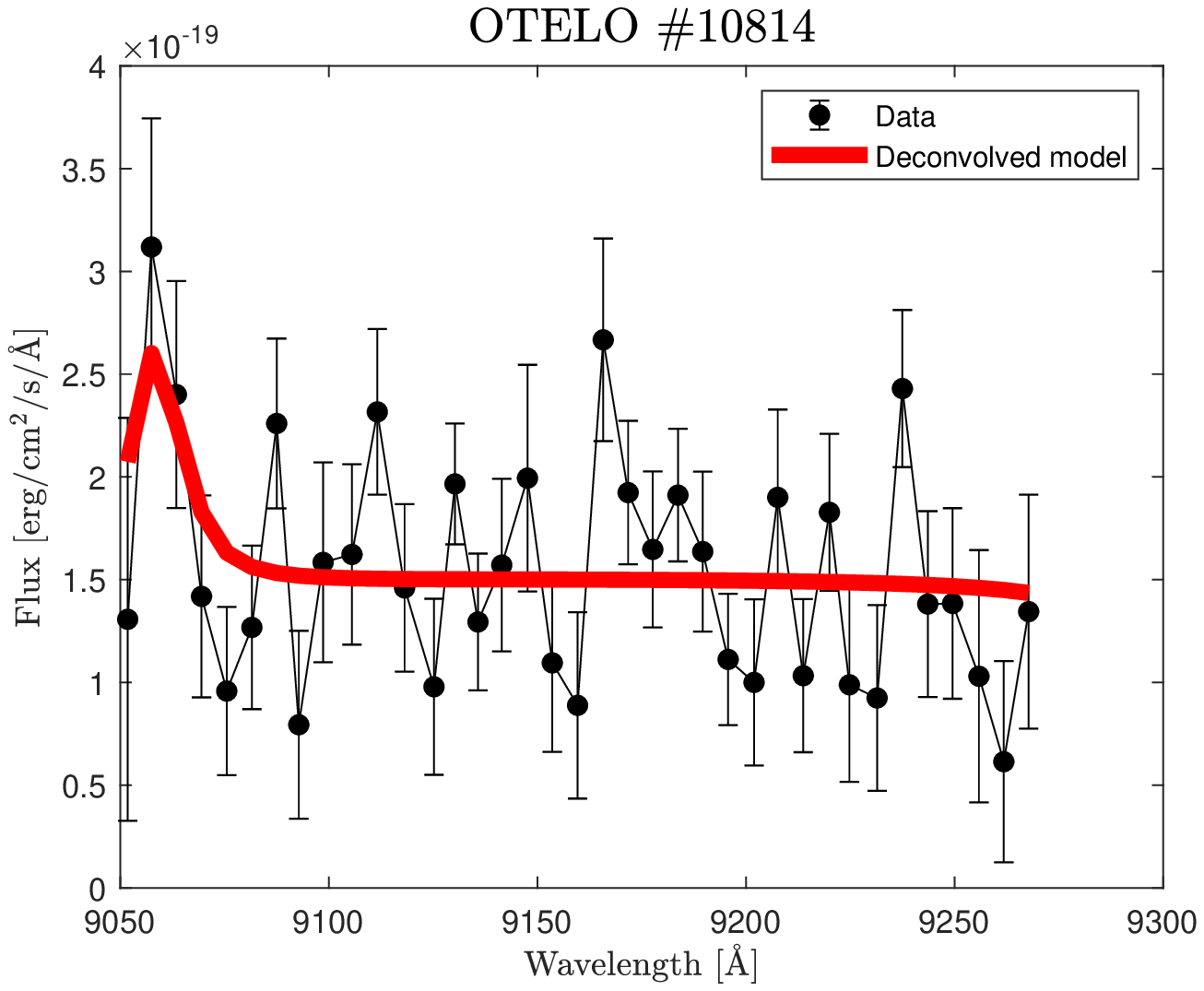}}
   \caption{Pseudo-spectra of the selected emitters. Black dots represent the measured pseudo-spectra, the red line is the best fitted deconvolved spectra.}
\end{figure*}

\end{document}